\documentclass{article}

\usepackage{arxiv}

\usepackage[utf8]{inputenc} 
\usepackage[T1]{fontenc}    
\usepackage{hyperref}       
\usepackage{url}            
\usepackage{booktabs}       
\usepackage{amsfonts}       
\usepackage{nicefrac}       
\usepackage{microtype}      
\usepackage{caption} 
\usepackage{lipsum}		
\usepackage{graphicx}
\usepackage[numbers]{natbib}
\usepackage{doi}
\usepackage{amsmath,amsfonts,amssymb}
\usepackage{mathtools}
\usepackage{graphicx}
\usepackage{setspace}
\usepackage{tocloft}
\usepackage{soul}
\usepackage{svg} 
\usepackage{siunitx} 
\usepackage{longtable} 
\usepackage{appendix} 
\usepackage[auth-lg]{authblk} 
\usepackage{verbatim} 
\usepackage{colortbl} 
\usepackage{caption} 
\usepackage{gensymb} 
\usepackage{xcolor} 
\definecolor{c1}{rgb}{0.6,0,1} 

\DeclareMathOperator\erf{erf}

\title{Cardiac mechanics modeling: recent developments and current challenges}

\author[a,b]{Aaron L. Brown}
\author[c,*]{Ju Liu}
\author[b,d,f]{Daniel B. Ennis}
\author[a,b,e,f]{Alison L. Marsden}

\affil[a]{Department of Mechanical Engineering, Stanford University, Stanford, CA, USA}
\affil[b]{Stanford Cardiovascular Institute, Stanford, CA, USA}
\affil[c]{Department of Mechanics and Aerospace Engineering, Southern University of Science and Technology, Shenzhen, Guangdong, China}
\affil[d]{Department of Radiology, Stanford University, Stanford, CA, USA}
\affil[e]{Department of Pediatrics (Cardiology), Stanford University, Stanford, CA, USA}
\affil[f]{Department of Bioengineering, Stanford University, Stanford, CA, USA}

\affil[*]{Corresponding author: liuj36@sustech.edu.cn}

\hypersetup{
pdftitle={Cardiac mechanics review paper},
pdfsubject={Cardiac mechanics},
pdfauthor={Aaron L. Brown},
pdfkeywords={Cardiovascular modeling, cardiac mechanics},
}

\begin{document}
\maketitle

\begin{abstract}
Patient-specific computational models of the heart are powerful tools for cardiovascular research and medicine, with demonstrated applications in treatment planning, device evaluation, and surgical decision-making. Yet constructing such models is inherently difficult, reflecting the extraordinary complexity of the heart itself. Numerous considerations are required, including reconstructing the anatomy from medical images, representing myocardial mesostructure, capturing material behavior, defining model geometry and boundary conditions, coupling multiple physics, and selecting numerical methods. Many of these choices involve a tradeoff between physiological fidelity and modeling complexity. In this review, we summarize recent advances and unresolved questions in each of these areas, with particular emphasis on cardiac tissue mechanics. We argue that clarifying which complexities are essential, and which can be safely simplified, will be key to enabling clinical translation of these models.
\end{abstract}


\section{Introduction} \label{sect:intro}

Patient-specific, multiphysics computational models of the heart hold the potential to revolutionize cardiovascular medicine by enabling the creation of digital twins—virtual replicas of an individual’s heart continuously updated with information from the physical twin \cite{thangaraj_cardiovascular_2024}. Cardiac digital twins can support clinicians in developing personalized treatment strategies tailored to each patient’s unique anatomy and physiology. Beyond clinical applications, digital twin technology offers powerful tools for advancing research, by guiding the design of next-generation therapies and devices and by deepening our understanding of disease mechanisms. This review outlines the major steps in constructing heart models, delves into open questions and current debates, and summarizes recent advancements, with a primary focus on cardiac tissue mechanics. 

To be clear about terminology, true digital heart twins, for which the bidirectional flow of information is essential, are rare in the literature. Most current efforts construct ``passive" digital twins \cite{coorey_health_2022}, or ``digital snapshots" \cite{sel_building_2024}, built from patient data, but not dynamically updated. These models nonetheless lay the essential groundwork for future true digital twins, while ongoing work is exploring continuous data collection through wearables and other sensors, as well as new frameworks for dynamic integration of data into the digital twin \cite{thangaraj_cardiovascular_2024, sel_building_2024}. For the remainder of this review, we use the term personalized computational heart model to refer to such precursor digital twins \cite{coorey_health_2022}, reserving the term digital twin itself for the aspirational tool of the future.

Computational cardiac modeling has already demonstrated significant utility in treatment planning, device evaluation, and surgical decision-making. Electromechanical simulations have been employed to evaluate novel pacing strategies and their effect on cardiac pump function, comparisons that are impractical to perform directly in clinical settings \cite{meiburg_comparison_2023}. In related work, patient-specific models were used to optimize cardiac resynchronization therapy (CRT) by simulating various lead placements and computing key biomarkers such as dP/dt\textsubscript{max}, ejection fraction, and stroke work \cite{capuano_personalized_2024}. In structural heart interventions, simulations have elucidated the mechanisms by which devices like the Parachute\textsuperscript{\textregistered} implant reduce end-diastolic myofiber stress and potentially reverse post-infarct remodeling \cite{lee_patient-specific_2014}. Additionally, hybrid experimental-computational platforms have supported the evaluation of mechanical circulatory support devices such as the Jarvik 2000 pump for single-ventricle circulations \cite{kung_hybrid_2019}. Virtual surgical planning using patient-specific CFD models has been applied to optimize graft geometry in coronary artery bypass surgery \cite{seo_computational_2022} and to assist in transcatheter aortic valve replacement planning \cite{dowling_first--human_2020}. Finally, even purely anatomical 3D models, without any underlying physics simulation, can aid surgical planning, for example when designing an intracardiac baffle in double outlet right ventricle cases \cite{vigil_modeling_2021}.

Beyond clinical applications, computational modeling provides a powerful framework for uncovering pathophysiological mechanisms that are otherwise difficult to probe. In hypertrophic cardiomyopathy (HCM), electrophysiology simulations have revealed that diffuse myocardial fibrosis significantly increases arrhythmogenic risk, suggesting that fibrosis burden may be a more informative predictor of sudden cardiac death than current clinical criteria \cite{ohara_personalized_2022}. Related work has evaluated multiple hypothesized contributors to impaired ventricular mechanics in HCM, including increased wall thickness, increased stiffness, decreased contractility, and fiber disarray \cite{kovacheva_causes_2021}. Multiscale modeling that incorporates metabolic function has revealed the subcellular mechanisms by which 2'-deoxy-ATP improves cardiac function in heart failure patients \cite{teitgen_multiscale_2024}. Multiphysics modeling has also been used to explore how atrial myopathy contributes to stroke, demonstrating that fibrosis-mediated impairment of left atrial motion alters blood flow and promotes stasis in the left atrial appendage \cite{gonzalo_multiphysics_2024}. In developmental cardiology, computational studies have shown that in embryonic zebrafish hearts, trabeculae enhance wall deformability, buffer fluid-induced stress, and structurally reinforce the myocardium \cite{cairelli_role_2024}. Finally, \emph{in silico} drug trials have begun to emerge as a promising application, with whole-heart models offering mechanistic insight into drug-induced electromechanical dysfunction \cite{sahli-costabal_classifying_2020, peirlinck_how_2022}.


\begin{figure} 
\centering 
\includegraphics[width=\linewidth]{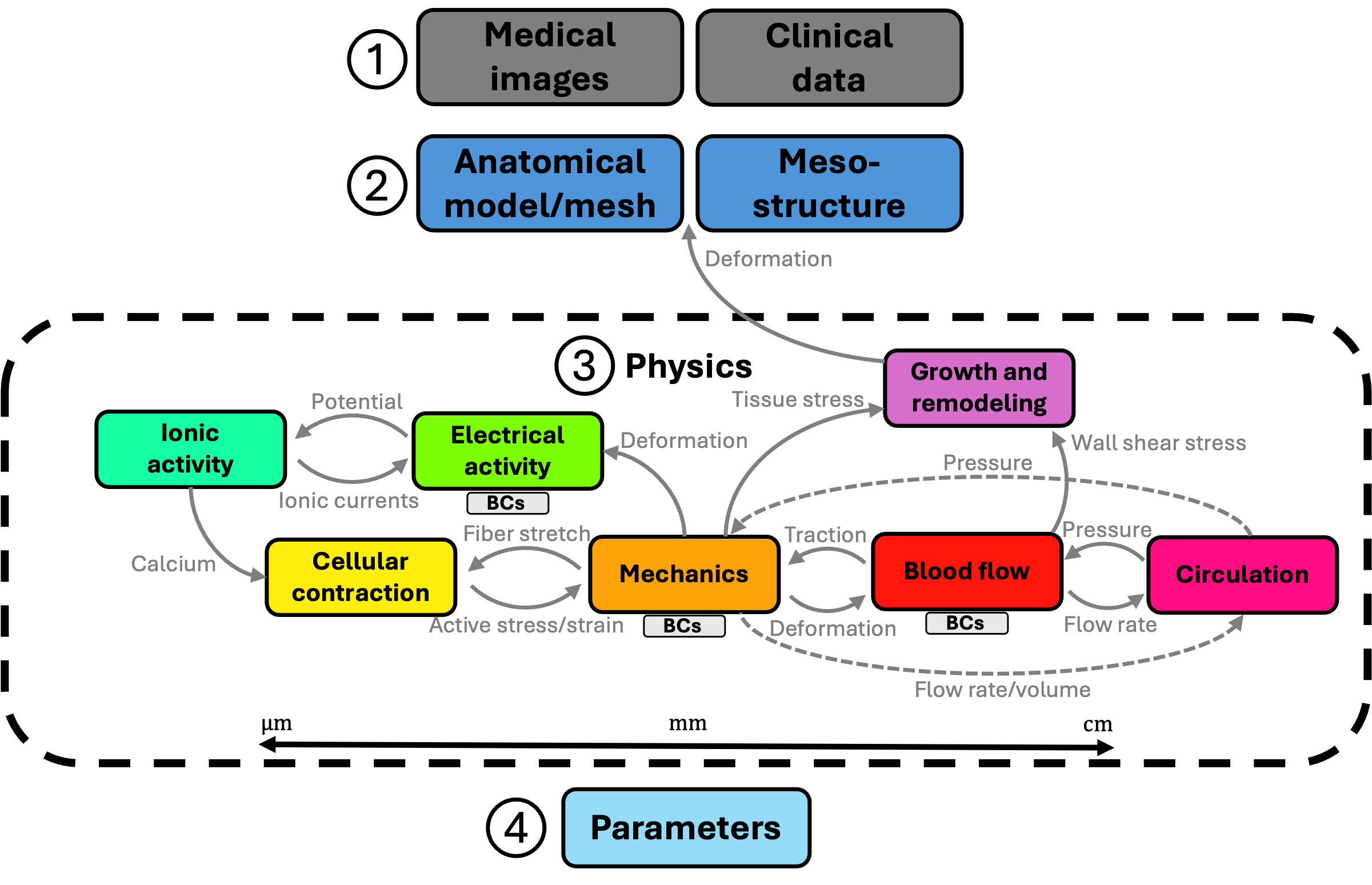} 
\caption{Summary of the major modeling considerations required when developing a personalized computational heart model. 1) The modeling process begins with medical images (CT or MRI), as well as other clinical data like an ECG or blood pressure measurements. 2) Next, a digital anatomic model of the heart is constructed by segmenting the medical image, then meshed in preparation for numerical simulation. The anisotropic mesostructure of the heart is also defined in this step. 3) Once the anatomical model and mesostructure are established, the relevant physics models must be selected and appropriately coupled, depending on the problem at hand. Accurate boundary conditions are essential. 4) Finally, the parameters of the computational model, which can number in the dozens, must be personalized to match patient data.} 
\label{fig:cardiac_modeling_summary} 
\end{figure}

The essential stages in constructing a personalized computational heart model are illustrated in Figure \ref{fig:cardiac_modeling_summary}. The process begins with a cardiovascular imaging exam using computed tomography (CT) or magnetic resonance imaging (MRI) to obtain comprehensive, often time-resolved anatomical images. Additional clinical data, such as electrocardiogram (ECG), blood pressure readings (either non-invasive cuff-based or invasive catheter-based), and ultrasound measures may also be collected (Figure \ref{fig:cardiac_modeling_summary} step 1). The CT or MRI images are segmented (in space and, if available, in time) to identify the myocardium, as well the four cardiac chambers -- left atrium (LA), right atrium (RA), left ventricle (LV), and right ventricle (RV) -- and other structures like the aorta and pulmonary arteries. Manual segmentation remains the gold standard, but machine learning has significantly accelerated this process \cite{kong_deep-learning_2021, alnasser_advancements_2024}. From the segmentation, typically a triangulated surface model is constructed, which is then volumetrically meshed using standard meshing software \cite{shi_personalized_2026, fedele_polygonal_2021} (Figure \ref{fig:cardiac_modeling_summary} step 2). A critical yet often overlooked decision involves determining the extent of the heart to include in the anatomical model; choices between left ventricular models, bi-ventricular models, and four chamber models, and around anatomical truncation and small-feature representation significantly impact both model fidelity and computational cost. The heart’s anisotropic mesostructure, commonly described as "myofibers" arranged in "sheetlets" \cite{wilson_myocardial_2022}, must also be accounted for, either directly from advanced cardiac MRI data \cite{dallarmellina_cardiac_2025} or, more commonly, using heuristic approaches known as rule-based methods \cite{piersanti_modeling_2021}. For simulations involving cardiac electrophysiology, it is also necessary to define the conduction system, particularly the Purkinje network, which plays a critical role in coordinated electrical activation \cite{sahli_costabal_generating_2016, barber_estimation_2021, camps_digital_2024}. Similarly, when modeling intracardiac blood flow, realistic representations of the heart valves are required. These can be derived directly from medical imaging data \cite{emendi_patient-specific_2021}, approximated using parametric geometries \cite{haj-ali_general_2012}, or constructed from first principles based on mechanical and physiological considerations \cite{kaiser_design-based_2021}.

Once the anatomical model and mesostructure are established, the relevant physics for a particular problem must be selected (Figure \ref{fig:cardiac_modeling_summary} step 3). The heart is inherently a multiphysics system. During a cardiac cycle, electrical signals propagate through the conduction system, triggering ionic exchanges that cause cellular contraction. This generates stress in the myocardial tissue, leading to mechanical deformation, which in turn drives blood flow within the heart and throughout the circulatory system. 
We emphasize that appropriate boundary conditions (BCs) for electrophysiology \cite{pagani_data_2021, africa_lifex-ep_2023}, tissue mechanics \cite{peirlinck_kinematic_2019, pfaller_importance_2019}, and blood flow \cite{vignon-clementel_outflow_2006} are critical for achieving physiological results. For some problems, it is also important to account for growth and remodeling (G\&R), which describes the gradual changes in cardiac anatomy and structure driven by long-term mechanical stimuli \cite{niestrawska_computational_2020}. Additional processes, such as perfusion of the myocardium \cite{michler_computationally_2013} and valve dynamics \cite{le_computational_2022}, may also be modeled. The implementation of these physics in a numerical framework presents significant technical challenges, especially when coupling different physical processes, and some physics may be ignored depending on the problem being studied.

Finally, the parameters of the model should be "personalized" with the goal of matching model outputs to patient-specific clinical data (Figure \ref{fig:cardiac_modeling_summary} step 4). Depending on the complexity of the model, there could be tens or hundreds of model parameters, ranging from electrical conductivity to constitutive model coefficients to mesostructural orientations. This can be extremely challenging, and much work has focused on novel optimization methods \cite{tonini_two_2025, shi_optimization_2024, strocchi_integrating_2025}, especially using neural network surrogates \cite{salvador_fast_2023, shi_heartsimsage_2025, caforio_physics-informed_2025, hofler_physics-informed_2025}, as well as sensitivity analyses \cite{levrero-florencio_sensitivity_2020, strocchi_cell_2023} and uncertainty quantification \cite{salvador_fast_2023, lazarus_sensitivity_2022, menon_personalized_2025, lee_probabilistic_2024}.

In addition to the modeling considerations discussed above, a cardiac modeler must also select appropriate numerical solvers and configure their settings; these choices significantly affect both accuracy and computational efficiency and should not be overlooked. Focusing on tissue mechanics, the first decision is whether to use finite element analysis (FEA), which remains the standard, or newer approaches such as isogeometric analysis (IGA) \cite{torre_isogeometric_2023} and meshless methods \cite{lluch_calibration_2020}. Within FEA, the element type (tetrahedral vs. hexahedral, linear vs. higher order) and the mesh resolution must be specified. For dynamic simulations, as opposed to static or quasistatic ones, one must also choose the time integration scheme (explicit vs. implicit) and time step size, which depend on the underlying physics and application. Additional considerations, including linear and nonlinear solvers, preconditioners, parallelization strategies, and multiphysics coupling schemes are also very important for efficient simulations. These issues are discussed in greater detail in Section \ref{sect:numerical_formulations}. We also refer the reader to a recent cardiac mechanics benchmark for further guidance \cite{arostica_software_2025}.

Several prior reviews have addressed different aspects of cardiac modeling. Avazmohammadi et al. (2019) \cite{avazmohammadi_contemporary_2019} reviewed constitutive models for cardiac mechanics, while Niederer et al. (2019) \cite{niederer_short_2019} explored models of active tension generation from the cellular to organ scale. Bracamonte et al. (2022) \cite{bracamonte_patient-specific_2022} examined inverse modeling techniques for patient-specific simulations, and Rodero et al. (2023) \cite{rodero_advancing_2023} discussed the current state of multiphysics modeling, focusing on clinical translation. Arzani et al. (2022) \cite{arzani_machine_2022} offered perspectives on the growing use of machine learning in physics-based models of cardiovascular biomechanics. This review will differ by focusing on recent developments and unresolved challenges, particularly in cardiac mechanics, that we believe are important to improving the physiological fidelity and clinical relevance of these models.

The subsequent sections explore the following topics. We begin with recent advances in patient-specific anatomical model construction, highlighting the use of machine learning to enhance efficiency (Section \ref{sect:anatomic_model_construction}). This is followed by a review of myocardial mesostructure, including current strategies for its integration into computational models and the open questions that remain (Section \ref{sect:myocardial_mesostructure}). We then turn to active and passive constitutive modeling of myocardium, with a focus on viscoelasticity, compressibility, and myofiber dispersion (Section \ref{sect:constitutive_modeling}). Next, we examine commonly-used cardiac geometries, the interplay among different cardiac structures, and the importance of mechanical boundary conditions (Section \ref{sect:anatomy_and_bcs}). We then review the benefits and challenges of multiphysics modeling, with a focus on the interactions among mechanics, electrophysiology, hemodynamics, and circulatory dynamics (Section \ref{sect:multiphysics_coupling}). Subsequently, we discuss important numerical challenges and considerations in cardiac multiphysics modeling (Section \ref{sect:numerical_formulations}). Finally, we conclude by highlighting additional important areas of research, including machine learning techniques to accelerate simulation, growth and remodeling, and patient-specific modeling, and offer our perspective on future directions for the field (Section \ref{sect:conclusions}).


\paragraph{Main takeaways}
\begin{itemize}
    \item \textbf{Anatomical model construction}: Creating accurate patient-specific anatomical heart models from medical images is labor-intensive when done manually. Machine learning methods appear essential for constructing models on clinically relevant timescales.

    \item \textbf{Myocardial mesostructure}: Myocardial mesostructure has a major impact on simulated mechanics, electrophysiology, and global function. Existing approaches oftentimes oversimplify the problem, which is complicated by inconsistencies in the literature. \emph{In vivo} cardiac diffusion tensor MRI (cDTI) offers a patient-specific alternative, especially in pathological cases, though the technology still requires significant refinement.

    \item \textbf{Constitutive modeling}: Myocardial viscosity, compressibility, and fiber dispersion are well-supported in theory and modeling, but more work need to be done to establish their importance to cardiac function and modeling.

    \item \textbf{Anatomical model and boundary conditions}: Mechanical interactions among heart structures, as well as between the heart and its surroundings, influence simulation results. Boundary conditions are also a critical modeling component and, while they have been studied extensively, further research is needed for greater physiological accuracy.

    \item \textbf{Multiphysics coupling}: The heart is inherently a multiphysics and multiscale organ, and high-fidelity models should ideally capture these complexities. However, such models are difficult to build and tune, and simpler models are often adequate for specific clinical or research applications.

    \item \textbf{Numerical considerations}: Numerical solvers underpin the entire modeling pipeline, with the finite element method most commonly used. A rich body of work exists on improving simulation efficiency, including developments in discretization, preconditioners, and multiphysics coupling. Many cardiovascular-specific solver platforms are now available.

    \item \textbf{Future outlook}: Much of the computational foundation for cardiac digital twins is already in place, but demonstrating their clinical value remains the next major challenge, particularly with regard to validation and regulatory acceptance. Tackling this will expose technical and practical limitations in current models, while also clarifying which complexities are truly necessary for clinical impact and which can be simplified or accelerated with modern tools like machine learning.
\end{itemize}
    
\section{Anatomical model construction} \label{sect:anatomic_model_construction}

\begin{figure} \centering \includegraphics[width=1.0\linewidth]{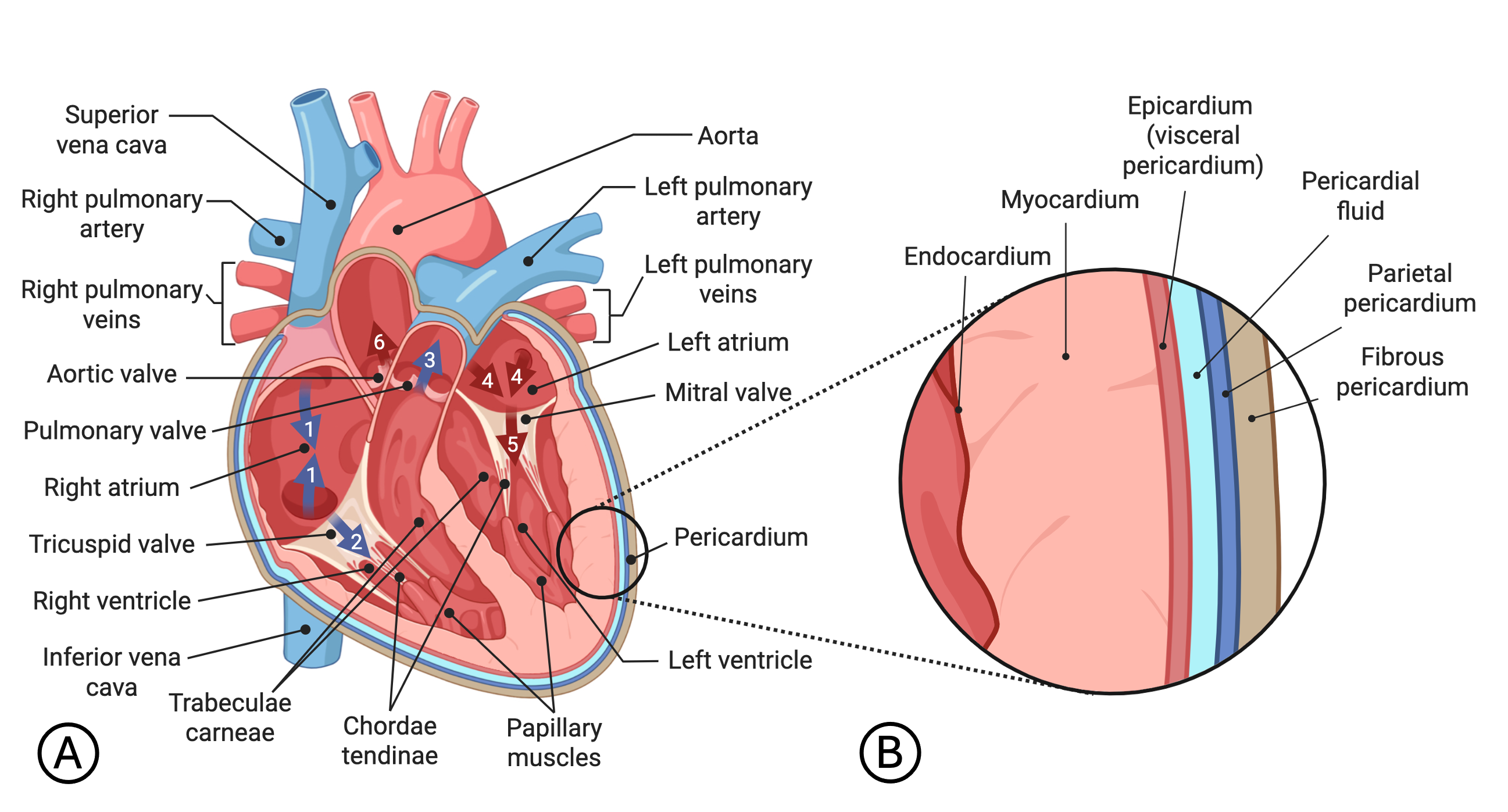} \caption{A) Major structures of the heart. B) Layers of the heart wall. Created in  \url{https://BioRender.com}. Adapted from ``The Heart: From the Organ to the Cell" template by Vinay Kara and Sally Kim on BioRender.} \label{fig:heart_anatomy} 
\end{figure}

The first step in any cardiac mechanics simulation is the development of a geometric model of the domain of interest. Before exploring specific techniques for constructing these models, we review the basic anatomy of the heart, as a thorough anatomical understanding underpins accurate model generation. An illustration of the heart with an inset showing the layers of the heart wall is given in Figure \ref{fig:heart_anatomy}.

Blood flow through the healthy heart follows a well-defined unidirectional sequence (Figure \ref{fig:heart_anatomy}A), beginning with systemic venous return from the superior and inferior vena cava into the RA (step 1). From the RA, blood crosses the tricuspid valve (TV) (step 2), filling the RV with blood (diastole). During RV contraction (systole), the TV closes to prevent backflow, and the pulmonary valve (PV) opens, allowing deoxygenated blood to be pumped through the left and right pulmonary arteries into the lungs (step 3). After gas exchange occurs in the pulmonary capillary network, oxygenated blood returns via the pulmonary veins to the LA (step 4). Blood then flows across the mitral valve (MV) (step 5), filling the LV (diastole). During LV contraction (systole), the MV closes and the aortic valve (AV) opens, and oxygenated blood is pushed into the aorta and throughout the body (step 6).

Each of the four cardiac valves opens and closes passively in response to pressure gradients across the chambers and vessels. The atrioventricular valves (MV and TV), located between each atrium and ventricle, are anchored to the myocardium by chordae tendineae and papillary muscles, which prevent leaflet prolapse under high systolic pressure. The semilunar valves (PV and AV) control outflow from the ventricles into the pulmonary artery and aorta, respectively. Systemic arterial pressures (approximately 120/80 mmHg \cite{luscher_what_2018}) are markedly higher than pulmonary pressures (approximately 20/10 mmHg \cite{mebazaa_acute_2004}), and the LV correspondingly has a much thicker myocardial wall than the RV \cite{matsukubo_echocardiographic_1977}. The LA and RA also have relatively thin walls and they provide a supporting function to ventricular pumping, comprised of three distinct phases: reservoir, conduit, and booster pump \cite{fedele_comprehensive_2023}. An additional feature of the heart is the so-called trabeculae carneae — a network of spongy, muscle-lined ridges found on the inner surfaces of the ventricles.

The heart wall consists of several distinct layers (Figure \ref{fig:heart_anatomy}B). The endocardium lines the interior surfaces of the heart chambers. Outside of this is the myocardium, a thick, muscular layer composed of a continuously branching syncytium of cardiomyocytes (contractile heart muscle cells) that perform the work of contraction. Surrounding the myocardium is the epicardium, which forms the outermost layer of heart tissue. The entire heart is enclosed in a stiff, fibrous pericardial sac. The epicardium (or visceral pericardium) is considered the inner layer of the pericardium. Between the epicardium and the parietal pericardium lies the pericardial cavity, which contains a thin layer of pericardial fluid. The outermost layer, or fibrous pericardium, is tethered to structures surrounding the heart, such as the lungs, diaphragm, sternum, ribs, aorta, and esophagus. As discussed in Section \ref{sect:anatomy_and_bcs}, the pericardium plays a significant role in constraining the motion of the heart \cite{pfaller_importance_2019}.

Early computational heart models relied on simplified, idealized geometries, which offered a convenient means to study ventricular function. For example, simplified cylindrical or elliptical geometries were used to approximate the LV, while combined elliptical models for LV and RV have also been developed \cite{peirlinck_precision_2021}. Other groups have made great progress using anatomically detailed, generic heart models. The Zygote heart \cite{zygote_media_group_inc_zygote_2014}, in particular, is commonly used \cite{fedele_comprehensive_2023, baillargeon_living_2014, alfonso_santiago_fully_2018}. Publicly available cohorts of heart models are also available \cite{strocchi_publicly_2020}. One-dimensional models that capture parameters like sarcomere length have also been applied for studies where only limited geometric representation is needed \cite{arts_adaptation_2005}. Although such idealized, generic, or reduced order models are beneficial for parametric studies and benchmarking simulations \cite{arostica_software_2025}, the morphology of the heart significantly influences its function (especially in pathological cases), and these models lack the anatomical precision needed for personalized heart simulations. 

Personalized computational heart models should instead incorporate detailed anatomical features obtained from medical imaging. Typically, this process begins with acquiring three-dimensional computed tomography (CT) or magnetic resonance imaging (MRI) scans, followed by segmentation to identify the myocardium and other structures. The gold standard remains manual segmentation, which requires an expert to manually label anatomical structures within the 3D image stack. The segmentation is then ``cleaned up" by removing features like trabeculae and papillary muscles, structures not usually included in the anatomical models due to their complexity. This process also involves smoothing to mitigate noise and artifacts introduced during segmentation. The resulting binary labeled image is then usually transformed into a surface mesh, followed by volumetric meshing to create a model suitable for finite element simulations \cite{shi_personalized_2026, fedele_polygonal_2021}, although the segmentation can also be directly meshed \cite{strocchi_semi-automatic_2024}. Volumetric meshing of complex geometries is research field in and of itself, but in most cases meshing can be done using established software libraries such as TetGen \cite{si_tetgen_2015}, the Computational Geometry Algorithms Library (CGAL) \cite{the_cgal_project_cgal_2024}, Gmsh \cite{geuzaine_gmsh_2009}, and Meshtool \cite{neic_automating_2020}. This standard pipeline is time-consuming and becomes increasingly challenging if multiple models must be generated from time-dependent data or for large patient cohorts \cite{strocchi_semi-automatic_2024}.

Given these limitations, recent advances in machine learning have aimed to streamline the geometric model construction process. These automated methods accelerate segmentation, enabling heart mesh generation in seconds versus hours, with virtually no decrease in accuracy compared to manual segmentations \cite{pak_patient-specific_2024}. In early work by \cite{zheng_four-chamber_2008}, the authors developed an automatic four-chamber segmentation method using machine learning to first identify the position and orientation of the heart, then estimate the non-rigid deformation of control points to match the image. \cite{romaszko_neural_2021} used a deep convolutional neural network (CNN) approach to directly generate LV meshes from MRI images. To reduce the learning challenge, they first performed principal component analysis (PCA) on a large population of LV meshes to construct a low-dimensional feature space. Then, they trained two CNNs: the first learned to produce segmentations of the LV myocardium and blood pool from images, and the second learned to generate LV geometries (encoded in the PCA space) from the segmentations. \cite{kong_learning_2023} used CNNs and graph convolutional networks (GCNs) to automatically generate whole-heart meshes from medical images by learning to deform a small number of control points on a template heart surface mesh. To enhance mesh quality for computational fluid dynamics (CFD) simulations, they also introduced novel mesh regularization loss terms. In subsequent work, \cite{narayanan_linflo-net_2024} refined the deformation process by instead learning a flow vector field that drives the mesh deformation. This modification ensures a diffeomorphic transformation and reduces the likelihood of mesh self-intersections. \cite{pak_patient-specific_2024} developed a similar deformation-based deep learning method capable of generating high quality FEA-ready volumetric meshes, not only of myocardium but even valve leaflets. Most of these machine learning models have been trained on datasets of normal anatomies, and pathological cases can present additional complexities. For example, congenital heart diseases (CHDs) often involve substantial variations not only in heart morphology but also topology; for instance, a ventricular septal defect (VSD), where a hole in the interventricular septum connects the left and right ventricles, transforms two cavities into one. Recognizing these challenges, \cite{kong_sdf4chd_2024} recently developed specialized machine learning methods based on signed distance fields to produce simulation-ready meshes for patients with CHD. 

To summarize, the heart’s complex anatomy presents significant challenges for computational modeling, motivating early studies to rely on idealized or generic geometries. Traditional workflows for patient-specific model generation, based on manual segmentation and meshing from medical images, are labor-intensive and time-consuming. However, recent advances in machine learning have rapidly transformed this landscape, enabling highly accurate, automated mesh generation. As interest in these techniques continues to grow, we should only expect further improvements in speed, accuracy, and applicability to increasingly complex anatomies and diverse patient populations.

We close this section by noting that, despite recent advances, \textit{in vivo} imaging cannot directly capture the unloaded configuration of the heart or its residual stresses, both of which influence mechanical behavior \cite{hadjicharalambous_investigating_2021}. Many studies simply assume an imaged configuration as the reference state, typically during diastole \cite{strocchi_simulating_2020, wang_modelling_2009, augustin_computationally_2021}, when pressures are relatively low. This is a simplification, however, and several methods have been proposed to approximate the unloaded configuration \cite{shi_optimization_2024, barnafi_reconstructing_2024}, often relying on iterative procedures. An alternative strategy is to compute a prestress field that balances the observed \textit{in vivo} pressure \cite{pfaller_importance_2019, hirschvogel_monolithic_2017}. Yet even these approaches remain incomplete, as they typically neglect residual stresses, which are likely present and may play an important role in cardiac mechanics \cite{pfaller_importance_2019, hadjicharalambous_investigating_2021}.

\section{Myocardial mesostructure} \label{sect:myocardial_mesostructure}

The myocardium, which constitutes the ``functional tissue'' \cite{holzapfel_constitutive_2009} of the heart, is anisotropic owing to a continuously branching syncytium of cardiomyocytes embedded in a network of collagen and elastin \cite{wilson_myocardial_2022}. Accurately accounting for the myocardial mesostructure is essential because it influences both cardiac deformation and the propagation of electrical signals \cite{shi_optimization_2024, eriksson_influence_2013, holz_transmural_2023, rodriguez-padilla_impact_2022}. Myocardial mesostructure is extensively characterized in both animal models and humans \cite{lombaert_human_2012}, using techniques such as histology, confocal microscopy, cardiac diffusion tensor MRI (DTMRI or cDTI), and computed tomography (CT). These imaging methods primarily quantify the local orientation of cardiomyocytes, which defines the “grain” of the myocardium. Additionally, cardiomyocytes tend to form clusters often referred to as “sheets,” “sheetlets,” “lamellae," or “myocardial aggregates” \cite{wilson_myocardial_2022, anderson_three-dimensional_2009, agger_assessing_2020}.

In cardiac mechanics simulations, myocardial mesostructure is typically modeled at the continuum scale using a local orthonormal coordinate system, with directions labeled as fiber ($\mathbf{f}$), sheetlet ($\mathbf{s}$), and sheetlet-normal ($\mathbf{n}$) \cite{legrice_laminar_1997}. In this framework, $\mathbf{f}$ represents the local orientation of cardiomyocytes, $\mathbf{n}$ is a vector normal to the myocardial sheetlets, and $\mathbf{s}$ is a vector in the plane of the sheetlet orthogonal to both $\mathbf{f}$ and $\mathbf{n}$ in the undeformed or reference configuration. In a computational model, $\mathbf{f}$ determines the local direction of cardiomyocyte contraction (and if simulating electrophysiology, the preferred electrical propagation direction), while $\mathbf{f}$, $\mathbf{s}$, and $\mathbf{n}$ may all factor into the passive constitutive and active contraction models. The specifics of passive constitutive and active contraction modeling are discussed in later sections; here, we focus on mesostructure and its integration into computational mechanics models.

\begin{figure}
    \centering
    \includegraphics[width=0.9\linewidth]{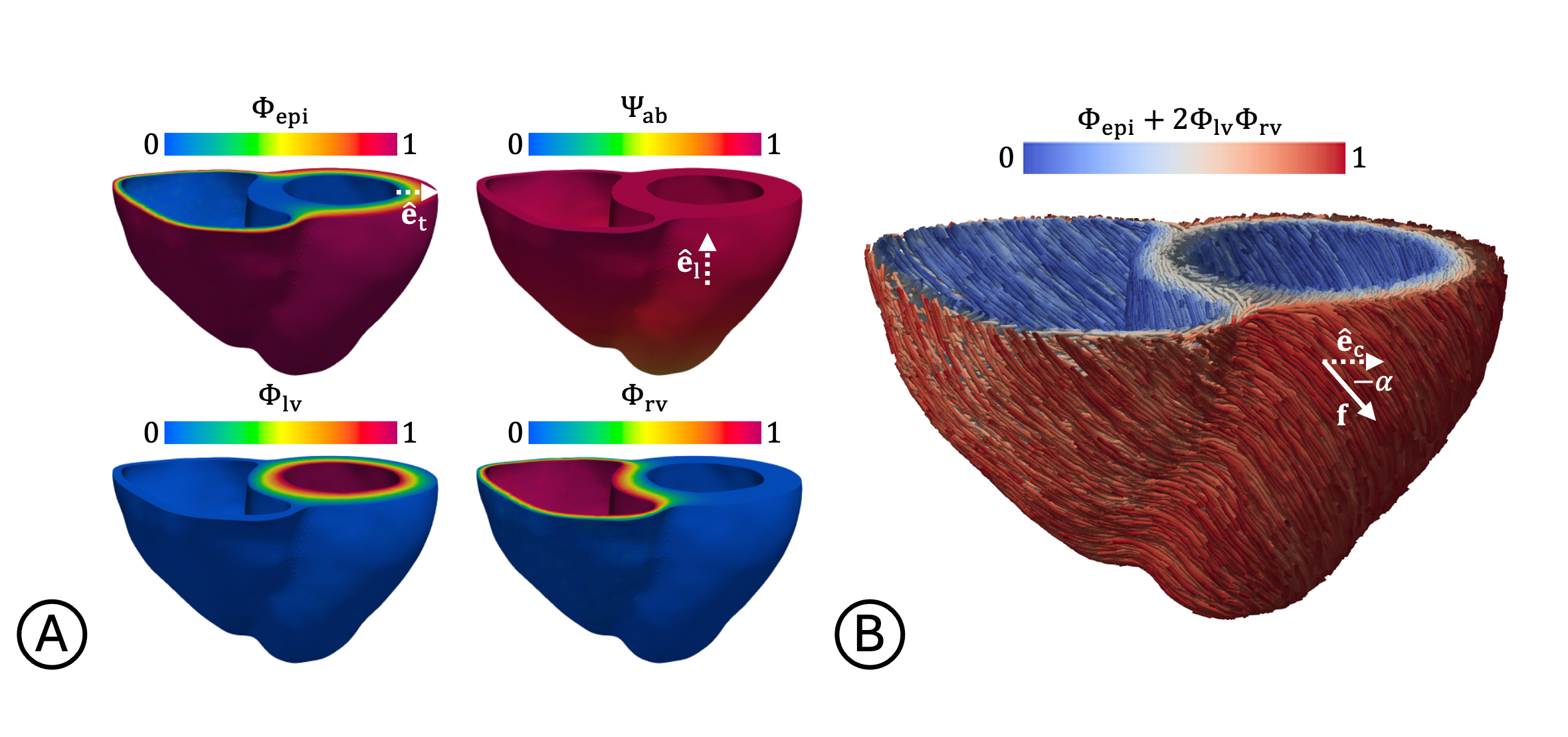}
    \caption{Using a Laplace-Dirichlet rule-based method (LDRBM) \cite{bayer_novel_2012} to generate myocardial mesostructure on a truncated biventricular computational model. A) Visualization of the four Laplace fields required in this method. Each is generated by solving Laplace's equation with Dirichlet boundary conditions on prescribed surfaces. $\Phi_\mathrm{epi}$ parameterizes the transmural depth, and its gradient defines the local transmural direction $\hat{\mathbf{e}}_\mathrm{t}$ (in the free walls). $\Psi_\mathrm{ab}$ parameterizes the apex-to-base distance \protect\footnotemark, and its gradient defines the local apicobasal or longitudinal direction $\hat{\mathbf{e}}_\mathrm{l}$. $\Phi_\mathrm{lv}$ and $\Phi_\mathrm{rv}$ are additional fields used in conjunction to parameterize the septum. B) Streamlines of the $\mathbf{f}$ field with a linear transmural variation from $\alpha = +60^\circ$ on the endocardium to $\alpha = -60^\circ$ on the epicardium, relative to the drawn local circumferential vector $\hat{\mathbf{e}}_\mathrm{c}$. The coloring by $\Phi_\mathrm{epi} + 2\Phi_\mathrm{lv}\Phi_\mathrm{rv}$ is chosen to aid the eye in visualizing the streamlines.}
    \label{fig:LDRBM}
\end{figure}

Experimental studies have revealed characteristic patterns in the mesostructure of normal human myocardium \cite{lombaert_human_2012}. In the ventricles, the fiber direction $\mathbf{f}$ varies through the wall thickness. Relative to the circumferential direction (which points counter-clockwise when viewed from the base of the heart), $\mathbf{f}$ shifts from a positive helix angle $\alpha$ on the endocardial surface (i.e., oriented toward the base) to a negative angle on the epicardial surface (i.e., oriented toward the apex). See Figure \ref{fig:LDRBM}B. Standard ranges are, in the LV free wall, $+60^\circ$ to $-60^\circ$, and in the RV free wall, $+90^\circ$ to $-25^\circ$ \cite{piersanti_modeling_2021}; it should be noted, however, that these are nominal values and there are significant variations across studies, methods, and individuals. The transmural profile of $\mathbf{f}$ is often modeled as either linear or tangent-like \cite{wilson_myocardial_2022, sommer_biomechanical_2015}. Recent observations suggest that the transmural variation is more complex in the interventricular septum, where two distinct fiber layers have been described \cite{rodriguez-padilla_impact_2022}. In the atria, the myocardial mesostructure is also more complex, featuring multiple distinct and overlapping fiber bundles \cite{pashakhanloo_myofiber_2016}. 

\footnotetext{The color gradient from apex to base is difficult to discern because $\Psi_\mathrm{ab}$ rises sharply from zero at the apex over a very short distance (not visible from this angle), reaching values close to one across most of the domain. This can be problematic in some applications. A common remedy, as in the universal ventricular coordinates framework \cite{bayer_universal_2018} is to smooth $\Phi_\textrm{ab}$ by mapping it onto a geodesic from apex to base.}

Based on these observations, ``rule-based methods'' (RBMs) have been developed to define myocardial mesostructure in arbitrary cardiac geometries without the need to incorporate generalized experimental data or patient-specific measurements. These methods define $\mathbf{f}$, $\mathbf{s}$, and $\mathbf{n}$ at any point in the domain based on empirical rules derived from normal myocardial mesostructural observations. Among the most widely used are Laplace-Dirichlet RBMs (LDRBMs), which employ numerical solutions to Laplace's equation to parameterize the myocardium. We illustrate the approach by applying the LDRBM of Bayer et al. \cite{bayer_novel_2012} on an example truncated biventricular geometry (Figure \ref{fig:LDRBM}). First, a local anatomic coordinate system is established at every point in the heart, consisting of radial or transmural ($\hat{\mathbf{e}}_\mathrm{t}$), apex-to-base or longitudinal ($\hat{\mathbf{e}}_\mathrm{l}$), and circumferential ($\hat{\mathbf{e}}_\mathrm{c}$) basis vectors.
The transmural direction is obtained by solving a Laplace problem with a Dirichlet boundary condition of zero on the endocardial surface and one on the epicardial surface.

\begin{align*}
    &\nabla^2 \Phi_\mathrm{epi} = 0,
    \\
    &\Phi_\mathrm{epi} = 0 \hspace{10pt} \text{on} \hspace{10pt} \Gamma_{\mathrm{endo}},
    \\
    &\Phi_\mathrm{epi} = 1 \hspace{10pt} \text{on} \hspace{10pt} \Gamma_{\mathrm{epi}}.
\end{align*}

The resulting ``temperature'' field $\Phi_\mathrm{epi}$ (Figure \ref{fig:LDRBM}A) serves as a parameterization of transmural depth (in the free walls), with its gradient defining the transmural direction $\hat{\mathbf{e}}_\mathrm{t}$. Likewise, $\hat{\mathbf{e}}_\mathrm{l}$ is defined as the gradient of an apicobasal field, $\Psi_\mathrm{ab}$, obtained by solving a Laplace problem with zero at the apex and one at the base. The circumferential direction $\hat{\mathbf{e}}_\mathrm{c}$ is then defined as the vector orthogonal to both $\hat{\mathbf{e}}_\mathrm{t}$ and $\hat{\mathbf{e}}_\mathrm{l}$. Subsequently, a local mesostructural coordinate system $\{\mathbf{f}$, $\mathbf{s}$, $\mathbf{n}\}$ is obtained by prescribed rotations relative to the anatomic coordinate system. $\mathbf{f}$, for example, is obtained by rotating $\hat{\mathbf{e}}_\mathrm{c}$ by an angle $\alpha$ around $\hat{\mathbf{e}}_\mathrm{t}$. Note that $\{\mathbf{f}$, $\mathbf{s}$, $\mathbf{n}\}$ is not properly defined in the septum because $\Phi_\mathrm{epi}$ does not appropriately parameterize the septum. In the Bayer method, this is resolved by generating additional Laplace fields to distinguish the LV ($\Phi_\mathrm{lv}$) from the RV ($\Phi_\mathrm{rv}$), which are used to define a second mesostructural coordinate system that is accurate in the septum. The two systems are then blended using bidirectional spherical linear interpolation, which ensures smooth transitions in fiber orientations across the LV, RV, and septal junctions. Full methodological details are provided in Bayer et al. \cite{bayer_novel_2012}.

Initially based on observations of the LV mesostructure and proposed for a biventricular model truncated in the basal plane (below the valve annuli) \cite{bayer_novel_2012}, subsequent LDRBMs have been adapted to better represent the mesostructure in other regions of the heart, such as the ventricular septum, outflow tracts, right ventricle, and atria \cite{piersanti_modeling_2021, doste_rule-based_2019}, and to model the transmural variation of $\mathbf{f}$ more accurately \cite{holz_transmural_2021}. Several LDRBMs were considered under a uniform framework and compared in electrophysiology simulations \cite{piersanti_modeling_2021} and electromechanical simulations \cite{holz_transmural_2023}.

However, recent discussions have highlighted ambiguities regarding myocardial mesostructure that may impact the accuracy of cardiac mechanics simulations. The review by \cite{agger_assessing_2020} underscores a lack of consensus on methodologies for quantifying cardiomyocyte orientations, including inconsistencies around reporting either helical angles or their projected angles, which measure the angle between the circumferential direction and the \textit{projection} of $\mathbf{f}$ onto the plane tangent to the epicardial surface. The authors recommend reporting the helical angle without projection. They also suggest angles be measured relative to epicardial curvature rather than the heart’s long axis, and that myocardial sheetlet orientations be assessed using the normal vector $\mathbf{n}$. Additionally, although computational studies often assume that $\mathbf{f}$ lies within the wall plane, physiological evidence suggests this is not always accurate and the transmural component (i.e. imbrication angle) of $\mathbf{f}$ should be considered \cite{wilson_myocardial_2022, bovendeerd_determinants_2009}. Finally, experimental work suggests that cardiomyocytes may exhibit large-scale connectivity and assume a toroidal topology \cite{axel_probing_2014}, a feature that is not currently considered in RBMs and may be especially important at the base and apex of the ventricles (top and bottom of the torus). The impact of this structure on cardiac mechanics has been investigated in the recent work by Osouli et al. \cite{osouli_heart_2025}.

Another issue involves histological and DTMRI studies that have observed two populations of sheetlet directions at a given myocardial location \cite{kung_presence_2011}, challenging the common assumption of a single $\mathbf{s}$ direction in myocardial constitutive models. The implications of this structure, described as forming a ``herringbone'' pattern \cite{wilson_myocardial_2022}, on constitutive and cardiac mechanics modeling needs further investigation. Related to these issues is an apparent error in the definition of the sheetlet angle $\beta$ in the well-known Bayer LDRBM \cite{bayer_novel_2012}, which appears to have propagated through the cardiac modeling literature. While the figure in \cite{bayer_novel_2012} suggests that $\beta$ is associated with a counterclockwise rotation about $\mathbf{f}$, the provided equation implements a clockwise rotation. 

Due to these issues, we believe that care should be taken when applying mesostructural values from literature to ensure that the data and computational model are consistent. Collaborative efforts bridging experimental and computational researchers could help address these issues.

Given the importance of mesostructure in cardiac electromechanics, the ideal cardiac digital twin should incorporate patient-specific mesostructure, which is not realistically achievable with LDRBMs limited by only a few tunable angle parameters. Computational studies have identified significantly different results when comparing the smooth and spatially homogeneous LDRBM-generated mesostructures versus experimentally observed mesostructure \cite{eriksson_influence_2013, guan_effect_2020}. Moreover, current LDRBMs are based primarily on data from healthy cases, though patient-specific models would be most valuable for diseased hearts, where mesostructure often differs markedly due to congenital or acquired heart diseases and remodeling processes \cite{wilson_myocardial_2022, garcia-canadilla_complex_2018, ghonim_myocardial_2017, ho_anatomy_2009, ma_myofibril_2022, planinc_comprehensive_2021}. To provide some degree of personalization, Washi et al. \cite{washio_ventricular_2016, washio_using_2020} proposed a novel fiber optimization algorithm, in which fibers are locally reoriented based on the predominant direction of active tension generation during isovolumic contraction, but this remains to be validated. To achieve truly patient-specific myocardial mesostructures, \emph{in vivo} cardiac diffusion tensor MRI (DTMRI or cDTI) is under active investigation \cite{wilson_myocardial_2022, toussaint_vivo_2013, nielles_assessment_2017, moulin_probing_2020, phipps_accelerated_2021}, but faces several challenges, such as long acquisition times, limited spatial resolution, and the challenges associated with compensating for bulk cardiac motion \cite{rodero_advancing_2023, stimm_personalization_2022}. At the same time, the emergence of 3D histological methods and cDTI is shifting the field away from simplified angular descriptions of mesostructural features toward vector- and tensor-field representations, which can be mapped directly onto computational meshes \cite{guan_effect_2020, stimm_personalization_2022}.

\section{Constitutive modeling} \label{sect:constitutive_modeling}
        

Experimental mechanical testing on animal and human samples has revealed that the myocardium is a non-linear, anisotropic, viscoelastic, history-dependent material \cite{sommer_biomechanical_2015}. Early work assumed transverse isotropy, but the modern consensus is for orthotropy \cite{tueni_structural_2023}. Most material testing has focused on the ventricles, in particular the LV, although recent work has identified regional differences in myocardial properties \cite{kakaletsis_right_2021, nemavhola_study_2021}. Existing constitutive models generally contain an isotropic term accounting for the non-fibrous underlying matrix \cite{holzapfel_constitutive_2009}, with additional tensile stiffness along $\mathbf{f}$ and sometimes $\mathbf{s}$ or $\mathbf{n}$, as well as shear stiffness arising from interactions between $\mathbf{f}$, $\mathbf{s}$, and $\mathbf{n}$ \cite{avazmohammadi_contemporary_2019}. In the following, we briefly describe some prominent constitutive models for the myocardium. 
    
We first discuss the transversely isotropic Guccione model \cite{guccione_passive_1991, guccione_finite_1995}:
    \begin{equation}
        \psi_{G} = \frac{C}{2}(e^Q - 1),
    \end{equation}
    where
    \begin{equation}
        Q = 2b_1 (E_{ff}^2 + E_{ss}^2 + E_{nn}^2)
        + b_2 E_{ff}^2 
        + b_3 (E_{ss}^2 + E_{nn}^2 + E_{ns}^2 + E_{sn}^2)
        + b_4 (E_{fs}^2 + E_{sf}^2 + E_{fn}^2 + E_{nf}^2),
    \end{equation}
    and
    \begin{equation} \label{eq:guccione}
        E_{ab} = \mathbf{a} \cdot \mathbf{E} \mathbf{b}; \hspace{5pt} \mathbf{a}, \mathbf{b} \in \{\mathbf{f}_0, \mathbf{s}_0, \mathbf{n}_0\},
    \end{equation}
    where $\mathbf{f}_0, \mathbf{s}_0, \mathbf{n}_0$ are the fiber, sheetlet, and sheetlet-normal directions in the reference configuration. 
    The material parameters are $C$, which scales the stiffness, and $b_1$, $b_2$, $b_3$ and $b_4$, which control the degree of strain stiffening in tension, compression, and shear. 
    This has been used recently in \cite{strocchi_integrating_2025, frohlich_numerical_2023, shavik_computational_2021}.

The Usyk model \cite{usyk_effect_2000, usyk_computational_2002} is an orthotropic variation of the Guccione model: 
    \begin{equation}
        \psi_{U} = \frac{C}{2}(e^Q - 1),
    \end{equation}
    where
    \begin{equation} \label{eq:usyk}
        Q = b_{ff} E_{ff}^2 
        + b_{ss} E_{ss}^2
        + b_{nn} E_{nn}^2
        + b_{fs} (E_{fs}^2 + E_{sf}^2)
        + b_{fn} (E_{fn}^2 + E_{nf}^2)
        + b_{ns} (E_{ns}^2 + E_{sn}^2).
    \end{equation}
    $E_{ab}$ is defined as in the Guccione model, and
    $C$ and $b_{ij}$, $i,j \in \{f,s,n\}$ are material parameters with similar interpretations as in the Guccione model. This has been used recently in \cite{fedele_comprehensive_2023, augustin_computationally_2021, barnafi_reconstructing_2024}.
    
Finally, the orthotropic, invariant-based Holzapfel-Ogden (HO) model \cite{holzapfel_constitutive_2009} has been widely adopted and takes the form\footnote{Note that there is a typo in the original Holzapfel-Ogden paper. The isotropic term should read $e^{b(I_1 - 3)} - 1$, but $-1$ is missing from the original paper \cite{avazmohammadi_contemporary_2019, hirschvogel_monolithic_2017}. However, because only the derivative of $\psi$ with respect to the deformation, i.e., the stress, is physically meaningful, the omission is inconsequential. Also, while the original paper does not include a Heaviside function, it contains an equivalent statement that the fiber and sheetlet terms should be included in the strain energy only when $I_{4,f} > 1$ and $I_{4,s} > 1$, respectively.}

    \begin{equation} \label{eq:HO_model}
    \psi_{HO} = \frac{a}{2b}
    \Big( 
    e^{b(I_1 - 3)} - 1
    \Big)
    +
    \frac{a_{fs}}{2b_{fs}}
    \Big(
    e^{b_{fs}I_{8,fs}^2} - 1
    \Big)
    +
    \sum_{i \in \{f,s\}}
    \chi(I_{4,i})
    \frac{a_i}{2b_i}
    \Big(
    e^{b_i(I_{4,i} - 1)^2} - 1
    \Big),
\end{equation}
where the $a$ parameters scale the stiffness, and the $b$ parameters control the degree of strain stiffening. $\chi(x)$ is a (potentially smoothed) Heaviside function centered at $x = 1$, which enforces the assumption that fibers support stress in tension only \cite{shi_optimization_2024}. This has been used recently in \cite{arostica_software_2025, guan_effect_2020, alharbi_3d-0d_2024}.

In the literature, there are various modifications to these models, such as using isochoric strain or strain invariants and adding a volumetric energy term in a nearly incompressible formulation \cite{arostica_software_2025, eriksson_modeling_2013, gultekin_orthotropic_2016}. 
Another example is the work of \cite{nolan_robust_2014}, who showed that when modeling compressible materials, the use of full (rather than isochoric) anisotropic invariants in an anisotropic material model like the HO model is necessary to properly capture anisotropic volumetric deformations \cite{shi_optimization_2024, mcevoy_compressibility_2018}. We also mention the work of \cite{vaverka_modification_2025}, who modified the sheetlet term in the HO model with an invariant that measures the change in area of a myocardial sheetlet, demonstrating significant improvement in the quality of the fit to experimental tension and shear data.

For a review of myocardial constitutive models, see \cite{avazmohammadi_contemporary_2019}. Here, we focus on three developments in constitutive modeling that have not been fully integrated into standard practice: viscosity, compressibility, and fiber dispersion.

\subsection{Viscosity}

The models discussed above were primarily developed under the assumption of an elastic myocardium, with no consideration for viscoelastic behavior. Although increased myocardial viscosity has been associated with heart disease \cite{caporizzo_need_2021}, it was believed by many to be ``not important on the time scale of the cardiac cycle" \cite{holzapfel_constitutive_2009}. Additionally, the modeling of viscoelasticity has been hindered by a lack of comprehensive experimental data \cite{holzapfel_constitutive_2009}. Moreover, the mechanisms behind myocardial viscoelasticity remain debated, with various theories suggesting contributions from viscous effects related to blood or extracellular fluid flow through the porous myocardium, intrinsic viscoelasticity of cardiomyocytes, and molecular friction among extracellular matrix constituents, particularly collagen \cite{wang_viscoelastic_2016, nordsletten_viscoelastic_2021}. However, there has been a recent resurgence of interest in the viscoelastic properties of the myocardium and their potential relevance to cardiac function in both healthy and diseased states \cite{caporizzo_need_2021}.

In particular, \cite{sommer_biomechanical_2015} compiled a dataset from planar biaxial extension and triaxial simple shear experiments on excised human ventricular myocardium, using strain rate modulation and stress relaxation tests to quantify viscoelastic behavior. Their findings revealed significant hysteresis even under quasistatic loading, indicative of substantial energy dissipation within the tissue. This data was used to develop and calibrate several viscoelastic models of the myocardium, constructed as extensions of existing hyperelastic models; these are discussed in the following paragraphs. However, given the \emph{ex vivo} nature of these experiments and the state of the tissue, it remains unclear how important these effects are for \emph{in vivo} modeling efforts.

\cite{gultekin_orthotropic_2016} proposed a viscoelastic model for the myocardium based on a three-dimensional, nonlinear analogue of the generalized Maxwell model. This model combines multiple Maxwell elements (spring and dashpot connected in series) in parallel with hyperelastic components, using four Maxwell elements to account for viscoelasticity in the isotropic ground matrix ($m$), as well as fiber ($f$), sheetlet ($s$), and fiber-sheetlet ($fs$) components. The viscous overstress for each component $\alpha$ is given by the following convolution integral expression: 
\begin{equation} \label{eq:viscous_stress_Maxwell}
    \mathbf{S}_{visc, \alpha} = \int_0^t \exp\Big( - \frac{t-s}{\tau_\alpha} \Big) 
    \beta_\alpha \dot{\mathbf{S}}^\infty_{iso, \alpha}
     ; \hspace{10pt} \alpha \in \{m, f, s, fs\},
\end{equation}
where $\beta_\alpha$ are four scaling factors for the viscous strength, $\tau_\alpha$ are the corresponding four viscous relaxation times, and $\dot{\mathbf{S}}^\infty_{iso, \alpha}$ are the rates of change of the corresponding hyperelastic contribution to the stress.

These parameters were calibrated to experimental data from \cite{sommer_biomechanical_2015}. They also provided details on an efficient implementation of their model within a FEA framework. In test cases, the model generally captured biaxial extension data well, but it could not fully replicate the hysteresis shape in shear experiments and showed discrepancies in stress relaxation experiments, particularly at early times. In simulations of an idealized LV subjected to time-varying pressure loading, the model predicted transmurally varying viscoelastic responses, with significantly more pronounced hysteresis on the endocardial surface.

A simplified version of the generalized Maxwell model, using a single Maxwell element to represent the viscoelasticity of the isotropic ground matrix, was used in a four-chamber heart model by \cite{tikenogullari_how_2022}. Their parameter sensitivity study concluded that ``viscous relaxation has a negligible effect on the overall behavior of the heart,'' as assessed by left ventricular pressure-volume loops and myocardial fiber strain. Essentially, while changes in the viscous parameters significantly altered the mechanical response, the effect was primarily to increase stiffness, so that a stiffer hyperelastic material was able yield similar results. However, they observed that viscosity led to a cycle-to-cycle shift in pressure-volume loops, suggesting that ``relaxation is significant between cardiac cycles, even though it is not significant within an individual cycle." This was attributed to the relatively long viscous relaxation time (on the order of $O(10)$ seconds) compared to the short duration of a single cardiac cycle ($<O(1)$ second).

Nordsletten et al. \cite{nordsletten_viscoelastic_2021} employed a fractional viscoelastic approach, implementing a fractional Zener model \cite{zhang_comparative_2021}, which describes the viscoelastic stress via a fractional differential equation: 

\begin{equation} \label{eq:viscous_stress_fractional_Zener}
    \mathbf{S}_{ve} + \delta D_t^\alpha \mathbf{S}_{ve} = D_t^\alpha \mathbf{S}_v,
\end{equation}

where $D_t^\alpha$ is the Caputo derivative, defined as

\begin{equation}
    D_t^\alpha f = \frac{1}{\Gamma(1-\alpha)} \int_0^t (t-s)^{-\alpha} \dot{f}(s)ds,
\end{equation}

with $\alpha \in [0,1]$ representing the order of the derivative. In this formulation, $\mathbf{S}_v$ is given by an expression that mimics the elastic stress of the Holzapfel-Ogden (HO) model. While similar in form to the convolution integral in Eq. \eqref{eq:viscous_stress_Maxwell}, whereas Eq. \eqref{eq:viscous_stress_Maxwell} models only a few discrete relaxation timescales for viscosity, Eq. \eqref{eq:viscous_stress_fractional_Zener} encodes a continuous distribution of relaxation timescales, which better reflects the multiscale nature of myocardial viscoelasticity. The distribution's shape is controlled by the parameters $\alpha$ and $\delta$. The new model demonstrated superior fitting to experimental data compared to both traditional hyperelastic models and the Maxwell-based approach of \cite{gultekin_orthotropic_2016}. Furthermore, the identifiability of the 11 model parameters was successfully demonstrated.

\cite{zhang_simulating_2023} then incorporated this fractional viscoelastic model into a finite element framework, addressing several numerical challenges associated with solving the fractional differential equation \cite{zhang_efficient_2020}. Their model was applied to an idealized LV coupled with a 0D circulation model. Comparisons with a hyperelastic model (Holzapfel-Ogden) revealed that the viscoelastic model dampened spurious oscillations and reduced stroke work, along with other small but noticeable differences in LV deformation, fiber stress, and myocardial pressure. They argued that viscoelasticity plays its most important role during the relaxation phase of the cardiac cycle.

In addition to these advanced viscoelastic models, simplified approaches that approximate some viscous effects have also been applied in the literature. These include a viscous pseudopotential model \cite{pfaller_importance_2019, arostica_software_2025, chapelle_energy-preserving_2012} 

\begin{equation}
    \mathbf{S}_{visc} = \frac{\partial}{\partial \dot{\mathbf{E}}} \psi_{visc}, \hspace{10pt} \psi_{visc} = \frac{\mu}{2} \text{tr}(\dot{\mathbf{E}}^2),
\end{equation}

a Newtonian viscous model \cite{sugiura_ut-heart_2022}

\begin{equation}
    \mathbf{S}_{visc} = \mu J \mathbf{F}^{-1} (\nabla \mathbf{v} + \nabla \mathbf{v}^T) \mathbf{F}^{-T},
\end{equation}

and Rayleigh damping \cite{frohlich_numerical_2023, alfonso_santiago_fully_2018}, which adds a damping term

\begin{equation}
    \mathbf{C}(\mathbf{u}) = \alpha \mathbf{M} + \beta \mathbf{K}(\mathbf{u})
\end{equation}

to the time-space discretized elastodynamics equation $\mathbf{M} \ddot{\mathbf{u}} + \mathbf{C}(\mathbf{u})\dot{\mathbf{u}} + \mathbf{K}(\mathbf{u}) = \mathbf{R}$ \cite{lafortune_coupled_2012}.

In summary, while significant progress has been made in developing and applying viscoelastic models for the myocardium, the importance of viscous effects at the relevant timescales remains under debate and warrants further investigation.

\subsection{Compressibility}

Most constitutive models and the majority of cardiac mechanics studies assume that myocardium, like other biological tissues with high water content, is ``essentially incompressible'' \cite{holzapfel_constitutive_2009, guccione_passive_1991, yin_compressibility_1996, avazmohammadi_integrated_2017}. Early experimental studies have been cited to justify this assumption of myocardial incompressibility \cite{tsuiki_direct_1980, vossoughi_compressibility_1980}. However, it is worth noting that \cite{vossoughi_compressibility_1980} is not readily available through internet searches, and we were unable to locate the original publication. \cite{yin_compressibility_1996} tested the hypothesis that myocardial perfusion through the coronary circulation changes throughout the cardiac cycle due to variations in myocardial stiffness, leading to bulk compressible behavior. In their cyclic biaxial experiments on perfused, \emph{ex vivo} ventricular septa from dogs, they observed that the volume of myocardial tissue occupied by vasculature changed by 2-4\% depending on perfusion pressure. Using noninvasive MRI, \cite{rodriguez_noninvasive_2006} measured myocardial volume changes below 2\%. Based on the evidence at the time, \cite{bistoquet_myocardial_2008} concluded in 2008 that ``the total myocardial volume changes no more than 4\% during a cardiac cycle,'' supporting the assumption of near incompressibility.

However, recent investigations have challenged this assumption. In an \emph{in vivo} study of canine hearts, \cite{ashikaga_changes_2008} quantified myocardial volume changes across different layers, observing a 4.1\% systolic reduction in the sub-epicardial layer, 6.8\% in the midwall, and 10.3\% in the sub-endocardial layer. They further hypothesized the "existence of blood-filled spaces within the myocardium" that could directly exchange blood with the ventricular lumen, as the observed volume change exceeded what could be accounted for by coronary blood flow. Interestingly, they also noted that volume reductions in early activated myocardial regions occurred before the rise in chamber pressure, suggesting that these changes were due to myofiber contraction rather than compression caused by elevated chamber pressure.

In another study, \cite{mcevoy_compressibility_2018} conducted \emph{ex vivo} tension and compression experiments on cylindrical specimens of porcine myocardium. They found a 4\% volume change under uniaxial tension at a stretch of 1.3. In confined compression, they observed a highly nonlinear reponse, with a stress of 4 kPa at 5\% volume strain, rising to nearly 50 kPa at 10\% strain. To model these experimental results, they replaced the isotropic term in the Holzapfel-Ogden (HO) model (Eq. \eqref{eq:HO_model}) with a six-parameter Yeoh model, which accurately captured the myocardium's slightly compressible behavior. They also developed a representative volume element (RVE) consisting of cardiomyocytes, extracellular matrix material, and cylindrical voids representing capillaries. Simulation of this RVE under confined compression revealed that only 42\% of the total volume change could be attributed to capillary volume reduction, suggesting that the compressibility of the solid tissue itself must contribute significantly to myocardial volume change.

\cite{soares_modeling_2017, avazmohammadi_vivo_2020} used sonomicrometry to measure left ventricular free wall deformation in an ovine heart. In healthy animals, they observed significant myocardial compressibility, with important regional and transmural variations. On average, they recorded a 15\% systolic volume reduction, with up to 25\% reduction at the endocardial surface, and slightly greater compressibility at the apex. After myocardial infarction, which eliminates blood perfusion and tissue contraction, the infarcted region showed much smaller systolic volume reductions, and in some cases, the infarcted tissue even increased in volume during systole. In a follow-up study, \cite{liu_impact_2021} compared compressible and incompressible models in organ-level cardiac mechanics simulations. Based on their experimental findings, they characterized the myocardium as nearly incompressible during diastole but compressible during systole. To account for this, they proposed a material model in which the volumetric penalty parameter varies as a negative linear function of the active stress magnitude, allowing for compressibility during systole when active stress is maximized \cite{soares_modeling_2017}. Their simulations demonstrated that incorporating compressibility improved the agreement with experimental data, particularly showing that an incompressible model led to exaggerated systolic wall thickening, reduced LV longitudinal shortening, and a much higher active stress required to match the experimental pressure-volume loop.

Myocardial compressibility was also shown to be reduced in patients with heart failure with reduced ejection fraction (HFrEF). \cite{kumar_cardiac_2021} used cardiac MRI and transthoracic echocardiography (TTE) to measure myocardial volume changes. They found that in healthy individuals, systolic myocardial volume was 87\% or 79\% of the diastolic volume, based on MRI and TTE measurements, respectively. In contrast, MRI data from HFrEF patients indicated that systolic myocardial volume equaled diastolic volume, suggesting that while healthy myocardium is compressible, myocardium in HFrEF patients is incompressible.

In conclusion, existing evidence suggests that the myocardium is compressible \emph{in vivo}, although the degree of compressibility and underlying mechanisms remain incompletely understood. Additionally, myocardial compressibility appears to be altered in diseased states, but this is also of uncertain functional consequence. More research is needed to fully elucidate the role of compressibility and determine the best way to model it in simulations.

We close by noting that it is generally easier to simulate compressible materials than incompressible ones, as incompressibility often requires special numerical treatments to avoid ill-conditioned matrices (Section \ref{sect:numerical_formulations}).

\subsection{Fiber Dispersion}

While myocardial mesostructure is often modeled as a continuously varying fiber field, myofibers $\mathbf{f}$ (and sheetlet orientations $\mathbf{s}$) exhibit significant dispersion at the mesoscale. Within an appropriately sized region of interest (ROI), these fibers are not perfectly aligned with the mean direction, but instead have orientations dispersed around it. For instance, \cite{karlon_regional_2000} analyzed 2D histological sections of rat septum and quantified dispersion as the standard deviation of fiber angle within a small tissue subregion (approximately 0.01 \si{mm^2}). Other studies estimated dispersion by fitting an analytic distribution, characterized by a dispersion parameter, to the histogram of fiber angles within a dataset of a given size \cite{sommer_biomechanical_2015, schriefl_quantitative_2012, schriefl_automated_2013, ahmad_region-specific_2018}. Notably, the ROI size is important \cite{lanir_multi-scale_2017}; \cite{schriefl_quantitative_2012} investigated the effect of ROI size on fiber distribution estimates and provided guidelines for finite element analysts to select mesh sizes that balance capturing spatial inhomogeneities with statistical robustness and computational feasibility. Dispersion can also be assessed using cDTI, where low fractional anisotropy correlates with high dispersion \cite{wilson_myocardial_2022}. Further work with cDTI could better characterize dispersion throughout the heart using more advanced acquisition methods and signal models.

Fiber dispersion varies significantly between individuals \cite{sommer_biomechanical_2015}. Moreover, enhanced fiber dispersion, or ``disarray'', is a hallmark of certain cardiac diseases \cite{garcia-canadilla_complex_2018, ghonim_myocardial_2017, ma_myofibril_2022} and can lead to impaired mechanical function \cite{karlon_regional_2000}. Simulations of cardiac mechanics have further demonstrated the importance of fiber dispersion \cite{guan_effect_2020, eriksson_modeling_2013}.

From a constitutive modeling perspective, fiber dispersion was first incorporated into myocardial models by \cite{eriksson_modeling_2013}. They extended the standard myocardial HO model (Eq. \eqref{eq:HO_model}) \cite{holzapfel_constitutive_2009} using a generalized structure tensor (GST) approach, originally developed to model collagen dispersion in arterial tissue \cite{gasser_hyperelastic_2005}. This approach has been applied in recent studies \cite{mojumder_computational_2023, tikenogullari_how_2022}. Briefly, the orientation probability density function at an angle $\Theta$ from the mean fiber or sheetlet direction was described by a normalized $\pi$-periodic von Mises distribution:

\begin{equation} 
  \rho_i(\Theta) = 4\sqrt{\frac{b_\rho}{2\pi}}\frac{\exp[b_\rho(\cos(2\Theta)+1)]}{(-i)\erf(\sqrt{2b_\rho})}, \hspace{10pt} i \in {f,s}, 
\end{equation}

where $b_\rho$ is a concentration parameter, and $\erf(x)$ is the error function. This rotationally symmetric distribution was fit to limited data in the literature at the time. Dispersion parameters $\kappa_f$ and $\kappa_s$ for fibers and sheetlets were computed as

\begin{equation} 
  \kappa_i = \frac{1}{4} \int_0^\pi \rho_i(\Theta) \sin^3\Theta d\Theta, 
\end{equation}

and the standard HO model was modified to account for dispersion

\begin{equation} \label{eq:HO_disperion_model}
    \psi_{HO} = \frac{a}{2b}
    \Big( 
    e^{b(I_1 - 3)} - 1
    \Big)
    +
    \frac{a_{fs}}{2b_{fs}}
    \Big(
    e^{b_{fs}I_{8,fs}^2} - 1
    \Big)
    +
    \sum_{i \in \{f,s\}}
    \chi(I_{4,i}^*)
    \frac{a_i}{2b_i}
    \Big(
    e^{b_i(I_{4,i}^* - 1)^2} - 1
    \Big).
\end{equation}


where the dispersion-modified fiber and sheetlet invariants $I_{4,i}^*$ are defined as

\begin{equation} 
  I_{4,i}^* = \kappa_i I_1 + (1 - 3\kappa_i) I_{4,i}, \hspace{10pt} i \in {f,s}. 
\end{equation}

Later, \cite{melnik_generalised_2018} incorporated dispersion into the $fs$ term as well. In addition, a non-symmetric, two-parameter dispersion model has also been developed, first for collagen fibers in arterial walls \cite{holzapfel_modelling_2015}, and later adapted for myocardium \cite{guan_modelling_2021}. This model assumes the probability density is the product of independent functions describing in-plane and out-of-plane dispersions.

The GST approach should be compared to the alternative angular integration (AI) or continuous fiber dispersion method, in which the contribution to the total strain energy of a single fiber at a given stretch is integrated over all directions or solid angles, weighted by the probability density \cite{holzapfel_modelling_2015}. The AI and GST approaches have been directly compared in \cite{cortes_characterizing_2010, holzapfel_fiber_2017, holzapfel_comparison_2017}, with particular attention paid to whether GST properly excludes fibers under compression and the computational cost of AI. To address these issues, \cite{li_modeling_2018} modified the GST approach with a general invariant that properly excludes compressed fibers. Later, the same authors applied a discrete fiber dispersion (DFD) approach based on the AI approach, and showed similar results to AI with a significant reduction in computational cost \cite{li_discrete_2018}.

Finally, while the role of dispersion in the passive myocardial response has received much attention, less is known about the effect of dispersion on myocardial contraction. Fiber dispersion appears to be at least partly responsible for significant cross-fiber stresses generated during contraction \cite{lin_multiaxial_1998, krishnamurthy_microstructurally_2016}, which is likely important to physiological deformation \cite{bovendeerd_determinants_2009, guan_effect_2020, piersanti_3d0d_2022}. This has been modeled in several studies by adding a cross-fiber active stress, which can depend on the degree of dispersion \cite{usyk_effect_2000, eriksson_modeling_2013, sack_construction_2018}.

In summary, there has been significant work in characterizing and modeling fiber dispersion, but its importance in cardiac mechanics, especially in diseased cases and with respect to active contraction, should be investigated further.

\section{Importance of anatomic model and mechanical boundary conditions} \label{sect:anatomy_and_bcs}

\begin{figure}
    \centering
    \includegraphics[width=0.9\linewidth]{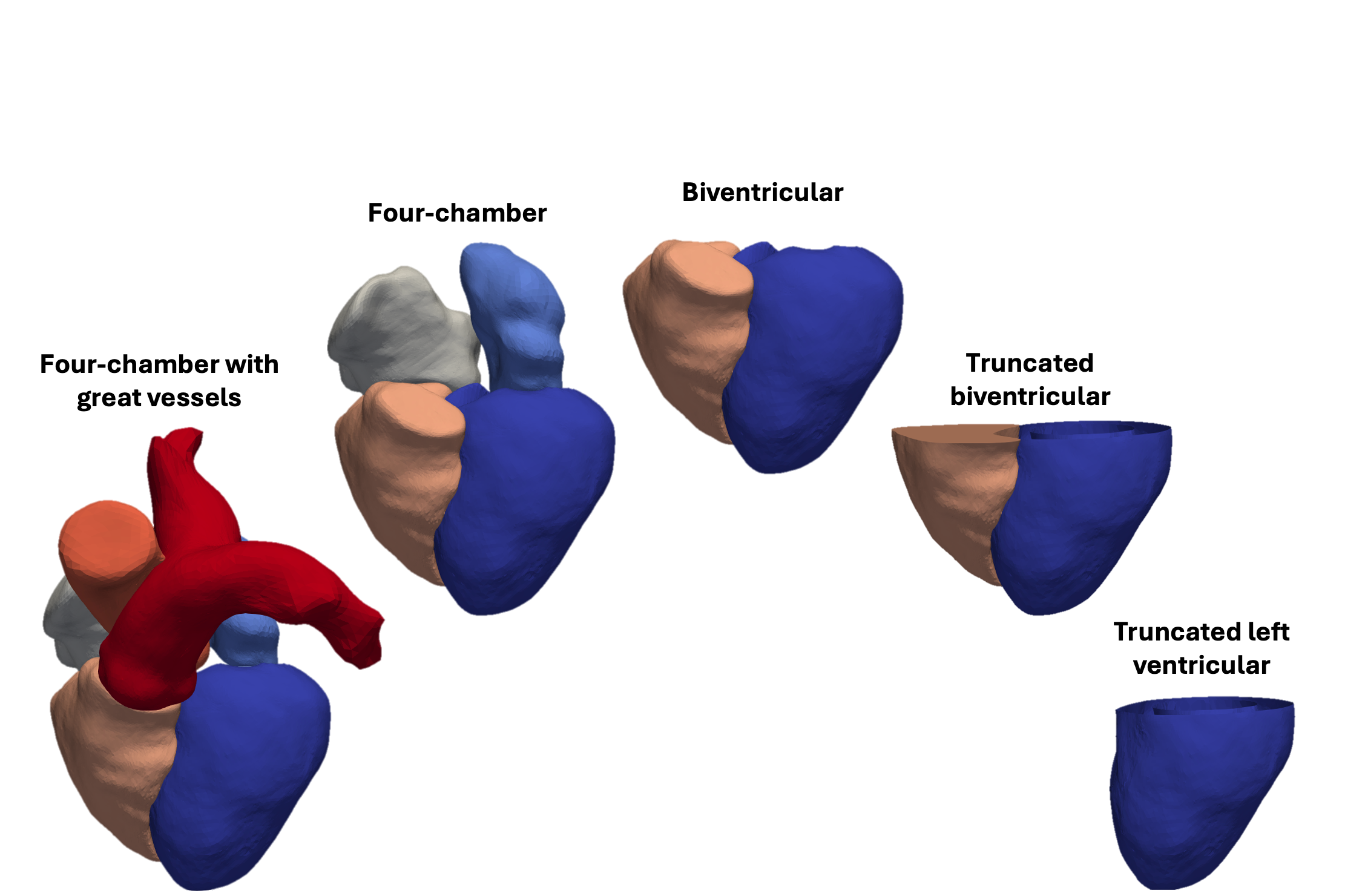}
    \caption{Representative cardiac geometries commonly used in the literature. Four-chamber models, sometimes including the great vessels, offer the most comprehensive anatomical representation but are difficult to reconstruct from medical imaging and substantially increase computational complexity. In contrast, biventricular (BiV) and LV models, sometimes truncated at the basal plane have been more widely used due to their reduced complexity, although the physiological consequences of excluding other cardiac structures remain poorly understood.}
    \label{fig:anatomic_models}
\end{figure}

A variety of anatomical models have been used in the cardiac modeling literature (Figure \ref{fig:anatomic_models}). Whole-heart models, which include all four chambers and sometimes the trunks of the aorta and pulmonary arteries, are the most realistic \cite{pfaller_importance_2019, strocchi_integrating_2025,  fedele_comprehensive_2023, gerach_electro-mechanical_2021, fritz_simulation_2014, ghebryal_effect_2023, feng_whole-heart_2024}, but constructing patient-specific models remains challenging due to limited medical image resolution in the thin-walled RV, atria, and basal portions of the ventricles \cite{asner_patient-specific_2017}. Additionally, whole-heart simulations tend to be more complex to develop and calibrate \cite{peirlinck_kinematic_2019}. Because of these challenges, biventricular (BiV) models that include only the LV and RV are common. Many BiV models are further simplified by truncating the geometry at the ``basal plane" \cite{peirlinck_kinematic_2019, hirschvogel_monolithic_2017}. An even further simplification is to include only the LV, which may also be truncated at the basal plane \cite{asner_patient-specific_2017}. These simplifications arise, in part, from the complexity of building models with multi-chamber geometries from coarsely sampled medical imaging data. They also stem from the expectation that the ventricles, particularly the LV, are the primary force generators underlying cardiac function. However, ``even subtle changes in cardiac anatomy can have a large impact on cardiac function" \cite{rodero_linking_2021}, so it follows that the choice of anatomic model may significantly influence the outcomes of cardiac mechanics simulations. This raises important questions, for example: How does the RV influence LV function, and vice versa? Through what mechanisms do the atria and ventricles interact, and how significant are these interactions?

Closely tied to the choice of anatomic model is the selection of boundary conditions (BCs), which are equally important in determining cardiac function \cite{niederer_importance_2009}. For example, when using a truncated BiV or LV model, the modeler must decide on an appropriate mechanical boundary condition for the basal surface. Ideally, this boundary condition should replicate the mechanical influence of the basal ventricular tissue and atria, to the degree that such an influence exists. 

In this section, we examine the mechanical interactions among various cardiac structures, highlighting how the omission of certain structures may impact simulation results. Additionally, we review commonly used boundary conditions designed to approximate, with varying degrees of fidelity, the effects of structures excluded from the computational domain.

\subsection{Mechanical interactions among structures of the heart}
The mechanical interaction between LV and RV, known in the literature as ventricular interdependence \cite{naeije_overloaded_2017}, occurs through direct and indirect mechanisms. The ventricles share a common septum, which provides direct coupling, while additional interactions arise via blood flow through the systemic and pulmonary circulations. Indeed, at steady-state operation without valve regurgitation or intracardiac shunting, LV and RV stroke volumes must be similar, or else blood would accumulate somewhere in the body \cite{mandapaka_simultaneous_2011}. Additionally, important interactions exist through the pericardium \cite{kim_computational_2023, santamore_theoretical_1990, kroeker_pericardium_2003, vaillant_influence_2025}. Both directions, LV to RV and RV to LV, have been shown to be important, especially in diseased cases \cite{damiano_significant_1991, petit_ventricular_2023}. 

Early work by \cite{chung_dynamic_1997} introduced a reduced-order mathematical model to study ventricular interdependence, modeling the LV free wall, RV free wall, septum, and pericardium with separate time-varying elastances. Their findings, consistent with experimental observations, include the following. Increased diastolic pressure in one ventricle reduces the compliance of the other, and the pericardium enhances the mechanical coupling between ventricles. Additionally, when the septum does not contract, it bulges into the RV during systole, creating a ``parasitic" effect -- RV pumping is enhanced at the expense of LV function. This effect is eliminated with a contracting septal wall. In recent work, \cite{kim_computational_2023} used a one-dimensional BiV model with pericardium and closed-loop circulatory model to investigate the effects of LV and RV systolic and diastolic dysfunction. Their results showed that LV dysfunction prompts the RV to compensate by increasing cardiac power output, whereas RV dysfunction reduces LV filling, a phenomenon that could be misinterpreted clinically as LV diastolic dysfunction (heart failure with preserved ejection fraction). Similarly, \cite{sack_investigating_2018} developed a patient-specific finite element model to investigate the nature of ventricular interdependence in the presence of a left ventricular assistance device (LVAD). They found that greater LVAD support reduces LV pressure and myofiber stress, shifts the RV PV loop to higher pressures and volumes, and causes abnormal bending of the septum into the LV.

Atrioventricular interactions occur through direct tissue coupling at the atrioventricular plane, as well as via blood flow exchanges across the mitral and tricuspid valves \cite{beyar_atrioventricular_1987}. As with ventricular interdependence, the pericardium may provide an additional interaction mechanism. The atria have conventionally been understood to support ventricular function in three distinct roles: serving as a reservoir of blood during ventricular systole, acting as a conduit in early diastole, and functioning as a booster pump in late ventricular diastole \cite{fedele_comprehensive_2023, blume_left_2011}. However, the ventricles, as the larger and stronger chambers, play an equally important role in atrial function. \cite{fritz_simulation_2014}, for example, used a four-chamber model to investigate the interactions among the ventricles, atria, and pericardium. They found that ventricular contraction is important for atrial filling, creating a suction of blood from the venae cavae and pulmonary veins into the atria. Similarly, \cite{strocchi_effect_2021} found that increasing ventricular fiber angles from $+40^\circ$/ $-40^\circ$ to $+60^\circ$/ $-60^\circ$ increased atrioventricular plane motion from \qty{1.0}{mm} to \qty{14.0}{mm} and increased LA maximum volume by \qty{8}{\percent}.

The atria influence ventricular function primarily through their role as booster pumps, though there is some conflicting evidence about the size of their impact. \cite{fedele_comprehensive_2023} compared whole-heart simulations with and without atrial contraction and found that without atrial contraction, ventricular end-diastolic volumes and peak systolic pressures decreased significantly. In contrast, \cite{land_influence_2018} found that while variations in atrial calcium transient models strongly affected atrial function, their impact on ventricular PV curves was relatively small. \cite{gerach_impact_2023} simulated LA ablation in a four-chamber model, assessing how altered electrical activation and increased stiffness from ablation scars affect cardiac function. With ablation, LV stroke volume decreased by only \qty{1.8}{\percent}, though they suggest the effect may be larger in older populations. Increased LA stiffness also reduced atrioventricular plane displacement, leading to lower peak systolic pressure in the LV, though the effect was modest. Notably, the RA and RV were essentially unaffected by LA ablation. Even an entirely passive atria may still influence ventricular mechanics. \cite{alfonso_santiago_fully_2018} compared a biventricular model with a whole-heart model and found that even non-contracting atria significantly affected ventricular deformation, specifically atrioventricular plane displacement and radial displacement of the basal portions of the LV and RV free walls. They attributed these effects to the added inertia of atrial tissue.

The existing evidence clearly demonstrates the significance of mechanical coupling among cardiac chambers, though the precise mechanisms and clinical importance of these interactions warrant further study. These considerations are particularly relevant for computational heart modelers, who must balance anatomical complexity with computational efficiency. In addition to the four chambers, it is important to consider the roles other cardiac structures, including the roots of the great vessels \cite{strocchi_simulating_2020, gerach_electro-mechanical_2021, land_influence_2018}, the trabeculae and papillary muscles \cite{fatemifar_computational_2019, serrani_influence_2013, vedula_effect_2016}, and epicardial adipose tissue \cite{pfaller_importance_2019, gerach_electro-mechanical_2021, rodriguez_structure_2017}.

\subsection{Mechanical boundary conditions}
Arguably the most important BC in cardiac mechanics simulations is the one applied to the endocardial surface of the heart chambers, representing the traction exerted by the blood on the heart wall. This traction is commonly modeled as a spatially-uniform pressure, either prescribed directly \cite{arostica_software_2025} or determined through a coupling with a 0D circulation model \cite{brown_modular_2024}. Alternatively, it can be explicitly computed in a fluid-structure interaction (FSI) simulation \cite{brenneisen_sequential_2021, davey_simulating_2024}. While this class of BC is relatively straightforward, the appropriate formulations of other BCs can be more ambiguous. Here, we discuss two important examples: BCs on the basal surface in truncated BiV and LV models and BCs on the epicardial surface to capture the effect of the pericardium, relevant to all anatomical models.

\paragraph{Basal BCs}
A variety of basal BCs have been proposed in the literature. It is important to recognize that \emph{in vivo} the basal aspects of the ventricles plunge towards the apex during contraction, while the apex remains relatively fixed in position. The apex does, however, rotate counter-clockwise relative to the base, which is largely held in place at the level of the atria by the blood vessels. Early approaches constrained the basal plane in the long-axis direction, while handling in-plane motion in different ways. \cite{peirlinck_kinematic_2019} compared five such basal BCs on truncated biventricular geometries, finding notable differences in global metrics and statistically significant, though minor and localized variations in tissue strain. They advised against fixing individual basal nodes, recommending instead BCs that constrain only the average in-plane motion of basal nodes, allowing circumferential and radial motion within the basal plane. However, given the crucial role of atrioventricular plane motion in ventricular function \cite{fritz_simulation_2014}, allowing long-axis motion of the basal plane is desirable. To achieve this, other researchers have used Robin BCs, conceptually a spring and dashpot in parallel \cite{augustin_computationally_2021, hirschvogel_monolithic_2017, brown_modular_2024}. These BCs introduce additional parameters describing the strength of the elastic (spring) and viscous (dashpot) contributions; nominal values can be found in a recent cardiac mechanics benchmark paper \cite{arostica_software_2025}. Interestingly, the previously described BCs lead to physical inconsistencies, including a net force exerted by the blood on the chamber wall and an inaccurate value of the work done by the blood pressure to inflate the heart chamber. To address these issues, \cite{regazzoni_machine_2020} proposed a so-called energy-consistent basal BC, which depends on the pressure being applied on the endocardial surface and permits global energy conservation when coupling to a 0D model of blood circulation \cite{piersanti_3d0d_2022, regazzoni_cardiac_2022}.

\begin{figure}
    \centering
    \includegraphics[width=0.9\linewidth]{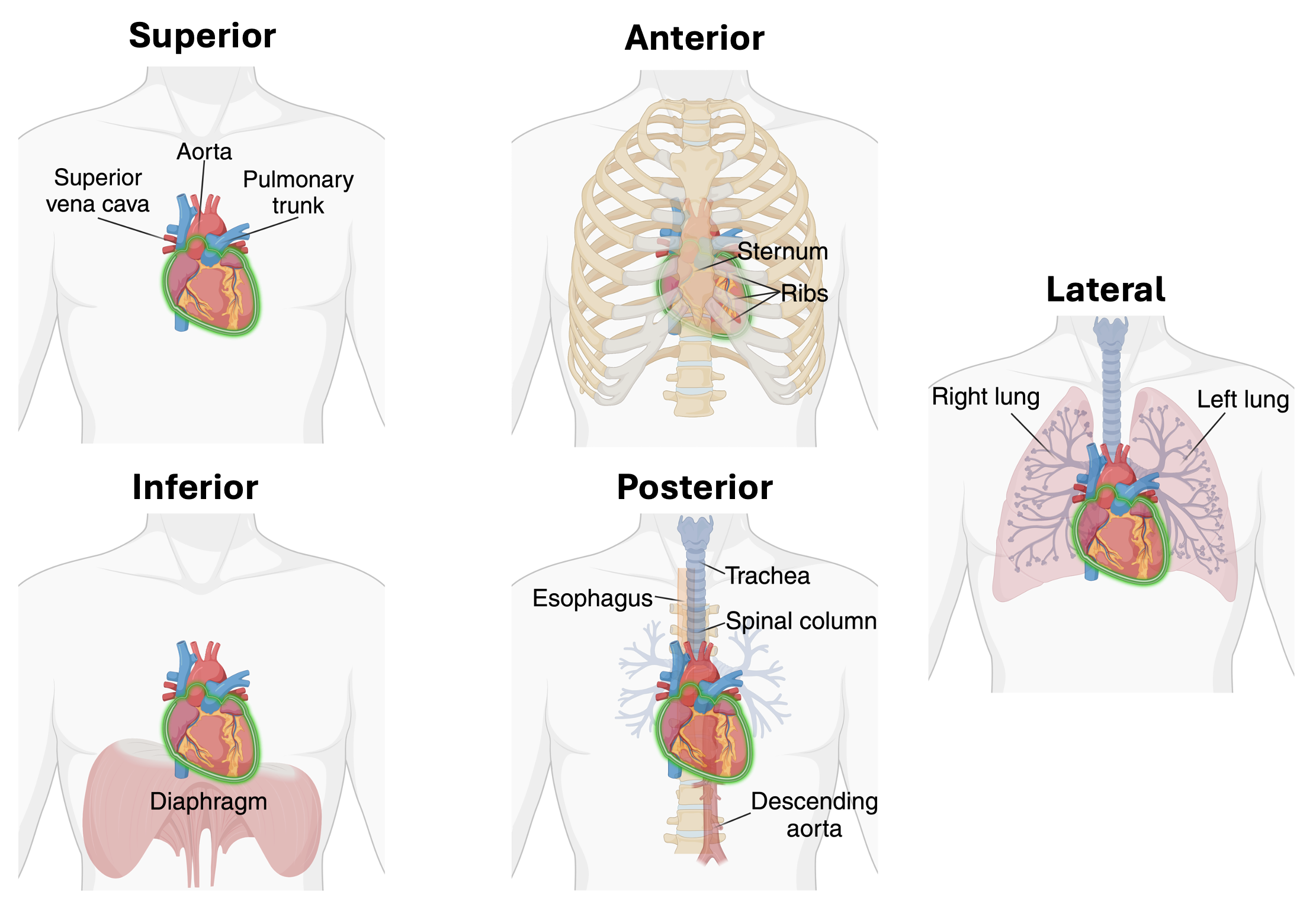}
    \caption{Illustrations showing the anatomic structures surrounding the heart, on the superior, inferior, anterior, posterior, and lateral sides. The pericardium is highlighted in green. Created in  \url{https://BioRender.com}.}
    \label{fig:heart_and_surrounding_anatomy}
\end{figure}

\paragraph{Pericardial BCs}
As discussed in Section \ref{sect:anatomic_model_construction}, the heart, as well as the roots of the great vessels, is enclosed in a fibrous sac called the pericardium \cite{pfaller_importance_2019}. A thin layer of pericardial fluid fills the space between the epicardium and the pericardium \cite{fritz_simulation_2014}, constraining relative motion in the normal direction while allowing tangential sliding with relatively low friction \cite{pfaller_importance_2019}. Between the myocardium and visceral pericardium are fat deposits known as \textit{epicardial} adipose tissue (EAT), which tend to occupy the grooves between the LV and RV, as well as between the ventricles and atria \cite{talman_epicardial_2014}. Additional fat deposits, known as \textit{paracardial} adipose tissue, surround up to \qty{80}{\percent} of the pericardium's outer surface (parietal pericardium) \cite{talman_epicardial_2014}. Beyond the pericardium are various neighboring anatomic structures, illustrated in Figure \ref{fig:heart_and_surrounding_anatomy}. Superiorly, the pericardium is continuous with the walls of the great vessels. Anteriorly, sternopericardial ligaments attach it to the sternum, and it may contact the fourth to sixth ribs on the left side. Inferiorly, the pericardiophrenic ligaments connect it with the diaphragm. Posteriorly, the pericardium is adjacent to structures such as the bronchi, esophagus, descending thoracic aorta, and vertebral column \cite{rodriguez_structure_2017}. Laterally, it is in contact with the lungs \cite{pfaller_importance_2019}. Given this complex anatomic environment, it is apparent that the tissue support on the epicardium is heterogeneous, and consequently very difficult to model.

Some groups have chosen to explicitly model the pericardium in cardiac simulations. \cite{fritz_simulation_2014} introduced a geometric model of the pericardium and surrounding tissue, solving a frictionless contact problem between the epicardial surface of a whole-heart mesh and the inner surface of a pericardial mesh. Performing simulations with and without pericardial support, they found that including the pericardium significantly influenced heart deformation and function, primarily by restricting motion of the outer contour of the heart, consistent with physiological observations. Additionally, pericardial support reduced systolic pressure and ejection fraction in both ventricles while enhancing atrioventricular plane displacement and consequently atrial filling. Similarly, \cite{mendiola_mechanical_2022} modeled the rat pericardium using shell finite elements and imposed frictionless sliding contact with the epicardium via a linear penalty method. Their study examined the effects of pericardial elasticity on cardiac function under pulmonary hypertension and simulated pericardiectomy by removing the pericardium entirely.

More commonly, however, the effect of the pericardium is incorporated through a BC on the epicardial surface, avoiding the computational expense of explicit contact modeling. Inspired by \cite{fritz_simulation_2014}, \cite{pfaller_importance_2019} introduced a widely used pericardial BC, formulated as a Robin BC acting only in the epicardial normal direction. This penalizes normal displacement while permitting free tangential sliding, reflecting the lubricating role of the pericardial fluid. Comparing simulations with and without the pericardial BC, they found that the pericardial BC improved agreement with MRI data in terms of atrioventricular plane displacement, atrial filling during ventricular systole, and LV and RV endocardial motion.

Building on this approach, \cite{strocchi_simulating_2020} introduced a spatially varying version of the pericardial Robin BC, further improving the agreement with image data. In their work, the stiffness parameter decreased from apex to base of the ventricles, based on observations of epicardial motion in CT images. This decrease in stiffness reflected the presence of EAT near the ventricular base, which locally reduces the constraining effect of the pericardium. Similarly, \cite{fedele_comprehensive_2023} assigned a higher stiffness value to epicardial regions directly contacting the pericardium and a much lower value to regions near the ventricular base and atrial appendages where EAT is more prominent. Further exploring this concept, \cite{ghebryal_effect_2023} varied the spatial heterogeneity of the pericardial BC and found that while local deformations were very sensitive to changes in BC parameters, global functional metrics such as stroke volume remained relatively unaffected. In recent work, \cite{jilberto_identification_2025} proposed a method to incorporate localized forces on the RV, particularly from the ribs and diaphragm, into an inverse mechanics framework, significantly improving reference configuration estimates for the RV.

In a comparative study, \cite{kraus_comparison_2023} evaluated five different pericardial modeling approaches on the same whole-heart mesh, including the sliding contact method of \cite{fritz_simulation_2014} and the constant and spatially-varying Robin BC approaches of \cite{pfaller_importance_2019}, \cite{strocchi_simulating_2020}, and \cite{fedele_comprehensive_2023}. Their results indicated that Robin BCs reduced LV twist compared to sliding contact models, likely due to their standard formulation being valid only for small epicardial rotations \cite{pfaller_importance_2019}. Additionally, they found Robin BCs to be more computationally demanding than the sliding contact approach, warranting further investigation.

An alternative approach to boundary conditions was explored by \cite{asner_patient-specific_2017, asner_estimation_2016}, who used boundary energy terms to weakly impose the motion of the epicardial and basal surfaces based on image data. Comparing their approach to traditional generic boundary conditions, such as those described previously, they found significant improvement in model accuracy. While a promising approach, the utility of image-based BCs in a clinical setting should be carefully considered, since they ``inherently do not accommodate for changes in mechanical environment upon performing (virtual) surgeries or (virtually) implanting a device" \cite{peirlinck_kinematic_2019}.

\section{Coupling tissue mechanics to other physics} \label{sect:multiphysics_coupling}


State-of-the-art heart models capture not only cardiac tissue mechanics, but also electrical signal propagation, subcellular force generation, and blood flow within the heart and throughout the broader circulatory system. In some cases, coupling these physical processes is essential to accurately represent key cardiac phenomena. However, such couplings can also introduce significant numerical challenges that require specialized solution strategies. Typically, multiphysics equations are coupled either monolithically or in a partitioned (segregated) manner: monolithic approaches offer greater robustness but can be complex to implement, whereas partitioned schemes are more modular and easier to develop but may suffer from numerical instabilities \cite{brown_modular_2024}. In this section, we review the physical processes commonly coupled with mechanics, the capabilities such couplings enable, their technical implementation, and the computational challenges they pose.

\subsection{Electrophysiology and cellular contraction} \label{subsect:electrophysiology_and_cellular_contraction}
The physical models governing cardiac electrophysiology and their interactions with mechanics are summarized in Figure \ref{fig:cardiac_modeling_summary}. Electrical activity in the heart is commonly modeled using a reaction-diffusion partial differential equation (PDE) for the transmembrane electric potential, resulting in a propagating wave solution \cite{trayanova_whole-heart_2011}. This formulation can represent differing conduction anisotropy ratios in the intracellular and extracellular spaces (bidomain model), or assume identical anisotropy in both domains (monodomain model) \cite{potse_comparison_2006}. From a numerical standpoint, the bidomain model is more challenging due to the need to solve an implicit equation for the extracellular potential at every timestep. Computationally efficient alternative formulations like the Eikonal and Reaction-Eikonal models have also been explored \cite{stella_fast_2022}.

Electrical activity is typically coupled with an ionic model: an ordinary differential equation (ODE) system describing the concentration dynamics of various ions that cross the cell membrane during depolarization and repolarization \cite{fedele_comprehensive_2023}. Distinct ionic models have been developed for specific cardiac cell types, including ventricular myocytes, atrial myocytes, and Purkinje cells \cite{sugiura_ut-heart_2022}. These ionic models also compute the intracellular calcium ion ($\textrm{Ca}^{2+}$) concentration, which is the primary input to a model for cellular contraction \cite{niederer_short_2019}. This contraction is linked to tissue-scale mechanics through one of three main strategies: active stress, active strain, or a combined approach known as Hill’s three-element model \cite{avazmohammadi_contemporary_2019}. Alternatively, electrical activity may simply yield an activation time, which triggers active contraction in a phenomenological modeling approach \cite{lluch_calibration_2020, rodero_advancing_2023, gurev_high-resolution_2015, fan_optimization_2022}. In many studies, a \textit{weak} or one-way coupling from electrophysiology to mechanics, which captures so-called excitation-contraction coupling \cite{radisic_influence_2025}, is used to efficiently reproduce the essential features of cardiac electromechanics \cite{strocchi_cell_2023, augustin_computationally_2021, stella_fast_2022, jung_integrated_2022}.

\textit{Strongly} coupled approaches to electromechanics  account for physiological feedbacks from mechanical deformation to both cellular contraction and electrical activity. Many models of contraction incorporate length-dependent activation, wherein stretched muscle fibers produce greater contractile force, an effect underlying the organ-level Frank-Starling mechanism \cite{niederer_short_2019}. Some models also incorporate a dependence on the rate of deformation \cite{regazzoni_biophysically_2020}. Recent studies have also examined the effects of mechanical deformation on electrophysiology, terms mechano-electric feedbacks (MEFs) \cite{gerach_differential_2024}, which include deformation-induced changes to the effective conductivity tensor and volume scaling of electrical current and potential \cite{alfonso_santiago_fully_2018, frohlich_numerical_2023}, as well as cell-level responses to stretch, such as the activation of stretch-sensitive ion channels \cite{radisic_influence_2025, quinn_cardiac_2021, salvador_role_2022}. MEFs are considered important in diseased cases \cite{rodero_advancing_2023, colli_franzone_effects_2017}, and are believed to explain phenomena like commotio cordis—ventricular fibrillation induced by a sudden blow to the chest—or its converse, the precordial thump, which can restore normal rhythm \cite{sahli_costabal_importance_2017}.

Simulating coupled electrophysiology and mechanics introduces several numerical challenges. Electrophysiology often demands finer mesh resolution and smaller timesteps than mechanics to accurately reproduce physiological conduction velocities \cite{sugiura_ut-heart_2022}. This discrepancy may necessitate variable interpolation between meshes and the use of specialized time-integration schemes \cite{torre_isogeometric_2023, fedele_comprehensive_2023, gerach_electro-mechanical_2021, salvador_intergrid_2020}, depending on the specific numerical approach. These challenges are discussed further in Section \ref{sect:numerical_formulations}. Moreover, mechanics to contraction feedbacks related to fiber stretch and stretch rate can destabilize staggered coupling schemes unless appropriate stabilization techniques are employed \cite{regazzoni_oscillation-free_2021}. For further detail on models of electrophysiology and their coupling to mechanics, we refer the reader to the reviews in \cite{clayton_models_2011, quarteroni_integrated_2017}.

\subsection{Blood flow}
 At the endocardial and valvular surfaces, the motion and stress are continuous between the tissue and blood \cite{le_computational_2022}, resulting in a strong coupling between the two domains. This interaction is fundamental: blood flow within the heart is predominantly determined by the motion of the heart walls and valve leaflets, while the blood, in turn, imposes significant stresses on cardiac tissue. The interplay between blood flow (fluid) and myocardial mechanics (structure) falls under the general category of fluid–structure interaction (FSI).

Two primary approaches dominate the modeling of cardiac FSI: the immersed boundary (IB) method \cite{feng_whole-heart_2024, davey_simulating_2024, verzicco_electro-fluid-mechanics_2022, viola_high-fidelity_2023} and the arbitrary Lagrangian–Eulerian (ALE) method \cite{sugiura_ut-heart_2022, bucelli_mathematical_2023, zingaro_electromechanics-driven_2024}. In brief, the IB method solves the fluid equations on a fixed (Eulerian) background mesh, incorporating the effects of structures such as the endocardium or valves through localized forcing terms on the right-hand side of the fluid momentum equation. Because the fluid and structural meshes do not conform at the interface, IB approaches often rely on adaptive mesh refinement to maintain accuracy near these boundaries.

By contrast, the ALE method uses a fluid mesh that conforms to and deforms with the moving structure, offering enhanced resolution at the fluid–structure interface. However, this benefit comes at the cost of potential mesh distortion, often requiring remeshing strategies to accommodate large deformations \cite{le_computational_2022}. A more comprehensive overview of cardiac FSI techniques and challenges can be found in \cite{le_computational_2022}.

Cardiac FSI models mark a significant advance toward developing cardiac digital twins, enabling the integration of mechanics, and often electrophysiology as well, with realistic simulations of blood flow within the cardiac chambers and across the valves. Blood flow modeling itself remains an expansive area of research with major clinical relevance. Nevertheless, FSI introduces unique computational challenges, which is why fully coupled, whole-heart FSI simulations have only recently become feasible.

\subsection{Circulatory dynamics}
While fluid–structure interaction (FSI) methods are frequently used to couple cardiac mechanics with blood flow within and proximal to the heart, they are rarely employed to model blood flow throughout the rest of the circulatory system. Instead, the systemic and pulmonary circulations are commonly represented using lumped parameter networks (LPNs), zero-dimensional (0D) models that analogize blood flow to electrical current \cite{quarteroni_geometric_2016} (1D models are also sometimes used \cite{caforio_coupling_2022}). This simplification allows for the computation of bulk hemodynamic quantities such as pressure and flow rate at a significantly reduced cost. Mathematically, LPNs are described by systems of differential-algebraic equations (DAEs) \cite{peiro_reduced_2009}, which can be derived by applying Kirchhoff’s first law to an equivalent electrical circuit.

These DAE systems can be coupled to the three-dimensional (3D) blood flow equations via inlet/outlet flow rate and pressure \cite{vignon-clementel_outflow_2006, quarteroni_geometric_2016, esmaily_moghadam_modular_2013}, or directly to the tissue mechanics equations using chamber flow rate or volume and endocardial pressure as coupling variables (indicated by the dashed lines in Figure \ref{fig:cardiac_modeling_summary}) \cite{augustin_computationally_2021, piersanti_3d0d_2022, brown_modular_2024, regazzoni_cardiac_2022, jafari_framework_2019, shavik_high_2018}. A wide range of LPN models have been proposed in the literature, from simple Windkessel-type representations \cite{westerhof_arterial_2009} to elaborate, closed-loop networks modeling the entire circulatory system \cite{regazzoni_cardiac_2022, kung_predictive_2013}.

Incorporating an LPN into a heart model enhances physiological fidelity by imposing preload and afterload \cite{regazzoni_cardiac_2022, vincent_understanding_2008}, and allows the study of cardiac responses to circulatory changes such as during exercise \cite{kung_simulation_2014}, under surgical conditions \cite{kaufmann_factors_2019}, or due to respiration \cite{van_de_bruaene_effect_2015}. For cardiac mechanics or FSI models that lack fully resolved valve dynamics, LPNs can also serve to model valve behavior, enabling physiologic reproduction of the cardiac cycle, especially the clear delineation of the four cardiac phases \cite{brown_modular_2024}. Furthermore, using an LPN, especially one representing the full circulatory loop, allows investigation of how cardiac function influences system-wide hemodynamics. Since patient-specific data are often available at various locations in the circulatory system, incorporating an LPN can support parameter tuning to improve personalization of cardiac models \cite{capuano_personalized_2024, salvador_fast_2023, kariya_personalized_2020}.

However, coupling heart models to LPNs introduces computational challenges akin to those encountered in FSI. Loosely coupled schemes can exhibit instabilities, rooted in the “balloon dilemma,” which must be addressed using specialized stabilization strategies \cite{regazzoni_stabilization_2025}. In strongly coupled approaches employing Newton-like solvers, the inclusion of an LPN modifies the structure of the tangent matrix, necessitating the use of carefully constructed preconditioners \cite{brown_modular_2024, hirschvogel_effective_2025, esmaily-moghadam_new_2013}.

\subsection{Other physical processes}
We conclude this section by noting that while electrophysiology, cellular contraction, tissue mechanics, blood flow, and circulatory dynamics represent the core physical processes modeled in multiphysics simulations of the heart, additional physiological phenomena have also been explored. Notable examples include myocardial perfusion \cite{barnafi_wittwer_multiscale_2022, fan_transmural_2021, munneke_myocardial_2023, zingaro_comprehensive_2023}, transport of blood species like oxygen \cite{kariya_personalized_2020}, metabolic/energetic function \cite{lopez_impaired_2020}, coagulation \cite{guerrero-hurtado_efficient_2023}, and the physiological regulation of arterial pressure via the baroreflex mechanism \cite{sharifi_multiscale_2024}. Future cardiac modeling approaches could be extended to include these phenomenon with increasing fidelity.

\section{Numerical formulations} \label{sect:numerical_formulations}
Numerical methods are essential for the investigation of cardiac mechanics, particularly under physiologically detailed and patient-specific settings. The inherently multiphysics and multiscale nature of cardiac modeling has driven significant research into advanced numerical methodologies. 

We begin by summarizing common numerical choices in the literature, focusing on FEA for simulating cardiac tissue mechanics. Tetrahedral elements are the most common, including linear \cite{fedele_comprehensive_2023, frohlich_numerical_2023, guan_modelling_2021, bucelli_mathematical_2023}, quadratic \cite{gerach_electro-mechanical_2021, shavik_high_2018}, and quadratic-linear in mixed displacement-pressure formulations \cite{hadjicharalambous_investigating_2021, asner_patient-specific_2017}, but
hexahedral elements have also been used \cite{capuano_personalized_2024, torre_isogeometric_2023, barnafi_comparative_2023}. For mechanics, average elements edge sizes are typically between \qty{1}{mm} and \qty{5}{mm} \cite{arostica_software_2025, fedele_comprehensive_2023, alfonso_santiago_fully_2018, strocchi_publicly_2020, hadjicharalambous_investigating_2021, stimm_personalization_2022, fritz_simulation_2014, feng_whole-heart_2024, bucelli_mathematical_2023}, resulting in $O(10^3)$ to $O(10^5)$ elements for a biventricular mesh \cite{arostica_software_2025}; a useful guideline is to have at least two elements through the wall thickness \cite{augustin_computationally_2021}. Time integration is usually implicit -- generalized-$\alpha$ \cite{augustin_computationally_2021, hirschvogel_monolithic_2017, brown_modular_2024}, Newmark \cite{frohlich_numerical_2023, gerach_electro-mechanical_2021, fritz_simulation_2014, sahli_costabal_importance_2017}, backward Euler \cite{capuano_personalized_2024, 
fedele_comprehensive_2023, regazzoni_cardiac_2022, barnafi_comparative_2023} -- with timestep sizes $O(1)$ \unit{ms}, but some studies have used explicit methods \cite{lluch_calibration_2020, alfonso_santiago_fully_2018, feng_whole-heart_2024}, which require smaller timestep sizes $O(10^{-2})$ \unit{ms}. Some groups forgo time integration entirely by performing a quasi-static analysis \cite{torre_isogeometric_2023, shavik_high_2018}.
Implicit schemes require the solution of a nonlinear algebraic system at each timestep, and Newton's method is almost universally used \cite{augustin_computationally_2021, frohlich_numerical_2023, fritz_simulation_2014, brown_modular_2024, regazzoni_cardiac_2022, bucelli_mathematical_2023,  barnafi_comparative_2023}. To solve the resulting linear system at each Newton iteration, the Generalized Minimal Residual method (GMRES) is common \cite{shi_optimization_2024, augustin_computationally_2021, frohlich_numerical_2023, 
regazzoni_cardiac_2022, bucelli_mathematical_2023, barnafi_comparative_2023}, but direct methods, like LU decomposition \cite{fritz_simulation_2014} have also been applied. Linear system preconditioners, including Balancing Domain Decomposition by Constraints \cite{barnafi_comparative_2023}, Algebraic Multigrid \cite{barnafi_comparative_2023}, Additive Schwarz \cite{shi_optimization_2024}, and incomplete LU with/without thresholding \cite{shi_optimization_2024, kariya_personalized_2020} have been used. Linear solvers and preconditioners are often used from packages like MUMPS, PETSc, and Trilinos \cite{arostica_software_2025}. Note that these solver settings apply only when simulating tissue mechanics. Other physics, notably electrophysiology, have different spatial and temporal discretization requirements \cite{krishnamoorthi_simulation_2014}, and can introduce particular nonlinear and linear solver challenges \cite{fedele_comprehensive_2023}. In the remainder of this section, we delve into these topics further, reviewing key modeling challenges in multiphysics heart modeling and the corresponding developments in numerical techniques designed to address them.

In modeling the passive mechanical behavior of myocardium, a major factor concerning the numerical design comes from the (quasi-)incompressibility of the material. This property essentially dictates the choice of the variational formulation and element technology. Conventionally, finite strain problems are formulated on the basis of the principle of stationary potential, wherein the bulk modulus is set to be substantially larger than the shear moduli to penalize the incompressibility condition \cite{fedele_comprehensive_2023,frohlich_numerical_2023, rossi_thermodynamically_2014}. However, that formulation, when paired with low-order elements, is known to suffer from mesh locking and pressure instability. To mitigate this issue, special element technologies have been developed, including the F-bar projection technique \cite{augustin_anatomically_2016,elguedj_b-bar_2008} and higher-order elements \cite{heisserer_volumetric_2008,yosibash_p-fems_2012}. In particular, the Hermite cubic element has been particularly favored in studying cardiac mechanics \cite{costa_three-dimensional_1996,kerckhoffs_coupling_2007,noauthor_continuity_nodate}. Nonetheless, a persistent drawback of the penalty-based formulation is the resulting near-singular system, which poses a significant challenge for linear solvers. In many cases, robust direct solvers are required to ensure numerical stability.

Alternatively, the incompressibility constraint can be enforced via a Lagrange multiplier, introduced as an additional field that is stress-like and sometimes simply referred to as the pressure. This leads to the so-called two-field variational or mixed formulation, in which both displacement and pressure are treated as primary unknowns. Unlike the pure displacement formulation, this approach enforces the incompressibility constraint equation directly without suffering from the singular discrete problem. Its mathematical form is of the saddle-point nature and requires the use an element pair that satisfies the celebrated inf-sup condition \cite{brezzi_mixed_1991}. Not all combinations of shape functions are inf-sup stable. The most common choice is a (tri-)quadratic interpolation for displacement and a (tri-)linear interpolation for the pressure, known as the Taylor-Hood element and denoted by $P_2/P_1$ for tetrahedra and $Q_2/Q_1$ for hexahedra \cite{hadjicharalambous_investigating_2021, asner_patient-specific_2017, gurev_high-resolution_2015,land_verification_2015,nobile_active_2012}. However, using Taylor-Hood elements necessitates generating a higher-order mesh for the kinematic field. Moreover, higher-order $C^0$-continuous Lagrange elements are not sufficiently robust for finite strain analysis \cite{lipton_robustness_2010}. Consequently, low-order element pairs are favored in practice. A widely used low-order choice is trilinear displacement paired with piecewise constant pressure, denoted as the $Q_1/Q_0$ or $Q_1/P_0$ element \cite{fritz_simulation_2014, land_verification_2015,gultekin_quasi-incompressible_2019}. Nevertheless, this element pair is marginally inf-sup stable, and may exhibit spurious pressure modes under certain loading conditions. Additionally, generating a full hexahedral mesh remains challenging for cardiac simulations \cite{gurev_high-resolution_2015}, making tetrahedral elements more desirable, especially for patient-specific geometries \cite{fedele_comprehensive_2023, frohlich_numerical_2023, guan_modelling_2021, bucelli_mathematical_2023}. Unfortunately, the $P_1/P_0$ element is not inf-sup stable and exhibits volumetric locking, although it has been applied in cardiac simulations \cite{augustin_anatomically_2016}. We refer readers to a systematic comparison of a variety of elements for finite strain analysis \cite{chamberland_comparison_2010}. Higher-order elements are appealing because of their superior accuracy per degrees-of-freedom. Recent developments have been made with inspiration from isogeometric analysis, and higher-order, higher-continuity splines have have been demonstrated to be a robust and accurate element technology in modeling tissue behavior \cite{torre_isogeometric_2023,liu_energy-stable_2019}. Additionally, meshless methods, originally proposed to circumvent mesh distortion and meshing complexities, have demonstrated promising performance in cardiac electromechanical simulations \cite{lluch_calibration_2020}. To conclude, we mention that the variational formulation is sometimes further extended by introducing the dilatation as an independent field, leading to a three-field variational formulation \cite{pezzuto_orthotropic_2014}.

The convenience of linear tetrahedral elements has spurred extensive research into stabilization strategies within the variational formulation. This line of development originated in computational fluid mechanics through the introduction of the Petrov-Galerkin formulation \cite{hughes_new_1986,franca_stabilized_1992}. It provides a foundation for circumventing the inf-sup condition by adding consistent stabilization terms. A rigorous analysis was performed that guarantees convergence using any combination of elements in the mixed formulation. Building on this, the variational multiscale (VMS) formulation was introduced as a unifying framework \cite{hughes_multiscale_1995}, in which the stabilization terms can be viewed as the subgrid scale models that approximate the impact of the unresolved parts of the solution on the discrete solution. In fluid dynamics, the VMS formulation not only offers a way of handling the pressure instability, but also provides an implicit large eddy simulation mechanism by modeling unresolved scales \cite{bazilevs_computational_2013,john_numerical_2010}. To date, the VMS formulation forms a cornerstone for patient-specific hemodynamic simulations \cite{updegrove_simvascular_2017,figueroa_blood_2017}. Importantly, because the VMS formulation is derived from a Galerkin framework, it is well-suited to fluid-structure interaction (FSI) where both fluid and solid subdomains are discretized using the finite element method \cite{bazilevs_computational_2013}. This compatibility has led to the widespread adoption of VMS-based FSI solvers in cardiac modeling, including simulations of ventricular mechanics \cite{quarteroni_integrated_2017, bucelli_mathematical_2023, tagliabue_fluid_2017,karabelas_towards_2018} and whole-heart modeling \cite{zingaro_electromechanics-driven_2024,karabelas_global_2022}.

A recent development in continuum modeling involves the use of the Gibbs free energy as the thermodynamic potential \cite{liu_unified_2018}. It leads to a unified continuum model capable of describing viscous fluid flow and finite strain deformation. This feature makes it appealing in designing numerical formulations for FSI as it enables consistent numerical treatment and monolithic coupling of the two sub-problems \cite{liu_fluid-structure_2020,sun_modeling_2025}. When applied to finite strain problems, this approach yields a mixed formulation distinct from the classical two-field or multi-field variational principles. It incorporates compressible and incompressible material behaviors in a single mixed formulation through the introduction of an isothermal compressibility factor. For fully incompressible materials, the constraint naturally emerges as a divergence-free condition for the velocity, identical to the constraint in the Navier-Stokes equations. An appealing feature of the formulation is its embedded nonlinear stability \cite{liu_energy-stable_2019}, which has been leveraged to construct structure-preserving time integrators for long-term elastodynamic simulations \cite{guan_structure-preserving_2023}. Furthermore, one may directly invoke the VMS formulation to introduce pressure stabilization for the mixed formulation, which allows equal-order interpolation for the kinematic and pressure fields \cite{liu_unified_2018}. The effectiveness of the approach has been demonstrated in studies of a fiber-reinforced anisotropic hyperelastic material \cite{liu_energy-stable_2019,liu_robust_2019}. Recently, this approach has been applied to cardiac mechanics \cite{shi_optimization_2024, arostica_software_2025}. 

Additional numerical challenges come from the spatial and temporal multiscale nature of cardiac electromechanics. As discussed in Section \ref{subsect:electrophysiology_and_cellular_contraction}, the electrophysiology problem is governed by a system of nonlinear reaction-diffusion equations, whose solution manifests as a propagating wave. During sinus rhythm, the action potential travels as a wavefront across the myocardium followed by a repolarization phase to return to its resting value. Capturing such wavefront dynamics imposes strict numerical requirements, as the steep gradients can easily give rise to spurious undershoots and overshoots. Consequently, numerical stability becomes a primary design concern. Typically, a relatively fine spatial resolution is needed to capture the rapid voltage upstroke and conduction for cardiac electrophysiology \cite{augustin_anatomically_2016}. Typical mesh element sizes are on the order of a few hundred microns. In contrast, the mechanical deformation of the ventricular myocardium is smoother, allowing for a coarser mesh resolution. A straightforward approach is to use a single mesh with resolution sufficiently high for electrophysiology. An obvious drawback is that this leads to an excessive number of degrees of freedom for the mechanics problem ($O(10^8)$ degrees-of-freedom for the human heart). To address this mismatch, several multiscale strategies have been proposed. One class of approaches employs adaptive mesh refinement to dynamically resolve the wavefront \cite{deuflhard_adaptive_2009}. Another class adopts a multi-mesh strategy, where separate meshes with different resolutions are used for the electrical and mechanical subproblems \cite{salvador_intergrid_2020, buist_deformable_2003,chapelle_numerical_2009}. In such cases, the electrophysiology data needs to be interpolated to the mechanics mesh during the simulation. If the two meshes are nested with the finer mesh generated by refining the coarse mesh, the interpolation across the two meshes is relatively straightforward \cite{capuano_personalized_2024, colli_franzone_numerical_2018}. However, generating such nested meshes demands extra care due to the geometrical and functional heterogeneity of the heart. If the two meshes are generated independently and do not share nodal alignment, more sophisticated projection techniques are necessary to transfer physical fields. For example, an intergrid transfer operator based on the radial basis function has been proposed, demonstrating superior accuracy and scalability \cite{salvador_intergrid_2020}.

The temporal multiscale phenomena also plays a critical role in the numerical treatment of cardiac electromechanics. The semi-discrete problem of electrophysiology exhibits significant stiffness in time \cite{spiteri_stiffness_2010}. To accurately resolve these fast transients, time marching must be performed using very small time step size, typically on the order of sub-milliseconds. In contrast, the mechanical responses can be adequately characterized with a time step size of around one millisecond, especially when using implicit time integration \cite{augustin_computationally_2021, hirschvogel_monolithic_2017, brown_modular_2024}. The temporal disparity strongly influences the design of coupling strategies, often requiring a balance between stability and computational efficiency. In monolithic approaches, one formulates a single large system of equations, encompassing both electrical and mechanical sub-problems, permitting stable solutions with larger time steps \cite{dal_fully_2013,wong_computational_2013}. However, solving the resulting large-scale nonlinear system efficiently necessitates careful design of robust linear solvers and preconditioners \cite{gerbi_monolithic_2019}. Moreover, a straightforward monolithic integration with a single global time step is often computationally prohibitive due to fine temporal resolution set by the electrophysiology. To address this, operator-splitting strategies are frequently employed \cite{sundnes_improved_2014}. In such schemes, the mechanical problem is advanced using a large time step, while the electrophysiology is integrated by subcycling using smaller time steps within each mechanical step. This approach effectively exploits the temporal scale separation and improves computational efficiency. Additionally, multiscale behavior is present within the electrophysiology model itself. Fortunately, the reaction–diffusion nature of the governing equations allows the stiff ionic reactions and the non-stiff diffusion terms to be integrated separately \cite{krishnamoorthi_numerical_2013}.

To conclude, we summarize prominent open-source software frameworks in the cardiac modeling community. Continuity, as a pioneering software environment for cardiac modeling, employs the finite element to model cardiac electromechanics \cite{noauthor_continuity_nodate}. Besides conventional finite element technology, it is also supports Hermite spline basis functions. Chaste is a highly object-oriented library for computational biology, featuring a well-validated cardiac modeling module \cite{pitt-francis_chaste_2009}. The finite element library \texttt{life}$^{\mathrm x}$ \cite{africa_lifex_2022}, built on top of deal.II \cite{arndt_dealii_nodate}, has undergone continuous development to incorporate detailed electrophysiology \cite{africa_lifex-ep_2023}, tissue mechanics \cite{africa_lifex-fiber_2023}, and blood flow \cite{africa_lifex-cfd_2024}. The SimVascular project \cite{updegrove_simvascular_2017} offers a complete pipeline for patient-specific cardiovascular modeling. Its new module, svMultiPhysics (\url{https://github.com/SimVascular/svMultiPhysics}
), builds on its Fortran-based predecessor svFSI \cite{zhu_svfsi_2022}, providing an efficient MPI-parallelized C++ solver with multiphysics capabilities, including electrophysiology and fluid–structure interaction. It also supports advanced linear algebra libraries such as Trilinos \cite{team_trilinos_nodate} and PETSc \cite{balay_petsc_2019}. Another notable framework is $\mathcal{C}$Heart, which offers comprehensive multiphysics capabilities specifically tailored for cardiac simulations \cite{lee_multiphysics_2016}. Other notable open-source multiphysics codes include 4C \cite{noauthor_4c_nodate}, written in C++ and originating from the in-house BACI code \cite{hirschvogel_monolithic_2017} at the Technical University of Munich, and Ambit \cite{hirschvogel_ambit_2024}, built on top of the Python-based FEniCS library. FEBio \cite{maas_febio_2012}, a multiphysics software tool for general biomechanics, has also been used to study the heart. For electrophysiology, openCARP \cite{plank_opencarp_2021} is a very popular open-source project. We also highlight SlicerHeart \cite{lasso_slicerheart_2022}, an open-source extension for 3D Slicer \cite{noauthor_3d_nodate} designed for advanced image processing, visualization, and (non-simulation) quantitative analysis for cardiovascular medicine. Alongside software packages, the community has developed benchmark problems to validate and compare numerical methods \cite{arostica_software_2025, land_verification_2015}, which are essential for evaluating the performance of different techniques.

\section{Concluding remarks} \label{sect:conclusions}

In this review, we discussed recent advances and open questions in the field of cardiac mechanics modeling, which we believe to be important to the development of cardiac digital twins as predictive clinical tools. In Section \ref{sect:anatomic_model_construction}, we first discussed the anatomy of the heart and the methods used to create cardiac anatomic models and meshes, which is typically the first step in any heart model. Historically, patient-specific models have required labor-intensive manual segmentation of medical images, but recent advances in machine learning are accelerating the process, in both healthy and diseased cases. Next, in Section \ref{sect:myocardial_mesostructure}, we reviewed the mesostructure of the heart, discussing how it is measured experimentally and modeled computationally, and highlighting some ambiguities we believe are important for modelers to be aware of. In Section \ref{sect:constitutive_modeling}, we summarized recent developments in three aspects of myocardial constitutive modeling -- viscosity, compressibility, and fiber dispersion, and discussed how they influence simulation results. We then reviewed commonly used anatomic model geometries and boundary conditions in Section \ref{sect:anatomy_and_bcs}. Excluding certain cardiac structures from the computational domain, and the boundary conditions necessitated by such exclusions, can have significant impacts on model predictions. In Section \ref{sect:multiphysics_coupling}, we briefly discussed how tissue mechanics is coupled to other physical processes in the heart, including electrophysiology, blood flow, and circulatory dynamics.  Adding these couplings usually increase the physiological fidelity of heart models, but inevitably introduces additional computational challenges. Finally, in Section \ref{sect:numerical_formulations}, we reviewed key topics in the design of numerical formulations for multiphysics simulations of the heart.

Several important topics fell outside the scope of this review, but we would nonetheless be remiss in not mentioning them. Much work has been done to accelerate simulations to make them applicable on clinical timescales. Many groups parallelize various aspects of their codes to run on high performance computing (HPC) clusters \cite{fedele_comprehensive_2023, alfonso_santiago_fully_2018, zhu_svfsi_2022} or with GPU acceleration \cite{viola_gpu_2023}. Other groups have targeted the underlying mathematical formulation \cite{hurtado_accelerating_2021} and preconditioners \cite{barnafi_comparative_2023, hirschvogel_effective_2025} to gain speedup. However, these efforts to accelerate simulations are dwarfed in comparison to machine learning approaches, which have seen explosive growth in recent years \cite{regazzoni_machine_2020, buoso_personalising_2021, cicci_efficient_2024, dabiri_application_2020, maso_talou_deep_2020, regazzoni_accelerating_2021, regazzoni_machine_2022, salvador_whole-heart_2024, motiwale_neural_2024}. Cardiac growth and remodeling, the process by which cardiac morphology and structure change in response to mechanical stimuli on long timescales \cite{rodero_advancing_2023, ambrosi_growth_2019}, is also an important component of predictive cardiac medicine under development. We point the reader to recent works in \cite{planinc_comprehensive_2021, lee_multiphysical_2023, gebauer_homogenized_2023,schwarz_fluidsolid-growth_2023, guan_updated_2023}, and in particular to the reviews in Niestrawska et al. \cite{niestrawska_computational_2020} and Holzapfel et al. \cite{holzapfel_biomechanics_2025}. Finally, patient-specific modeling, encompassing methods to determine parameter values or other simulation inputs to match continuously-collected experimental or clinical data, are essential to making cardiac digital twins a reality \cite{thangaraj_cardiovascular_2024, sel_building_2024}. We highlight recent works that aim to determine constitutive parameters \cite{shi_optimization_2024, lazarus_sensitivity_2022, lluch_calibration_2020, asner_estimation_2016, borowska_bayesian_2022, miller_implementation_2021}, unloaded configurations \cite{shi_optimization_2024, barnafi_reconstructing_2024}, electrophysiology parameters \cite{capuano_personalized_2024, salvador_digital_2024, kariya_personalized_2020}, cellular contraction parameters \cite{jung_integrated_2022}, and circulatory model parameters \cite{salvador_fast_2023, lluch_calibration_2020}. Sensitivity analyses and parameter estimation for multiphysics heart models can be found in recent works \cite{shi_personalized_2026, strocchi_integrating_2025, salvador_fast_2023, wang_calibration_2025}. Reviews of patient-specific modeling can be found in Bracamonte et al. \cite{bracamonte_patient-specific_2022}, Peirlinck et al. \cite{peirlinck_precision_2021}, and Schwarz et al. \cite{schwarz_beyond_2023}.

As demonstrated throughout this review, much of the computational groundwork for cardiac digital twins has already been laid, providing a strong foundation for future development. The pressing challenge now is to establish their clinical utility, an ambitious goal that will face hurdles in both validation \cite{wang_calibration_2025} and regulatory approval \cite{rodero_advancing_2023}. Establishing the utility of cardiac digital twins is likely to reveal the key limitations that still constrain model accuracy and robustness. These may include, for instance, incomplete representations of myocardial mesostructure, imprecise pericardial boundary conditions, limitations in constitutive modeling, and suboptimal strategies for parameter estimation. Beyond these technical issues, practical barriers, such as the time required to translate medical imaging data into actionable simulation results, may further hinder clinical adoption. At the same time, rigorous validation could help clarify which model complexities are essential for clinical relevance, and which may be safely simplified. It remains an open question whether highly detailed, multiphysics, multiscale whole-heart models are necessary for most use cases, or whether simplified models, possibly accelerated by machine learning methods, offer a more pragmatic path forward. To move the field forward, community-wide efforts such as a computational medicine challenge for cardiac modeling, akin to the 2015 International Aneurysm CFD Challenge \cite{valen-sendstad_real-world_2018}, could play a pivotal role in consolidating modeling practices and benchmarking performance.

\section*{Acknowledgements}
AB acknowledges the National Institutes of Health (grant numbers 5R01HL159970 and 5R01HL129727). JL acknowledges the National Natural Science Foundation of China (grant numbers 12172160, 12472201) and Shenzhen Science and Technology Program (grant number JCYJ20220818100600002). DE and AM acknowledge the National Institutes of Health (grant number 1R01HL173845). AM additionally acknowledges the National Science Foundation (Scientific Software Integration grant number 2310909), the Additional Ventures Foundation, and the Advanced Research Projects Agency for Health (ARPA-H).

\bibliographystyle{unsrtnat}
\bibliography{references}  

@article{michler_computationally_2013,
	title = {A computationally efficient framework for the simulation of cardiac perfusion using a multi-compartment {Darcy} porous-media flow model},
	volume = {29},
	copyright = {Copyright © 2012 John Wiley \& Sons, Ltd.},
	issn = {2040-7947},
	url = {https://onlinelibrary.wiley.com/doi/abs/10.1002/cnm.2520},
	doi = {10.1002/cnm.2520},
	language = {en},
	number = {2},
	urldate = {2025-06-28},
	journal = {International Journal for Numerical Methods in Biomedical Engineering},
	author = {Michler, C. and Cookson, A. N. and Chabiniok, R. and Hyde, E. and Lee, J. and Sinclair, M. and Sochi, T. and Goyal, A. and Vigueras, G. and Nordsletten, D. A. and Smith, N. P},
	year = {2013},
	pages = {217--232},
}

@article{peirlinck_kinematic_2019,
	title = {Kinematic boundary conditions substantially impact in silico ventricular function},
	volume = {35},
	copyright = {© 2018 John Wiley \& Sons, Ltd.},
	issn = {2040-7947},
	url = {https://onlinelibrary.wiley.com/doi/abs/10.1002/cnm.3151},
	doi = {10.1002/cnm.3151},
	language = {en},
	number = {1},
	urldate = {2024-11-15},
	journal = {International Journal for Numerical Methods in Biomedical Engineering},
	author = {Peirlinck, Mathias and Sack, Kevin L. and De Backer, Pieter and Morais, Pedro and Segers, Patrick and Franz, Thomas and De Beule, Matthieu},
	year = {2019},
	pages = {e3151},
}

@article{pagani_data_2021,
	title = {Data integration for the numerical simulation of cardiac electrophysiology},
	volume = {44},
	copyright = {© 2021 The Authors. Pacing and Clinical Electrophysiology published by Wiley Periodicals LLC},
	issn = {1540-8159},
	url = {https://onlinelibrary.wiley.com/doi/abs/10.1111/pace.14198},
	doi = {10.1111/pace.14198},
	language = {en},
	number = {4},
	urldate = {2025-06-28},
	journal = {Pacing and Clinical Electrophysiology},
	author = {Pagani, Stefano and Dede', Luca and Manzoni, Andrea and Quarteroni, Alfio},
	year = {2021},
	pages = {726--736},
}

@article{gerbi_monolithic_2019,
	title = {A monolithic algorithm for the simulation of cardiac electromechanics in the human left ventricle},
	volume = {1},
	copyright = {2019 The Author(s)},
	issn = {2640-3501},
	url = {http://www.aimspress.com/article/doi/10.3934/Mine.2018.1.1},
	doi = {10.3934/Mine.2018.1.1},
	language = {en},
	number = {1},
	urldate = {2025-05-11},
	journal = {Mathematics in Engineering},
	author = {Gerbi, Antonello and Dede', Luca and Quarteroni, Alfio},
	year = {2019},
	pages = {1--37},
}

@article{tikenogullari_how_2022,
	title = {How viscous is the beating heart? {Insights} from a computational study},
	volume = {70},
	issn = {1432-0924},
	shorttitle = {How viscous is the beating heart?},
	url = {https://doi.org/10.1007/s00466-022-02180-z},
	doi = {10.1007/s00466-022-02180-z},
	language = {en},
	number = {3},
	urldate = {2024-11-15},
	journal = {Computational Mechanics},
	author = {Tikenogullari, Oguz Ziya and Costabal, Francisco Sahli and Yao, Jiang and Marsden, Alison and Kuhl, Ellen},
	month = sep,
	year = {2022},
	pages = {565--579},
}

@article{mebazaa_acute_2004,
	title = {Acute right ventricular failure—from pathophysiology to new treatments},
	volume = {30},
	copyright = {2004 Springer-Verlag},
	issn = {1432-1238},
	url = {https://link-springer-com.laneproxy.stanford.edu/article/10.1007/s00134-003-2025-3},
	doi = {10.1007/s00134-003-2025-3},
	language = {en},
	number = {2},
	urldate = {2025-04-15},
	journal = {Intensive Care Medicine},
	publisher = {Springer Berlin Heidelberg},
	author = {Mebazaa, Alexandre and Karpati, Peter and Renaud, Estelle and Algotsson, Lars},
	month = feb,
	year = {2004},
	pages = {185--196},
}

@inproceedings{kraus_comparison_2023,
	title = {Comparison of {Pericardium} {Modeling} {Approaches} for {Mechanical} {Whole} {Heart} {Simulations}},
	volume = {50},
	issn = {2325-887X},
	url = {https://ieeexplore.ieee.org/abstract/document/10363870},
	doi = {10.22489/CinC.2023.150},
	urldate = {2024-12-11},
	booktitle = {2023 {Computing} in {Cardiology}},
	author = {Krauß, Jonathan and Gerach, Tobias and Loewe, Axel},
	month = oct,
	year = {2023},
	pages = {1--4},
}

@inproceedings{ghebryal_effect_2023,
	title = {Effect of {Varying} {Pericardial} {Boundary} {Conditions} on {Whole} {Heart} {Function}: {A} {Computational} {Study}},
	isbn = {978-3-031-35302-4},
	shorttitle = {Effect of {Varying} {Pericardial} {Boundary} {Conditions} on {Whole} {Heart} {Function}},
	doi = {10.1007/978-3-031-35302-4_56},
	language = {en},
	booktitle = {Functional {Imaging} and {Modeling} of the {Heart}},
	publisher = {Springer Nature Switzerland},
	author = {Ghebryal, Justina and Rodero, Cristobal and Barrows, Rosie K. and Strocchi, Marina and Roney, Caroline H. and Augustin, Christoph M. and Plank, Gernot and Niederer, Steven A.},
	editor = {Bernard, Olivier and Clarysse, Patrick and Duchateau, Nicolas and Ohayon, Jacques and Viallon, Magalie},
	year = {2023},
	pages = {545--554},
}

@article{heisserer_volumetric_2008,
	title = {On volumetric locking-free behaviour of p-version finite elements under finite deformations},
	volume = {24},
	copyright = {Copyright © 2007 John Wiley \& Sons, Ltd.},
	issn = {1099-0887},
	url = {https://onlinelibrary.wiley.com/doi/abs/10.1002/cnm.1008},
	doi = {10.1002/cnm.1008},
	language = {en},
	number = {11},
	urldate = {2025-05-11},
	journal = {Communications in Numerical Methods in Engineering},
	author = {Heisserer, Ulrich and Hartmann, Stefan and Düster, Alexander and Yosibash, Zohar},
	year = {2008},
	pages = {1019--1032},
}

@article{westerhof_arterial_2009,
	title = {The arterial {Windkessel}},
	volume = {47},
	copyright = {International Federation for Medical and Biological Engineering 2009},
	issn = {01400118},
	url = {https://www.proquest.com/docview/211490078/abstract/CFC7CEDB521A43E5PQ/1},
	doi = {10.1007/s11517-008-0359-2},
	language = {English},
	number = {2},
	urldate = {2025-04-05},
	journal = {Medical and Biological Engineering and Computing},
	publisher = {Springer Nature B.V.},
	author = {Westerhof, Nico and Lankhaar, Jan-willem and Westerhof, Berend E.},
	month = feb,
	year = {2009},
	pages = {131--41},
}

@article{washio_ventricular_2016,
	title = {Ventricular fiber optimization utilizing the branching structure},
	volume = {32},
	copyright = {Copyright © 2015 John Wiley \& Sons, Ltd.},
	issn = {2040-7947},
	url = {https://onlinelibrary.wiley.com/doi/abs/10.1002/cnm.2753},
	doi = {10.1002/cnm.2753},
	language = {en},
	number = {7},
	urldate = {2025-04-04},
	journal = {International Journal for Numerical Methods in Biomedical Engineering},
	author = {Washio, Takumi and Yoneda, Kazunori and Okada, Jun-ichi and Kariya, Taro and Sugiura, Seiryo and Hisada, Toshiaki},
	year = {2016},
	pages = {e02753},
}

@misc{wang_calibration_2025,
	title = {Calibration and validation strategy for electromechanical cardiac digital twins},
	copyright = {© 2025, Posted by Cold Spring Harbor Laboratory. This pre-print is available under a Creative Commons License (Attribution-NonCommercial 4.0 International), CC BY-NC 4.0, as described at http://creativecommons.org/licenses/by-nc/4.0/},
	url = {https://www.biorxiv.org/content/10.1101/2025.03.06.638897v1},
	doi = {10.1101/2025.03.06.638897},
	language = {en},
	urldate = {2025-05-08},
	publisher = {bioRxiv},
	author = {Wang, Zhinuo Jenny and Holmes, Maxx and Doste, Ruben and Camps, Julia and Margara, Francesca and Vazquez, Mariano and Rodriguez, Blanca},
	month = mar,
	year = {2025},
}

@article{vaillant_influence_2025,
	title = {Influence of pericardium on ventricular mechanical interdependence in an isolated biventricular working pig heart model},
	volume = {603},
	copyright = {© 2024 The Author(s). The Journal of Physiology published by John Wiley \& Sons Ltd on behalf of The Physiological Society.},
	issn = {1469-7793},
	url = {https://onlinelibrary.wiley.com/doi/abs/10.1113/JP286259},
	doi = {10.1113/JP286259},
	language = {en},
	number = {2},
	urldate = {2025-05-08},
	journal = {The Journal of Physiology},
	author = {Vaillant, Fanny and Abell, Emma and Bear, Laura R. and Caluori, Guido and Belterman, Charly and Coronel, Ruben and Ploux, Sylvain and Santos, Pierre Dos},
	year = {2025},
	pages = {285--300},
}

@article{updegrove_simvascular_2017,
	title = {{SimVascular}: {An} {Open} {Source} {Pipeline} for {Cardiovascular} {Simulation}},
	volume = {45},
	copyright = {2016 Biomedical Engineering Society},
	issn = {1573-9686},
	shorttitle = {{SimVascular}},
	url = {https://link.springer.com/article/10.1007/s10439-016-1762-8},
	doi = {10.1007/s10439-016-1762-8},
	language = {en},
	number = {3},
	urldate = {2025-05-11},
	journal = {Annals of Biomedical Engineering},
	publisher = {Springer US},
	author = {Updegrove, Adam and Wilson, Nathan M. and Merkow, Jameson and Lan, Hongzhi and Marsden, Alison L. and Shadden, Shawn C.},
	month = mar,
	year = {2017},
	pages = {525--541},
}

@article{tonini_two_2025,
	title = {Two {New} {Calibration} {Techniques} of {Lumped}-{Parameter} {Mathematical} {Models} for the {Cardiovascular} {System}},
	volume = {126},
	copyright = {© 2024 The Author(s). International Journal for Numerical Methods in Engineering published by John Wiley \& Sons Ltd.},
	issn = {1097-0207},
	url = {https://onlinelibrary.wiley.com/doi/abs/10.1002/nme.7648},
	doi = {10.1002/nme.7648},
	language = {en},
	number = {1},
	urldate = {2025-04-15},
	journal = {International Journal for Numerical Methods in Engineering},
	author = {Tonini, Andrea and Regazzoni, Francesco and Salvador, Matteo and Dede', Luca and Scrofani, Roberto and Fusini, Laura and Cogliati, Chiara and Pontone, Gianluca and Vergara, Christian and Quarteroni, Alfio},
	year = {2025},
	pages = {e7648},
}

@article{tagliabue_fluid_2017,
	title = {Fluid dynamics of an idealized left ventricle: the extended {Nitsche}'s method for the treatment of heart valves as mixed time varying boundary conditions},
	volume = {85},
	copyright = {Copyright © 2017 John Wiley \& Sons, Ltd.},
	issn = {1097-0363},
	shorttitle = {Fluid dynamics of an idealized left ventricle},
	url = {https://onlinelibrary.wiley.com/doi/abs/10.1002/fld.4375},
	doi = {10.1002/fld.4375},
	language = {en},
	number = {3},
	urldate = {2025-05-11},
	journal = {International Journal for Numerical Methods in Fluids},
	author = {Tagliabue, A. and Dede', L. and Quarteroni, A.},
	year = {2017},
	pages = {135--164},
}

@article{sun_modeling_2025,
	title = {Modeling {Fibrous} {Tissue} in {Vascular} {Fluid}–{Structure} {Interaction}: {A} {Morphology}-{Based} {Pipeline} and {Biomechanical} {Significance}},
	volume = {41},
	copyright = {© 2024 John Wiley \& Sons Ltd.},
	issn = {2040-7947},
	shorttitle = {Modeling {Fibrous} {Tissue} in {Vascular} {Fluid}–{Structure} {Interaction}},
	url = {https://onlinelibrary.wiley.com/doi/abs/10.1002/cnm.3892},
	doi = {10.1002/cnm.3892},
	language = {en},
	number = {1},
	urldate = {2025-05-11},
	journal = {International Journal for Numerical Methods in Biomedical Engineering},
	author = {Sun, Yujie and Huang, Jiayi and Lu, Qingshuang and Yue, Xinhai and Huang, Xuanming and He, Wei and Shi, Yun and Liu, Ju},
	year = {2025},
	pages = {e3892},
}

@article{spiteri_stiffness_2010,
	title = {Stiffness {Analysis} of {Cardiac} {Electrophysiological} {Models}},
	volume = {38},
	copyright = {2010 Biomedical Engineering Society},
	issn = {1573-9686},
	url = {https://link.springer.com/article/10.1007/s10439-010-0100-9},
	doi = {10.1007/s10439-010-0100-9},
	language = {en},
	number = {12},
	urldate = {2025-05-11},
	journal = {Annals of Biomedical Engineering},
	publisher = {Springer US},
	author = {Spiteri, Raymond J. and Dean, Ryan C.},
	month = dec,
	year = {2010},
	pages = {3592--3604},
}

@article{rodriguez_noninvasive_2006,
	title = {Noninvasive measurement of myocardial tissue volume change during systolic contraction and diastolic relaxation in the canine left ventricle},
	volume = {55},
	copyright = {Published 2006 Wiley-Liss, Inc.},
	issn = {1522-2594},
	url = {https://onlinelibrary.wiley.com/doi/abs/10.1002/mrm.20786},
	doi = {10.1002/mrm.20786},
	language = {en},
	number = {3},
	urldate = {2025-06-30},
	journal = {Magnetic Resonance in Medicine},
	author = {Rodriguez, Ignacio and Ennis, Daniel B. and Wen, Han},
	year = {2006},
	pages = {484--490},
}

@article{stella_fast_2022,
	title = {A fast cardiac electromechanics model coupling the {Eikonal} and the nonlinear mechanics equations},
	volume = {32},
	issn = {0218-2025},
	url = {https://www.worldscientific.com/doi/10.1142/S021820252250035X},
	doi = {10.1142/S021820252250035X},
	number = {08},
	urldate = {2025-04-03},
	journal = {Mathematical Models and Methods in Applied Sciences},
	publisher = {World Scientific Publishing Co.},
	author = {Stella, Simone and Regazzoni, Francesco and Vergara, Christian and Dede', Luca and Quarteroni, Alfio},
	month = jul,
	year = {2022},
	pages = {1531--1556},
}

@article{salvador_whole-heart_2024,
	title = {Whole-heart electromechanical simulations using {Latent} {Neural} {Ordinary} {Differential} {Equations}},
	volume = {7},
	copyright = {2024 The Author(s)},
	issn = {2398-6352},
	url = {https://www.nature.com/articles/s41746-024-01084-x},
	doi = {10.1038/s41746-024-01084-x},
	language = {en},
	number = {1},
	urldate = {2024-10-23},
	journal = {npj Digital Medicine},
	publisher = {Nature Publishing Group},
	author = {Salvador, Matteo and Strocchi, Marina and Regazzoni, Francesco and Augustin, Christoph M. and Dede', Luca and Niederer, Steven A. and Quarteroni, Alfio},
	month = apr,
	year = {2024},
	pages = {1--8},
}

@article{salvador_role_2022,
	title = {The role of mechano-electric feedbacks and hemodynamic coupling in scar-related ventricular tachycardia},
	volume = {142},
	issn = {0010-4825},
	url = {https://www.sciencedirect.com/science/article/pii/S0010482521009975},
	doi = {10.1016/j.compbiomed.2021.105203},
	urldate = {2024-10-23},
	journal = {Computers in Biology and Medicine},
	author = {Salvador, Matteo and Regazzoni, Francesco and Pagani, Stefano and Dede', Luca and Trayanova, Natalia and Quarteroni, Alfio},
	month = mar,
	year = {2022},
	pages = {105203},
}

@article{salvador_fast_2023,
	title = {Fast and robust parameter estimation with uncertainty quantification for the cardiac function},
	volume = {231},
	issn = {0169-2607},
	url = {https://www.sciencedirect.com/science/article/pii/S016926072300069X},
	doi = {10.1016/j.cmpb.2023.107402},
	urldate = {2024-11-04},
	journal = {Computer Methods and Programs in Biomedicine},
	author = {Salvador, Matteo and Regazzoni, Francesco and Dede', Luca and Quarteroni, Alfio},
	month = apr,
	year = {2023},
	pages = {107402},
}

@article{salvador_intergrid_2020,
	title = {An intergrid transfer operator using radial basis functions with application to cardiac electromechanics},
	volume = {66},
	issn = {1432-0924},
	url = {https://doi.org/10.1007/s00466-020-01861-x},
	doi = {10.1007/s00466-020-01861-x},
	language = {en},
	number = {2},
	urldate = {2024-10-23},
	journal = {Computational Mechanics},
	author = {Salvador, Matteo and Dede', Luca and Quarteroni, Alfio},
	month = aug,
	year = {2020},
	pages = {491--511},
}

@article{regazzoni_biophysically_2020,
	title = {Biophysically detailed mathematical models of multiscale cardiac active mechanics},
	volume = {16},
	issn = {1553-7358},
	url = {https://journals.plos.org/ploscompbiol/article?id=10.1371/journal.pcbi.1008294},
	doi = {10.1371/journal.pcbi.1008294},
	language = {en},
	number = {10},
	urldate = {2024-10-22},
	journal = {PLOS Computational Biology},
	publisher = {Public Library of Science},
	author = {Regazzoni, Francesco and Dede', Luca and Quarteroni, Alfio},
	month = oct,
	year = {2020},
	pages = {e1008294},
}

@article{regazzoni_machine_2022,
	title = {A machine learning method for real-time numerical simulations of cardiac electromechanics},
	volume = {393},
	issn = {0045-7825},
	url = {https://www.sciencedirect.com/science/article/pii/S004578252200144X},
	doi = {10.1016/j.cma.2022.114825},
	urldate = {2024-10-22},
	journal = {Computer Methods in Applied Mechanics and Engineering},
	author = {Regazzoni, F. and Salvador, M. and Dede', L. and Quarteroni, A.},
	month = apr,
	year = {2022},
	pages = {114825},
}

@article{regazzoni_cardiac_2022,
	title = {A cardiac electromechanical model coupled with a lumped-parameter model for closed-loop blood circulation},
	volume = {457},
	issn = {0021-9991},
	url = {https://www.sciencedirect.com/science/article/pii/S0021999122001450},
	doi = {10.1016/j.jcp.2022.111083},
	urldate = {2024-10-23},
	journal = {Journal of Computational Physics},
	author = {Regazzoni, F. and Salvador, M. and Africa, P. C. and Fedele, M. and Dede', L. and Quarteroni, A.},
	month = may,
	year = {2022},
	pages = {111083},
}

@article{regazzoni_machine_2020,
	title = {Machine learning of multiscale active force generation models for the efficient simulation of cardiac electromechanics},
	volume = {370},
	issn = {0045-7825},
	url = {https://www.sciencedirect.com/science/article/pii/S0045782520304539},
	doi = {10.1016/j.cma.2020.113268},
	urldate = {2024-10-23},
	journal = {Computer Methods in Applied Mechanics and Engineering},
	author = {Regazzoni, F. and Dede', L. and Quarteroni, A.},
	month = oct,
	year = {2020},
	pages = {113268},
}

@article{radisic_influence_2025,
	title = {Influence of cellular mechano-calcium feedback in numerical models of cardiac electromechanics},
	volume = {445},
	issn = {0045-7825},
	url = {https://www.sciencedirect.com/science/article/pii/S004578252500444X},
	doi = {10.1016/j.cma.2025.118172},
	urldate = {2025-08-29},
	journal = {Computer Methods in Applied Mechanics and Engineering},
	author = {Radišić, Irena and Regazzoni, Francesco and Bucelli, Michele and Pagani, Stefano and Dede', Luca and Quarteroni, Alfio},
	month = oct,
	year = {2025},
	pages = {118172},
}

@article{piersanti_3d0d_2022,
	title = {{3D}–{0D} closed-loop model for the simulation of cardiac biventricular electromechanics},
	volume = {391},
	issn = {0045-7825},
	url = {https://www.sciencedirect.com/science/article/pii/S0045782522000251},
	doi = {10.1016/j.cma.2022.114607},
	urldate = {2024-10-23},
	journal = {Computer Methods in Applied Mechanics and Engineering},
	author = {Piersanti, Roberto and Regazzoni, Francesco and Salvador, Matteo and Corno, Antonio F. and Dede', Luca and Vergara, Christian and Quarteroni, Alfio},
	month = mar,
	year = {2022},
	pages = {114607},
}

@article{piersanti_modeling_2021,
	title = {Modeling cardiac muscle fibers in ventricular and atrial electrophysiology simulations},
	volume = {373},
	issn = {00457825},
	url = {https://linkinghub.elsevier.com/retrieve/pii/S0045782520306538},
	doi = {10.1016/j.cma.2020.113468},
	language = {en},
	urldate = {2024-11-06},
	journal = {Computer Methods in Applied Mechanics and Engineering},
	author = {Piersanti, Roberto and Africa, Pasquale C. and Fedele, Marco and Vergara, Christian and Dede', Luca and Corno, Antonio F. and Quarteroni, Alfio},
	month = jan,
	year = {2021},
	pages = {113468},
}

@article{fedele_comprehensive_2023,
	title = {A comprehensive and biophysically detailed computational model of the whole human heart electromechanics},
	volume = {410},
	issn = {0045-7825},
	url = {https://www.sciencedirect.com/science/article/pii/S0045782523001068},
	doi = {10.1016/j.cma.2023.115983},
	urldate = {2024-10-19},
	journal = {Computer Methods in Applied Mechanics and Engineering},
	author = {Fedele, Marco and Piersanti, Roberto and Regazzoni, Francesco and Salvador, Matteo and Africa, Pasquale Claudio and Bucelli, Michele and Zingaro, Alberto and Dede', Luca and Quarteroni, Alfio},
	month = may,
	year = {2023},
	pages = {115983},
}

@article{bucelli_mathematical_2023,
	title = {A mathematical model that integrates cardiac electrophysiology, mechanics, and fluid dynamics: {Application} to the human left heart},
	volume = {39},
	copyright = {© 2023 The Authors. International Journal for Numerical Methods in Biomedical Engineering published by John Wiley \& Sons Ltd.},
	issn = {2040-7947},
	shorttitle = {A mathematical model that integrates cardiac electrophysiology, mechanics, and fluid dynamics},
	url = {https://onlinelibrary.wiley.com/doi/abs/10.1002/cnm.3678},
	doi = {10.1002/cnm.3678},
	language = {en},
	number = {3},
	urldate = {2024-10-22},
	journal = {International Journal for Numerical Methods in Biomedical Engineering},
	author = {Bucelli, Michele and Zingaro, Alberto and Africa, Pasquale Claudio and Fumagalli, Ivan and Dede', Luca and Quarteroni, Alfio},
	year = {2023},
	pages = {e3678},
}

@article{barnafi_wittwer_multiscale_2022,
	title = {A {Multiscale} {Poromechanics} {Model} {Integrating} {Myocardial} {Perfusion} and the {Epicardial} {Coronary} {Vessels}},
	volume = {82},
	issn = {0036-1399},
	url = {https://epubs.siam.org/doi/abs/10.1137/21M1424482},
	doi = {10.1137/21M1424482},
	number = {4},
	urldate = {2024-10-23},
	journal = {SIAM Journal on Applied Mathematics},
	publisher = {Society for Industrial and Applied Mathematics},
	author = {Barnafi Wittwer, Nicolás Alejandro and Gregorio, Simone Di and Dede', Luca and Zunino, Paolo and Vergara, Christian and Quarteroni, Alfio},
	month = aug,
	year = {2022},
	pages = {1167--1193},
}

@article{arostica_software_2025,
	title = {A software benchmark for cardiac elastodynamics},
	volume = {435},
	issn = {0045-7825},
	url = {https://www.sciencedirect.com/science/article/pii/S0045782524007394},
	doi = {10.1016/j.cma.2024.117485},
	urldate = {2024-12-19},
	journal = {Computer Methods in Applied Mechanics and Engineering},
	author = {Aróstica, Reidmen and Nolte, David and Brown, Aaron and Gebauer, Amadeus and Karabelas, Elias and Jilberto, Javiera and Salvador, Matteo and Bucelli, Michele and Piersanti, Roberto and Osouli, Kasra and Augustin, Christoph and Finsberg, Henrik and Shi, Lei and Hirschvogel, Marc and Pfaller, Martin and Africa, Pasquale Claudio and Gsell, Matthias and Marsden, Alison and Nordsletten, David and Regazzoni, Francesco and Plank, Gernot and Sundnes, Joakim and Dede', Luca and Peirlinck, Mathias and Vedula, Vijay and Wall, Wolfgang and Bertoglio, Cristóbal},
	month = feb,
	year = {2025},
	pages = {117485},
}

@article{africa_lifex-ep_2023,
	title = {lifex-ep: a robust and efficient software for cardiac electrophysiology simulations},
	volume = {24},
	issn = {1471-2105},
	shorttitle = {lifex-ep},
	url = {https://doi.org/10.1186/s12859-023-05513-8},
	doi = {10.1186/s12859-023-05513-8},
	number = {1},
	urldate = {2025-05-11},
	journal = {BMC Bioinformatics},
	author = {Africa, Pasquale Claudio and Piersanti, Roberto and Regazzoni, Francesco and Bucelli, Michele and Salvador, Matteo and Fedele, Marco and Pagani, Stefano and Dede', Luca and Quarteroni, Alfio},
	month = oct,
	year = {2023},
	pages = {389},
}

@article{africa_lifex-fiber_2023,
	title = {lifex-fiber: an open tool for myofibers generation in cardiac computational models},
	volume = {24},
	issn = {1471-2105},
	shorttitle = {lifex-fiber},
	url = {https://doi.org/10.1186/s12859-023-05260-w},
	doi = {10.1186/s12859-023-05260-w},
	language = {en},
	number = {1},
	urldate = {2024-10-23},
	journal = {BMC Bioinformatics},
	author = {Africa, Pasquale Claudio and Piersanti, Roberto and Fedele, Marco and Dede', Luca and Quarteroni, Alfio},
	month = apr,
	year = {2023},
	pages = {143},
}

@article{africa_lifex-cfd_2024,
	title = {lifex-cfd: {An} open-source computational fluid dynamics solver for cardiovascular applications},
	volume = {296},
	issn = {0010-4655},
	shorttitle = {lifex-cfd},
	url = {https://www.sciencedirect.com/science/article/pii/S0010465523003843},
	doi = {10.1016/j.cpc.2023.109039},
	urldate = {2025-05-11},
	journal = {Computer Physics Communications},
	author = {Africa, Pasquale Claudio and Fumagalli, Ivan and Bucelli, Michele and Zingaro, Alberto and Fedele, Marco and Dede', Luca and Quarteroni, Alfio},
	month = mar,
	year = {2024},
	pages = {109039},
}

@article{potse_comparison_2006,
	title = {A {Comparison} of {Monodomain} and {Bidomain} {Reaction}-{Diffusion} {Models} for {Action} {Potential} {Propagation} in the {Human} {Heart}},
	volume = {53},
	issn = {1558-2531},
	url = {https://ieeexplore.ieee.org/document/4015619/?arnumber=4015619},
	doi = {10.1109/TBME.2006.880875},
	number = {12},
	urldate = {2025-04-03},
	journal = {IEEE Transactions on Biomedical Engineering},
	author = {Potse, Mark and Dube, Bruno and Richer, Jacques and Vinet, Alain and Gulrajani, Ramesh M.},
	month = dec,
	year = {2006},
	pages = {2425--2435},
}

@article{liu_energy-stable_2019,
	title = {An energy-stable mixed formulation for isogeometric analysis of incompressible hyperelastodynamics},
	volume = {120},
	copyright = {© 2019 John Wiley \& Sons, Ltd.},
	issn = {1097-0207},
	url = {https://onlinelibrary.wiley.com/doi/abs/10.1002/nme.6165},
	doi = {10.1002/nme.6165},
	language = {en},
	number = {8},
	urldate = {2025-05-11},
	journal = {International Journal for Numerical Methods in Engineering},
	author = {Liu, Ju and Marsden, Alison L. and Tao, Zhen},
	year = {2019},
	pages = {937--963},
}

@article{lee_probabilistic_2024,
	title = {A probabilistic neural twin for treatment planning in peripheral pulmonary artery stenosis},
	volume = {40},
	copyright = {© 2024 John Wiley \& Sons Ltd.},
	issn = {2040-7947},
	url = {https://onlinelibrary.wiley.com/doi/abs/10.1002/cnm.3820},
	doi = {10.1002/cnm.3820},
	language = {en},
	number = {5},
	urldate = {2025-08-26},
	journal = {International Journal for Numerical Methods in Biomedical Engineering},
	author = {Lee, John D. and Richter, Jakob and Pfaller, Martin R. and Szafron, Jason M. and Menon, Karthik and Zanoni, Andrea and Ma, Michael R. and Feinstein, Jeffrey A. and Kreutzer, Jacqueline and Marsden, Alison L. and Schiavazzi, Daniele E.},
	year = {2024},
	pages = {e3820},
}

@article{lazarus_sensitivity_2022,
	title = {Sensitivity analysis and inverse uncertainty quantification for the left ventricular passive mechanics},
	volume = {21},
	copyright = {2022 The Author(s)},
	issn = {1617-7940},
	url = {https://link.springer.com/article/10.1007/s10237-022-01571-8},
	doi = {10.1007/s10237-022-01571-8},
	language = {en},
	number = {3},
	urldate = {2025-03-24},
	journal = {Biomechanics and Modeling in Mechanobiology},
	publisher = {Springer Berlin Heidelberg},
	author = {Lazarus, Alan and Dalton, David and Husmeier, Dirk and Gao, Hao},
	month = jun,
	year = {2022},
	pages = {953--982},
}

@article{lafortune_coupled_2012,
	title = {Coupled electromechanical model of the heart: {Parallel} finite element formulation},
	volume = {28},
	copyright = {Copyright © 2012 John Wiley \& Sons, Ltd.},
	issn = {2040-7947},
	shorttitle = {Coupled electromechanical model of the heart},
	url = {https://onlinelibrary.wiley.com/doi/abs/10.1002/cnm.1494},
	doi = {10.1002/cnm.1494},
	language = {en},
	number = {1},
	urldate = {2024-11-18},
	journal = {International Journal for Numerical Methods in Biomedical Engineering},
	author = {Lafortune, Pierre and Arís, Ruth and Vázquez, Mariano and Houzeaux, Guillaume},
	year = {2012},
	pages = {72--86},
}

@article{kung_presence_2011,
	title = {The presence of two local myocardial sheet populations confirmed by diffusion tensor {MRI} and histological validation},
	volume = {34},
	copyright = {Copyright © 2011 Wiley Periodicals, Inc.},
	issn = {1522-2586},
	url = {https://onlinelibrary.wiley.com/doi/abs/10.1002/jmri.22725},
	doi = {10.1002/jmri.22725},
	language = {en},
	number = {5},
	urldate = {2025-06-30},
	journal = {Journal of Magnetic Resonance Imaging},
	author = {Kung, Geoffrey L. and Nguyen, Tom C. and Itoh, Aki and Skare, Stefan and Ingels Jr, Neil B. and Miller, D. Craig and Ennis, Daniel B.},
	year = {2011},
	pages = {1080--1091},
}

@article{kaiser_design-based_2021,
	title = {A design-based model of the aortic valve for fluid-structure interaction},
	volume = {20},
	copyright = {2021 The Author(s), under exclusive licence to Springer-Verlag GmbH Germany, part of Springer Nature},
	issn = {1617-7940},
	url = {https://link.springer.com/article/10.1007/s10237-021-01516-7},
	doi = {10.1007/s10237-021-01516-7},
	language = {en},
	number = {6},
	urldate = {2025-07-01},
	journal = {Biomechanics and Modeling in Mechanobiology},
	publisher = {Springer Berlin Heidelberg},
	author = {Kaiser, Alexander D. and Shad, Rohan and Hiesinger, William and Marsden, Alison L.},
	month = dec,
	year = {2021},
	pages = {2413--2435},
}

@article{hirschvogel_monolithic_2017,
	title = {A monolithic {3D}-{0D} coupled closed-loop model of the heart and the vascular system: {Experiment}-based parameter estimation for patient-specific cardiac mechanics},
	volume = {33},
	copyright = {Copyright © 2016 John Wiley \& Sons, Ltd.},
	issn = {2040-7947},
	shorttitle = {A monolithic {3D}-{0D} coupled closed-loop model of the heart and the vascular system},
	url = {https://onlinelibrary.wiley.com/doi/abs/10.1002/cnm.2842},
	doi = {10.1002/cnm.2842},
	language = {en},
	number = {8},
	urldate = {2024-11-07},
	journal = {International Journal for Numerical Methods in Biomedical Engineering},
	author = {Hirschvogel, Marc and Bassilious, Marina and Jagschies, Lasse and Wildhirt, Stephen M. and Gee, Michael W.},
	year = {2017},
	pages = {e2842},
}

@article{gultekin_orthotropic_2016,
	title = {An orthotropic viscoelastic model for the passive myocardium: continuum basis and numerical treatment},
	volume = {19},
	issn = {1025-5842},
	shorttitle = {An orthotropic viscoelastic model for the passive myocardium},
	url = {https://doi.org/10.1080/10255842.2016.1176155},
	doi = {10.1080/10255842.2016.1176155},
	number = {15},
	urldate = {2024-11-14},
	journal = {Computer Methods in Biomechanics and Biomedical Engineering},
	publisher = {Taylor \& Francis},
	author = {Gültekin, Osman and Sommer, Gerhard and Holzapfel, Gerhard A.},
	month = nov,
	year = {2016},
	pages = {1647--1664},
}

@article{geuzaine_gmsh_2009,
	title = {Gmsh: {A} 3-{D} finite element mesh generator with built-in pre- and post-processing facilities},
	volume = {79},
	copyright = {Copyright © 2009 John Wiley \& Sons, Ltd.},
	issn = {1097-0207},
	shorttitle = {Gmsh},
	url = {https://onlinelibrary.wiley.com/doi/abs/10.1002/nme.2579},
	doi = {10.1002/nme.2579},
	language = {en},
	number = {11},
	urldate = {2025-08-27},
	journal = {International Journal for Numerical Methods in Engineering},
	author = {Geuzaine, Christophe and Remacle, Jean-François},
	year = {2009},
	pages = {1309--1331},
}

@article{gerach_electro-mechanical_2021,
	title = {Electro-{Mechanical} {Whole}-{Heart} {Digital} {Twins}: {A} {Fully} {Coupled} {Multi}-{Physics} {Approach}},
	volume = {9},
	copyright = {http://creativecommons.org/licenses/by/3.0/},
	issn = {2227-7390},
	shorttitle = {Electro-{Mechanical} {Whole}-{Heart} {Digital} {Twins}},
	url = {https://www.mdpi.com/2227-7390/9/11/1247},
	doi = {10.3390/math9111247},
	language = {en},
	number = {11},
	urldate = {2024-10-23},
	journal = {Mathematics},
	publisher = {Multidisciplinary Digital Publishing Institute},
	author = {Gerach, Tobias and Schuler, Steffen and Fröhlich, Jonathan and Lindner, Laura and Kovacheva, Ekaterina and Moss, Robin and Wülfers, Eike Moritz and Seemann, Gunnar and Wieners, Christian and Loewe, Axel},
	month = jan,
	year = {2021},
	pages = {1247},
}

@article{gerach_differential_2024,
	title = {Differential effects of mechano-electric feedback mechanisms on whole-heart activation, repolarization, and tension},
	volume = {602},
	copyright = {© 2024 The Authors. The Journal of Physiology published by John Wiley \& Sons Ltd on behalf of The Physiological Society.},
	issn = {1469-7793},
	url = {https://onlinelibrary.wiley.com/doi/abs/10.1113/JP285022},
	doi = {10.1113/JP285022},
	language = {en},
	number = {18},
	urldate = {2025-08-29},
	journal = {The Journal of Physiology},
	author = {Gerach, Tobias and Loewe, Axel},
	year = {2024},
	pages = {4605--4624},
}

@article{gebauer_homogenized_2023,
	title = {A homogenized constrained mixture model of cardiac growth and remodeling: analyzing mechanobiological stability and reversal},
	volume = {22},
	copyright = {2023 The Author(s)},
	issn = {1617-7940},
	shorttitle = {A homogenized constrained mixture model of cardiac growth and remodeling},
	url = {https://link.springer.com/article/10.1007/s10237-023-01747-w},
	doi = {10.1007/s10237-023-01747-w},
	language = {en},
	number = {6},
	urldate = {2025-04-08},
	journal = {Biomechanics and Modeling in Mechanobiology},
	publisher = {Springer Berlin Heidelberg},
	author = {Gebauer, Amadeus M. and Pfaller, Martin R. and Braeu, Fabian A. and Cyron, Christian J. and Wall, Wolfgang A.},
	month = dec,
	year = {2023},
	pages = {1983--2002},
}

@article{frohlich_numerical_2023,
	title = {Numerical evaluation of elasto-mechanical and visco-elastic electro-mechanical models of the human heart},
	volume = {46},
	copyright = {© 2024 The Authors. GAMM - Mitteilungen published by Wiley-VCH GmbH.},
	issn = {1522-2608},
	url = {https://onlinelibrary.wiley.com/doi/abs/10.1002/gamm.202370010},
	doi = {10.1002/gamm.202370010},
	language = {en},
	number = {3-4},
	urldate = {2024-10-23},
	journal = {GAMM-Mitteilungen},
	author = {Fröhlich, Jonathan and Gerach, Tobias and Krauß, Jonathan and Loewe, Axel and Stengel, Laura and Wieners, Christian},
	year = {2023},
	pages = {e202370010},
}

@incollection{figueroa_blood_2017,
	title = {Blood {Flow}},
	copyright = {Copyright © 2017 John Wiley \& Sons, Ltd. All rights reserved.},
	isbn = {978-1-119-17681-7},
	url = {https://onlinelibrary.wiley.com/doi/abs/10.1002/9781119176817.ecm2068},
	doi = {10.1002/9781119176817.ecm2068},
	language = {en},
	urldate = {2025-05-11},
	booktitle = {Encyclopedia of {Computational} {Mechanics} {Second} {Edition}},
	publisher = {John Wiley \& Sons, Ltd},
	author = {Figueroa, C. Alberto and Taylor, Charles A. and Marsden, Alison L.},
	year = {2017},
	pages = {1--31},
}

@article{fedele_polygonal_2021,
	title = {Polygonal surface processing and mesh generation tools for the numerical simulation of the cardiac function},
	volume = {37},
	copyright = {© 2021 The Authors. International Journal for Numerical Methods in Biomedical Engineering published by John Wiley \& Sons Ltd.},
	issn = {2040-7947},
	url = {https://onlinelibrary.wiley.com/doi/abs/10.1002/cnm.3435},
	doi = {10.1002/cnm.3435},
	language = {en},
	number = {4},
	urldate = {2025-08-27},
	journal = {International Journal for Numerical Methods in Biomedical Engineering},
	author = {Fedele, Marco and Quarteroni, Alfio},
	year = {2021},
	pages = {e3435},
}

@article{fatemifar_computational_2019,
	title = {Computational {Modeling} of {Human} {Left} {Ventricle} to {Assess} the {Effects} of {Trabeculae} {Carneae} on the {Diastolic} and {Systolic} {Functions}},
	volume = {141},
	issn = {0148-0731},
	url = {https://doi.org/10.1115/1.4043831},
	doi = {10.1115/1.4043831},
	number = {9},
	urldate = {2025-03-19},
	journal = {Journal of Biomechanical Engineering},
	author = {Fatemifar, Fatemeh and Feldman, Marc D. and Clarke, Geoffrey D. and Finol, Ender A. and Han, Hai-Chao},
	month = aug,
	year = {2019},
}

@article{esmaily-moghadam_new_2013,
	title = {A new preconditioning technique for implicitly coupled multidomain simulations with applications to hemodynamics},
	volume = {52},
	copyright = {2013 Springer-Verlag Berlin Heidelberg},
	issn = {1432-0924},
	url = {https://link.springer.com/article/10.1007/s00466-013-0868-1},
	doi = {10.1007/s00466-013-0868-1},
	language = {en},
	number = {5},
	urldate = {2025-04-05},
	journal = {Computational Mechanics},
	publisher = {Springer Berlin Heidelberg},
	author = {Esmaily-Moghadam, Mahdi and Bazilevs, Yuri and Marsden, Alison L.},
	month = nov,
	year = {2013},
	pages = {1141--1152},
}

@article{eriksson_modeling_2013,
	title = {Modeling the dispersion in electromechanically coupled myocardium},
	volume = {29},
	copyright = {Copyright © 2013 John Wiley \& Sons, Ltd.},
	issn = {2040-7947},
	url = {https://onlinelibrary.wiley.com/doi/abs/10.1002/cnm.2575},
	doi = {10.1002/cnm.2575},
	language = {en},
	number = {11},
	urldate = {2024-11-12},
	journal = {International Journal for Numerical Methods in Biomedical Engineering},
	author = {Eriksson, Thomas S. E. and Prassl, Anton J. and Plank, Gernot and Holzapfel, Gerhard A.},
	year = {2013},
	pages = {1267--1284},
}

@article{emendi_patient-specific_2021,
	title = {Patient-{Specific} {Bicuspid} {Aortic} {Valve} {Biomechanics}: {A} {Magnetic} {Resonance} {Imaging} {Integrated} {Fluid}–{Structure} {Interaction} {Approach}},
	volume = {49},
	copyright = {2020 Biomedical Engineering Society},
	issn = {1573-9686},
	shorttitle = {Patient-{Specific} {Bicuspid} {Aortic} {Valve} {Biomechanics}},
	url = {https://link.springer.com/article/10.1007/s10439-020-02571-4},
	doi = {10.1007/s10439-020-02571-4},
	language = {en},
	number = {2},
	urldate = {2025-07-01},
	journal = {Annals of Biomedical Engineering},
	publisher = {Springer International Publishing},
	author = {Emendi, Monica and Sturla, Francesco and Ghosh, Ram P. and Bianchi, Matteo and Piatti, Filippo and Pluchinotta, Francesca R. and Giese, Daniel and Lombardi, Massimo and Redaelli, Alberto and Bluestein, Danny},
	month = feb,
	year = {2021},
	pages = {627--641},
}

@article{davey_simulating_2024,
	title = {Simulating cardiac fluid dynamics in the human heart},
	volume = {3},
	issn = {2752-6542},
	url = {https://doi.org/10.1093/pnasnexus/pgae392},
	doi = {10.1093/pnasnexus/pgae392},
	number = {10},
	urldate = {2024-10-23},
	journal = {PNAS Nexus},
	author = {Davey, Marshall and Puelz, Charles and Rossi, Simone and Smith, Margaret Anne and Wells, David R and Sturgeon, Gregory M and Segars, W Paul and Vavalle, John P and Peskin, Charles S and Griffith, Boyce E},
	month = oct,
	year = {2024},
	pages = {392},
}

@article{cicci_efficient_2024,
	title = {Efficient approximation of cardiac mechanics through reduced-order modeling with deep learning-based operator approximation},
	volume = {40},
	copyright = {© 2023 The Authors. International Journal for Numerical Methods in Biomedical Engineering published by John Wiley \& Sons Ltd.},
	issn = {2040-7947},
	url = {https://onlinelibrary.wiley.com/doi/abs/10.1002/cnm.3783},
	doi = {10.1002/cnm.3783},
	language = {en},
	number = {1},
	urldate = {2024-10-23},
	journal = {International Journal for Numerical Methods in Biomedical Engineering},
	author = {Cicci, Ludovica and Fresca, Stefania and Manzoni, Andrea and Quarteroni, Alfio},
	year = {2024},
	pages = {e3783},
}

@article{caforio_physics-informed_2025,
	title = {Physics-informed neural network estimation of material properties in soft tissue nonlinear biomechanical models},
	volume = {75},
	copyright = {2024 The Author(s)},
	issn = {1432-0924},
	url = {https://link.springer.com/article/10.1007/s00466-024-02516-x},
	doi = {10.1007/s00466-024-02516-x},
	language = {en},
	number = {2},
	urldate = {2025-07-02},
	journal = {Computational Mechanics},
	publisher = {Springer Berlin Heidelberg},
	author = {Caforio, Federica and Regazzoni, Francesco and Pagani, Stefano and Karabelas, Elias and Augustin, Christoph and Haase, Gundolf and Plank, Gernot and Quarteroni, Alfio},
	month = feb,
	year = {2025},
	pages = {487--513},
}

@article{buist_deformable_2003,
	title = {A {Deformable} {Finite} {Element} {Derived} {Finite} {Difference} {Method} for {Cardiac} {Activation} {Problems}},
	volume = {31},
	copyright = {2003 Biomedical Engineering Society},
	issn = {1573-9686},
	url = {https://link.springer.com/article/10.1114/1.1567283},
	doi = {10.1114/1.1567283},
	language = {en},
	number = {5},
	urldate = {2025-05-11},
	journal = {Annals of Biomedical Engineering},
	publisher = {Kluwer Academic Publishers-Plenum Publishers},
	author = {Buist, Martin and Sands, Gregory and Hunter, Peter and Pullan, Andrew},
	month = may,
	year = {2003},
	pages = {577--588},
}

@article{bracamonte_patient-specific_2022,
	title = {Patient-{Specific} {Inverse} {Modeling} of {In} {Vivo} {Cardiovascular} {Mechanics} with {Medical} {Image}-{Derived} {Kinematics} as {Input} {Data}: {Concepts}, {Methods}, and {Applications}},
	volume = {12},
	copyright = {http://creativecommons.org/licenses/by/3.0/},
	issn = {2076-3417},
	shorttitle = {Patient-{Specific} {Inverse} {Modeling} of {In} {Vivo} {Cardiovascular} {Mechanics} with {Medical} {Image}-{Derived} {Kinematics} as {Input} {Data}},
	url = {https://www.mdpi.com/2076-3417/12/8/3954},
	doi = {10.3390/app12083954},
	language = {en},
	number = {8},
	urldate = {2024-10-10},
	journal = {Applied Sciences},
	publisher = {Multidisciplinary Digital Publishing Institute},
	author = {Bracamonte, Johane H. and Saunders, Sarah K. and Wilson, John S. and Truong, Uyen T. and Soares, Joao S.},
	month = jan,
	year = {2022},
	pages = {3954},
}

@article{borowska_bayesian_2022,
	title = {Bayesian optimisation for efficient parameter inference in a cardiac mechanics model of the left ventricle},
	volume = {38},
	copyright = {© 2022 The Authors. International Journal for Numerical Methods in Biomedical Engineering published by John Wiley \& Sons Ltd.},
	issn = {2040-7947},
	url = {https://onlinelibrary.wiley.com/doi/abs/10.1002/cnm.3593},
	doi = {10.1002/cnm.3593},
	language = {en},
	number = {5},
	urldate = {2024-10-22},
	journal = {International Journal for Numerical Methods in Biomedical Engineering},
	author = {Borowska, Agnieszka and Gao, Hao and Lazarus, Alan and Husmeier, Dirk},
	year = {2022},
	pages = {e3593},
}

@book{bazilevs_computational_2013,
	edition = {1},
	title = {Computational {Fluid}–{Structure} {Interaction}: {Methods} and {Applications}},
	url = {https://onlinelibrary.wiley.com/doi/10.1002/9781118483565},
	doi = {10.1002/9781118483565},
	urldate = {2025-05-11},
	publisher = {John Wiley \& Sons, Ltd},
	author = {Bazilevs, Y. and Takizawa, K. and Tezduyar, T.E.},
	year = {2013},
}

@article{avazmohammadi_contemporary_2019,
	title = {A {Contemporary} {Look} at {Biomechanical} {Models} of {Myocardium}},
	volume = {21},
	issn = {1523-9829, 1545-4274},
	url = {https://www.annualreviews.org/content/journals/10.1146/annurev-bioeng-062117-121129},
	doi = {10.1146/annurev-bioeng-062117-121129},
	language = {en},
	urldate = {2024-10-10},
	journal = {Annual Review of Biomedical Engineering},
	publisher = {Annual Reviews},
	author = {Avazmohammadi, Reza and Soares, João S. and Li, David S. and Raut, Samarth S. and Gorman, Robert C. and Sacks, Michael S.},
	month = jun,
	year = {2019},
	pages = {417--442},
}

@article{anderson_three-dimensional_2009,
	title = {The three-dimensional arrangement of the myocytes in the ventricular walls},
	volume = {22},
	issn = {1098-2353},
	url = {https://onlinelibrary.wiley.com/doi/abs/10.1002/ca.20645},
	doi = {10.1002/ca.20645},
	language = {en},
	number = {1},
	urldate = {2025-06-29},
	journal = {Clinical Anatomy},
	author = {Anderson, Robert H. and Smerup, Morten and Sanchez-Quintana, Damian and Loukas, Marios and Lunkenheimer, Paul P.},
	year = {2009},
	pages = {64--76},
}

@article{alfonso_santiago_fully_2018,
	title = {Fully coupled fluid‐electro‐mechanical model of the human heart for supercomputers},
	volume = {34},
	doi = {10.1002/cnm.3140},
	number = {12},
	journal = {International Journal for Numerical Methods in Biomedical Engineering},
	author = {{Alfonso Santiago} and {Jazmín Aguado-Sierra} and {Miguel Zavala-Aké} and Doste-Beltran, Ruben and {Samuel Gómez} and {Ruth Arís} and {Juan Carlos Cajas} and {Eva Casoni} and {Mariano Vázquez}},
	month = dec,
	year = {2018},
	doi = {10.1002/cnm.3140},
}

@article{agger_assessing_2020,
	title = {Assessing {Myocardial} {Architecture}: {The} {Challenges} and {Controversies}},
	volume = {7},
	copyright = {http://creativecommons.org/licenses/by/3.0/},
	issn = {2308-3425},
	shorttitle = {Assessing {Myocardial} {Architecture}},
	url = {https://www.mdpi.com/2308-3425/7/4/47},
	doi = {10.3390/jcdd7040047},
	language = {en},
	number = {4},
	urldate = {2024-11-04},
	journal = {Journal of Cardiovascular Development and Disease},
	publisher = {Multidisciplinary Digital Publishing Institute},
	author = {Agger, Peter and Stephenson, Robert S.},
	month = dec,
	year = {2020},
	pages = {47},
}

@misc{noauthor_4c_nodate,
	title = {{4C}: {A} {Comprehensive} {Multiphysics} {Simulation} {Framework}},
	url = {https://www.4c-multiphysics.org},
}

@book{the_cgal_project_cgal_2024,
	edition = {6.0.1},
	title = {{CGAL} {User} and {Reference} {Manual}},
	url = {https://doc.cgal.org/6.0.1/Manual/packages.html},
	publisher = {CGAL Editorial Board},
	author = {The CGAL Project},
	year = {2024},
}

@article{gonzalo_multiphysics_2024,
	title = {Multiphysics simulations reveal haemodynamic impacts of patient-derived fibrosis-related changes in left atrial tissue mechanics},
	volume = {602},
	issn = {1469-7793},
	url = {https://onlinelibrary.wiley.com/doi/abs/10.1113/JP287011},
	doi = {10.1113/JP287011},
	language = {en},
	number = {24},
	urldate = {2025-04-18},
	journal = {The Journal of Physiology},
	author = {Gonzalo, Alejandro and Augustin, Christoph M. and Bifulco, Savannah F. and Telle, Åshild and Chahine, Yaacoub and Kassar, Ahmad and Guerrero-Hurtado, Manuel and Durán, Eduardo and Martínez-Legazpi, Pablo and Flores, Oscar and Bermejo, Javier and Plank, Gernot and Akoum, Nazem and Boyle, Patrick M. and del Alamo, Juan C.},
	year = {2024},
	pages = {6789--6812},
}

@article{elguedj_b-bar_2008,
	title = {B-bar and {F}-bar projection methods for nearly incompressible linear and non-linear elasticity and plasticity using higher-order {NURBS} elements},
	volume = {197},
	issn = {0045-7825},
	url = {https://www.sciencedirect.com/science/article/pii/S0045782508000248},
	doi = {10.1016/j.cma.2008.01.012},
	number = {33},
	urldate = {2025-05-11},
	journal = {Computer Methods in Applied Mechanics and Engineering},
	author = {Elguedj, T. and Bazilevs, Y. and Calo, V. M. and Hughes, T. J. R.},
	month = jun,
	year = {2008},
	pages = {2732--2762},
}

@article{melnik_generalised_2018,
	title = {A generalised structure tensor model for the mixed invariant {I8}},
	volume = {107},
	issn = {0020-7462},
	url = {https://www.sciencedirect.com/science/article/pii/S002074621830163X},
	doi = {10.1016/j.ijnonlinmec.2018.08.018},
	urldate = {2024-12-02},
	journal = {International Journal of Non-Linear Mechanics},
	author = {Melnik, Andrey V. and Luo, Xiaoyu and Ogden, Ray W.},
	month = dec,
	year = {2018},
	pages = {137--148},
}

@article{zingaro_electromechanics-driven_2024,
	title = {An electromechanics-driven fluid dynamics model for the simulation of the whole human heart},
	volume = {504},
	issn = {00219991},
	url = {https://linkinghub.elsevier.com/retrieve/pii/S0021999124001347},
	doi = {10.1016/j.jcp.2024.112885},
	language = {en},
	urldate = {2026-02-21},
	journal = {Journal of Computational Physics},
	author = {Zingaro, Alberto and Bucelli, Michele and Piersanti, Roberto and Regazzoni, Francesco and Dede', Luca and Quarteroni, Alfio},
	month = may,
	year = {2024},
	pages = {112885},
}

@article{strocchi_integrating_2025,
	title = {Integrating {Imaging} and {Invasive} {Pressure} {Data} into a {Multiscale} {Whole}-{Heart} {Model}},
	volume = {148},
	issn = {0148-0731},
	url = {https://doi.org/10.1115/1.4069497},
	doi = {10.1115/1.4069497},
	number = {051001},
	urldate = {2026-02-03},
	journal = {Journal of Biomechanical Engineering},
	author = {Strocchi, Marina and Augustin, Christoph M. and Gsell, Matthias A. F. and Rinaldi, Christopher A. and Vigmond, Edward J. and Plank, Gernot and Oates, Chris J. and Wilkinson, Richard D. and Niederer, Steven A.},
	month = nov,
	year = {2025},
}

@article{stimm_personalization_2022,
	title = {Personalization of biomechanical simulations of the left ventricle by in-vivo cardiac {DTI} data: {Impact} of fiber interpolation methods},
	volume = {13},
	issn = {1664-042X},
	shorttitle = {Personalization of biomechanical simulations of the left ventricle by in-vivo cardiac {DTI} data},
	url = {https://www.frontiersin.org/journals/physiology/articles/10.3389/fphys.2022.1042537/full},
	doi = {10.3389/fphys.2022.1042537},
	language = {English},
	urldate = {2024-12-11},
	journal = {Frontiers in Physiology},
	publisher = {Frontiers},
	author = {Stimm, Johanna and Nordsletten, David A. and Jilberto, Javiera and Miller, Renee and Berberoğlu, Ezgi and Kozerke, Sebastian and Stoeck, Christian T.},
	month = nov,
	year = {2022},
}

@article{guan_effect_2020,
	title = {Effect of myofibre architecture on ventricular pump function by using a neonatal porcine heart model: from {DT}-{MRI} to rule-based methods},
	volume = {7},
	shorttitle = {Effect of myofibre architecture on ventricular pump function by using a neonatal porcine heart model},
	url = {https://royalsocietypublishing.org/doi/10.1098/rsos.191655},
	doi = {10.1098/rsos.191655},
	number = {4},
	urldate = {2024-11-22},
	journal = {Royal Society Open Science},
	publisher = {Royal Society},
	author = {Guan, Debao and Yao, Jiang and Luo, Xiaoyu and Gao, Hao},
	month = apr,
	year = {2020},
	pages = {191655},
}

@article{shi_personalized_2026,
	title = {Personalized multiscale modeling of left atrial mechanics and blood flow},
	volume = {448},
	issn = {0045-7825},
	url = {https://www.sciencedirect.com/science/article/pii/S004578252500684X},
	doi = {10.1016/j.cma.2025.118412},
	urldate = {2025-10-03},
	journal = {Computer Methods in Applied Mechanics and Engineering},
	author = {Shi, Lei and Gan, Boyang and Chen, Ian Y. and Vedula, Vijay},
	month = jan,
	year = {2026},
	pages = {118412},
}

@article{osouli_heart_2025,
	title = {Heart in a knot: unraveling the impact of the nested tori myofiber architecture on ventricular mechanics},
	volume = {24},
	issn = {1617-7940},
	shorttitle = {Heart in a knot},
	url = {https://doi.org/10.1007/s10237-025-01995-y},
	doi = {10.1007/s10237-025-01995-y},
	language = {en},
	number = {5},
	urldate = {2025-09-25},
	journal = {Biomechanics and Modeling in Mechanobiology},
	author = {Osouli, Kasra and De Gaetano, Francesco and Costantino, Maria Laura and Peirlinck, Mathias},
	month = oct,
	year = {2025},
	pages = {1815--1835},
}

@article{holz_transmural_2021,
	title = {A {Transmural} {Path} {Model} {Improves} the {Definition} of the {Orthotropic} {Tissue} {Structure} in {Heart} {Simulations}},
	volume = {144},
	issn = {0148-0731},
	url = {https://doi.org/10.1115/1.4052219},
	doi = {10.1115/1.4052219},
	number = {031002},
	urldate = {2024-11-07},
	journal = {Journal of Biomechanical Engineering},
	author = {Holz, David and Duong, Minh Tuan and Martonová, Denisa and Alkassar, Muhannad and Leyendecker, Sigrid},
	month = oct,
	year = {2021},
}

@article{haj-ali_general_2012,
	title = {A general three-dimensional parametric geometry of the native aortic valve and root for biomechanical modeling},
	volume = {45},
	issn = {0021-9290},
	url = {https://www.sciencedirect.com/science/article/pii/S0021929012004125},
	doi = {10.1016/j.jbiomech.2012.07.017},
	number = {14},
	urldate = {2025-09-01},
	journal = {Journal of Biomechanics},
	author = {Haj-Ali, Rami and Marom, Gil and Ben Zekry, Sagit and Rosenfeld, Moshe and Raanani, Ehud},
	month = sep,
	year = {2012},
	pages = {2392--2397},
}

@book{team_trilinos_nodate,
	title = {The {Trilinos} {Project} {Website}},
	author = {Team, The Trilinos Project},
}

@misc{noauthor_3d_nodate,
	title = {{3D} {Slicer} image computing platform},
	url = {https://slicer.org/},
	urldate = {2025-08-29},
	journal = {3D Slicer},
}

@article{lasso_slicerheart_2022,
	title = {{SlicerHeart}: {An} open-source computing platform for cardiac image analysis and modeling},
	volume = {9},
	issn = {2297-055X},
	shorttitle = {{SlicerHeart}},
	url = {https://www.frontiersin.org/journals/cardiovascular-medicine/articles/10.3389/fcvm.2022.886549/full},
	doi = {10.3389/fcvm.2022.886549},
	language = {English},
	urldate = {2025-08-29},
	journal = {Frontiers in Cardiovascular Medicine},
	publisher = {Frontiers},
	author = {Lasso, Andras and Herz, Christian and Nam, Hannah and Cianciulli, Alana and Pieper, Steve and Drouin, Simon and Pinter, Csaba and St-Onge, Samuelle and Vigil, Chad and Ching, Stephen and Sunderland, Kyle and Fichtinger, Gabor and Kikinis, Ron and Jolley, Matthew A.},
	month = sep,
	year = {2022},
}

@article{maas_febio_2012,
	title = {{FEBio}: {Finite} {Elements} for {Biomechanics}},
	volume = {134},
	issn = {0148-0731},
	shorttitle = {{FEBio}},
	url = {https://doi.org/10.1115/1.4005694},
	doi = {10.1115/1.4005694},
	number = {011005},
	urldate = {2025-08-29},
	journal = {Journal of Biomechanical Engineering},
	author = {Maas, Steve A. and Ellis, Benjamin J. and Ateshian, Gerard A. and Weiss, Jeffrey A.},
	month = feb,
	year = {2012},
}

@article{hirschvogel_ambit_2024,
	title = {Ambit – {A} {FEniCS}-based cardiovascular multi-physicssolver},
	volume = {9},
	copyright = {http://creativecommons.org/licenses/by/4.0/},
	issn = {2475-9066},
	url = {https://joss.theoj.org/papers/10.21105/joss.05744},
	doi = {10.21105/joss.05744},
	language = {en},
	number = {93},
	urldate = {2025-08-29},
	journal = {Journal of Open Source Software},
	author = {Hirschvogel, Marc},
	month = jan,
	year = {2024},
	pages = {5744},
}

@article{plank_opencarp_2021,
	title = {The \textit{{openCARP}} simulation environment for cardiac electrophysiology},
	volume = {208},
	issn = {0169-2607},
	url = {https://www.sciencedirect.com/science/article/pii/S0169260721002972},
	doi = {10.1016/j.cmpb.2021.106223},
	urldate = {2025-08-29},
	journal = {Computer Methods and Programs in Biomedicine},
	author = {Plank, Gernot and Loewe, Axel and Neic, Aurel and Augustin, Christoph and Huang, Yung-Lin and Gsell, Matthias A. F. and Karabelas, Elias and Nothstein, Mark and Prassl, Anton J. and Sánchez, Jorge and Seemann, Gunnar and Vigmond, Edward J.},
	month = sep,
	year = {2021},
	pages = {106223},
}

@article{colli_franzone_effects_2017,
	title = {Effects of mechanical feedback on the stability of cardiac scroll waves: {A} bidomain electro-mechanical simulation study},
	volume = {27},
	issn = {1054-1500, 1089-7682},
	shorttitle = {Effects of mechanical feedback on the stability of cardiac scroll waves},
	url = {https://pubs.aip.org/cha/article/27/9/093905/342097/Effects-of-mechanical-feedback-on-the-stability-of},
	doi = {10.1063/1.4999465},
	language = {en},
	number = {9},
	urldate = {2025-08-29},
	journal = {Chaos: An Interdisciplinary Journal of Nonlinear Science},
	author = {Colli Franzone, P. and Pavarino, L. F. and Scacchi, S.},
	month = sep,
	year = {2017},
	pages = {093905},
}

@article{tueni_structural_2023,
	title = {On the structural origin of the anisotropy in the myocardium: {Multiscale} modeling and analysis},
	volume = {138},
	issn = {1751-6161},
	shorttitle = {On the structural origin of the anisotropy in the myocardium},
	url = {https://www.sciencedirect.com/science/article/pii/S1751616122005057},
	doi = {10.1016/j.jmbbm.2022.105600},
	urldate = {2025-08-29},
	journal = {Journal of the Mechanical Behavior of Biomedical Materials},
	author = {Tueni, Nicole and Allain, Jean-Marc and Genet, Martin},
	month = feb,
	year = {2023},
	pages = {105600},
}

@article{neic_automating_2020,
	title = {Automating image-based mesh generation and manipulation tasks in cardiac modeling workflows using {Meshtool}},
	volume = {11},
	issn = {2352-7110},
	url = {https://www.sciencedirect.com/science/article/pii/S235271101930295X},
	doi = {10.1016/j.softx.2020.100454},
	urldate = {2025-08-27},
	journal = {SoftwareX},
	author = {Neic, Aurel and Gsell, Matthias A. F. and Karabelas, Elias and Prassl, Anton J. and Plank, Gernot},
	month = jan,
	year = {2020},
	pages = {100454},
}

@article{camps_digital_2024,
	title = {Digital twinning of the human ventricular activation sequence to {Clinical} 12-lead {ECGs} and magnetic resonance imaging using realistic {Purkinje} networks for in silico clinical trials},
	volume = {94},
	issn = {1361-8415},
	url = {https://www.sciencedirect.com/science/article/pii/S1361841524000331},
	doi = {10.1016/j.media.2024.103108},
	urldate = {2025-08-27},
	journal = {Medical Image Analysis},
	author = {Camps, Julia and Berg, Lucas Arantes and Wang, Zhinuo Jenny and Sebastian, Rafael and Riebel, Leto Luana and Doste, Ruben and Zhou, Xin and Sachetto, Rafael and Coleman, James and Lawson, Brodie and Grau, Vicente and Burrage, Kevin and Bueno-Orovio, Alfonso and Weber dos Santos, Rodrigo and Rodriguez, Blanca},
	month = may,
	year = {2024},
	pages = {103108},
}

@article{sel_building_2024,
	title = {Building {Digital} {Twins} for {Cardiovascular} {Health}: {From} {Principles} to {Clinical} {Impact}},
	volume = {13},
	issn = {2047-9980},
	shorttitle = {Building {Digital} {Twins} for {Cardiovascular} {Health}},
	url = {https://www.ahajournals.org/doi/10.1161/JAHA.123.031981},
	doi = {10.1161/JAHA.123.031981},
	language = {en},
	number = {19},
	urldate = {2025-08-27},
	journal = {Journal of the American Heart Association},
	author = {Sel, Kaan and Osman, Deen and Zare, Fatemeh and Masoumi Shahrbabak, Sina and Brattain, Laura and Hahn, Jin‐Oh and Inan, Omer T. and Mukkamala, Ramakrishna and Palmer, Jeffrey and Paydarfar, David and Pettigrew, Roderic I. and Quyyumi, Arshed A. and Telfer, Brian and Jafari, Roozbeh},
	month = oct,
	year = {2024},
	pages = {e031981},
}

@article{coorey_health_2022,
	title = {The health digital twin to tackle cardiovascular disease—a review of an emerging interdisciplinary field},
	volume = {5},
	copyright = {2022 The Author(s)},
	issn = {2398-6352},
	url = {https://www.nature.com/articles/s41746-022-00640-7},
	doi = {10.1038/s41746-022-00640-7},
	language = {en},
	number = {1},
	urldate = {2025-08-27},
	journal = {npj Digital Medicine},
	publisher = {Nature Publishing Group},
	author = {Coorey, Genevieve and Figtree, Gemma A. and Fletcher, David F. and Snelson, Victoria J. and Vernon, Stephen Thomas and Winlaw, David and Grieve, Stuart M. and McEwan, Alistair and Yang, Jean Yee Hwa and Qian, Pierre and O’Brien, Kieran and Orchard, Jessica and Kim, Jinman and Patel, Sanjay and Redfern, Julie},
	month = aug,
	year = {2022},
	pages = {126},
}

@article{menon_personalized_2025,
	title = {Personalized and uncertainty-aware coronary hemodynamics simulations: {From} {Bayesian} estimation to improved multi-fidelity uncertainty quantification},
	volume = {271},
	issn = {0169-2607},
	shorttitle = {Personalized and uncertainty-aware coronary hemodynamics simulations},
	url = {https://www.sciencedirect.com/science/article/pii/S0169260725003682},
	doi = {10.1016/j.cmpb.2025.108951},
	urldate = {2025-08-26},
	journal = {Computer Methods and Programs in Biomedicine},
	author = {Menon, Karthik and Zanoni, Andrea and Khan, M. Owais and Geraci, Gianluca and Nieman, Koen and Schiavazzi, Daniele E. and Marsden, Alison L.},
	month = nov,
	year = {2025},
	pages = {108951},
}

@article{kong_deep-learning_2021,
	title = {A deep-learning approach for direct whole-heart mesh reconstruction},
	volume = {74},
	issn = {1361-8415},
	url = {https://www.sciencedirect.com/science/article/pii/S136184152100267X},
	doi = {10.1016/j.media.2021.102222},
	urldate = {2025-08-26},
	journal = {Medical Image Analysis},
	author = {Kong, Fanwei and Wilson, Nathan and Shadden, Shawn},
	month = dec,
	year = {2021},
	pages = {102222},
}

@article{vigil_modeling_2021,
	title = {Modeling {Tool} for {Rapid} {Virtual} {Planning} of the {Intracardiac} {Baffle} in {Double} {Outlet} {Right} {Ventricle}},
	volume = {111},
	issn = {0003-4975},
	url = {https://www.ncbi.nlm.nih.gov/pmc/articles/PMC8154721/},
	doi = {10.1016/j.athoracsur.2021.02.058},
	number = {6},
	urldate = {2025-08-26},
	journal = {The Annals of thoracic surgery},
	author = {Vigil, Chad and Lasso, Andras and Ghosh, Reena and Pinter, Csaba and Cianciulli, Alana and Nam, Hannah H. and Abid, Ashraful and Herz, Christian and Mascio, Christopher E. and Chen, Jonathan and Fuller, Stephanie and Whitehead, Kevin and Jolley, Matthew A.},
	month = jun,
	year = {2021},
	pages = {2078--2083},
}

@article{guccione_finite_1995,
	title = {Finite element stress analysis of left ventricular mechanics in the beating dog heart},
	volume = {28},
	copyright = {https://www.elsevier.com/tdm/userlicense/1.0/},
	issn = {00219290},
	url = {https://linkinghub.elsevier.com/retrieve/pii/0021929094001743},
	doi = {10.1016/0021-9290(94)00174-3},
	language = {en},
	number = {10},
	urldate = {2025-08-18},
	journal = {Journal of Biomechanics},
	author = {Guccione, Julius M. and Costa, Kevin D. and McCulloch, Andrew D.},
	month = oct,
	year = {1995},
	pages = {1167--1177},
}

@misc{shi_heartsimsage_2025,
	title = {{HeartSimSage}: {Attention}-{Enhanced} {Graph} {Neural} {Networks} for {Accelerating} {Cardiac} {Mechanics} {Modeling}},
	shorttitle = {{HeartSimSage}},
	url = {http://arxiv.org/abs/2504.18968},
	doi = {10.48550/arXiv.2504.18968},
	urldate = {2025-08-14},
	publisher = {arXiv},
	author = {Shi, Lei and Chen, Yurui and Vedula, Vijay},
	month = apr,
	year = {2025},
	note = {arXiv:2504.18968 [physics]},
}

@article{mandapaka_simultaneous_2011,
	title = {Simultaneous {Measurement} of {Left} and {Right} {Ventricular} {Volumes} and {Ejection} {Fraction} {During} {Dobutamine} {Stress} {Cardiovascular} {Magnetic} {Resonance}},
	volume = {35},
	issn = {0363-8715},
	url = {https://www.ncbi.nlm.nih.gov/pmc/articles/PMC3666038/},
	doi = {10.1097/RCT.0b013e31822abbcd},
	number = {5},
	urldate = {2025-08-01},
	journal = {Journal of computer assisted tomography},
	author = {Mandapaka, Sangeeta and Hamilton, Craig A. and Morgan, Timothy M. and Hundley, W. Gregory},
	year = {2011},
	pages = {614--617},
}

@inproceedings{jilberto_identification_2025,
	address = {Cham},
	title = {Identification of the {Unloaded} {Heart} {Configuration} {Including} {External} {Interactions}},
	isbn = {978-3-031-94559-5},
	doi = {10.1007/978-3-031-94559-5_30},
	language = {en},
	booktitle = {Functional {Imaging} and {Modeling} of the {Heart}},
	publisher = {Springer Nature Switzerland},
	author = {Jilberto, Javiera and Nordsletten, David},
	editor = {Chabiniok, Radomír and Zou, Qing and Hussain, Tarique and Nguyen, Hoang H. and Zaha, Vlad G. and Gusseva, Maria},
	year = {2025},
	pages = {331--342},
}

@misc{hofler_physics-informed_2025,
	title = {Physics-informed neural network estimation of active material properties in time-dependent cardiac biomechanical models},
	url = {http://arxiv.org/abs/2505.03382},
	doi = {10.48550/arXiv.2505.03382},
	urldate = {2025-07-02},
	publisher = {arXiv},
	author = {Höfler, Matthias and Regazzoni, Francesco and Pagani, Stefano and Karabelas, Elias and Augustin, Christoph and Haase, Gundolf and Plank, Gernot and Caforio, Federica},
	month = may,
	year = {2025},
	note = {arXiv:2505.03382 [cs]},
}

@article{kroeker_pericardium_2003,
	title = {Pericardium modulates left and right ventricular stroke  volumes to compensate for sudden changes in atrial volume},
	volume = {284},
	issn = {0363-6135},
	url = {https://journals.physiology.org/doi/full/10.1152/ajpheart.00613.2002},
	doi = {10.1152/ajpheart.00613.2002},
	number = {6},
	urldate = {2025-07-02},
	journal = {American Journal of Physiology-Heart and Circulatory Physiology},
	publisher = {American Physiological Society},
	author = {Kroeker, Carol A. Gibbons and Shrive, Nigel G. and Belenkie, Israel and Tyberg, John V.},
	month = jun,
	year = {2003},
	pages = {H2247--H2254},
}

@article{barber_estimation_2021,
	title = {Estimation of {Personalized} {Minimal} {Purkinje} {Systems} {From} {Human} {Electro}-{Anatomical} {Maps}},
	volume = {40},
	issn = {1558-254X},
	url = {https://ieeexplore.ieee.org/abstract/document/9405625},
	doi = {10.1109/TMI.2021.3073499},
	number = {8},
	urldate = {2025-07-01},
	journal = {IEEE Transactions on Medical Imaging},
	author = {Barber, Fernando and Langfield, Peter and Lozano, Miguel and García-Fernández, Ignacio and Duchateau, Josselin and Hocini, Mélèze and Haïssaguerre, Michel and Vigmond, Edward and Sebastian, Rafael},
	month = aug,
	year = {2021},
	pages = {2182--2194},
}

@article{sahli_costabal_generating_2016,
	series = {Cardiovascular {Biomechanics} in {Health} and {Disease}},
	title = {Generating {Purkinje} networks in the human heart},
	volume = {49},
	issn = {0021-9290},
	url = {https://www.sciencedirect.com/science/article/pii/S0021929015007332},
	doi = {10.1016/j.jbiomech.2015.12.025},
	number = {12},
	urldate = {2025-07-01},
	journal = {Journal of Biomechanics},
	author = {Sahli Costabal, Francisco and Hurtado, Daniel E. and Kuhl, Ellen},
	month = aug,
	year = {2016},
	pages = {2455--2465},
}

@article{lee_patient-specific_2014,
	title = {Patient-specific finite element modeling of the {Cardiokinetix} {Parachute}® device: effects on left ventricular wall stress and function},
	volume = {52},
	issn = {1741-0444},
	shorttitle = {Patient-specific finite element modeling of the {Cardiokinetix} {Parachute}® device},
	url = {https://doi.org/10.1007/s11517-014-1159-5},
	doi = {10.1007/s11517-014-1159-5},
	language = {en},
	number = {6},
	urldate = {2025-06-30},
	journal = {Medical \& Biological Engineering \& Computing},
	author = {Lee, Lik Chuan and Ge, Liang and Zhang, Zhihong and Pease, Matthew and Nikolic, Serjan D. and Mishra, Rakesh and Ratcliffe, Mark B. and Guccione, Julius M.},
	month = jun,
	year = {2014},
	pages = {557--566},
}

@article{thangaraj_cardiovascular_2024,
	title = {Cardiovascular care with digital twin technology in the era of generative artificial intelligence},
	volume = {45},
	issn = {0195-668X},
	url = {https://doi.org/10.1093/eurheartj/ehae619},
	doi = {10.1093/eurheartj/ehae619},
	number = {45},
	urldate = {2025-06-30},
	journal = {European Heart Journal},
	author = {Thangaraj, Phyllis M and Benson, Sean H and Oikonomou, Evangelos K and Asselbergs, Folkert W and Khera, Rohan},
	month = dec,
	year = {2024},
	pages = {4808--4821},
}

@article{dowling_first--human_2020,
	title = {First-in-{Human} {Experience} {With} {Patient}-{Specific} {Computer} {Simulation} of {TAVR} in {Bicuspid} {Aortic} {Valve} {Morphology}},
	volume = {13},
	issn = {1936-8798},
	url = {https://www.sciencedirect.com/science/article/pii/S1936879819316115},
	doi = {10.1016/j.jcin.2019.07.032},
	number = {2},
	urldate = {2025-06-30},
	journal = {JACC: Cardiovascular Interventions},
	author = {Dowling, Cameron and Firoozi, Sami and Brecker, Stephen J.},
	month = jan,
	year = {2020},
	pages = {184--192},
}

@article{kung_hybrid_2019,
	title = {A {Hybrid} {Experimental}-{Computational} {Modeling} {Framework} for {Cardiovascular} {Device} {Testing}},
	volume = {141},
	issn = {0148-0731},
	url = {https://doi.org/10.1115/1.4042665},
	doi = {10.1115/1.4042665},
	number = {051012},
	urldate = {2025-06-30},
	journal = {Journal of Biomechanical Engineering},
	author = {Kung, Ethan and Farahmand, Masoud and Gupta, Akash},
	month = mar,
	year = {2019},
}

@article{seo_computational_2022,
	title = {Computational evaluation of venous graft geometries in coronary artery bypass surgery},
	volume = {34},
	issn = {1043-0679},
	url = {https://www.ncbi.nlm.nih.gov/pmc/articles/PMC8429518/},
	doi = {10.1053/j.semtcvs.2021.03.007},
	number = {2},
	urldate = {2025-06-30},
	journal = {Seminars in thoracic and cardiovascular surgery},
	author = {Seo, Jongmin and Ramachandra, Abhay B. and Boyd, Jack and Marsden, Alison L. and Kahn, Andrew M.},
	year = {2022},
	pages = {521--532},
}

@article{schwarz_beyond_2023,
	title = {Beyond {CFD}: {Emerging} methodologies for predictive simulation in cardiovascular health and disease},
	volume = {4},
	issn = {2688-4089},
	shorttitle = {Beyond {CFD}},
	url = {https://doi.org/10.1063/5.0109400},
	doi = {10.1063/5.0109400},
	number = {1},
	urldate = {2025-06-30},
	journal = {Biophysics Reviews},
	author = {Schwarz, Erica L. and Pegolotti, Luca and Pfaller, Martin R. and Marsden, Alison L.},
	month = jan,
	year = {2023},
	pages = {011301},
}

@article{sahli-costabal_classifying_2020,
	title = {Classifying {Drugs} by their {Arrhythmogenic} {Risk} {Using} {Machine} {Learning}},
	volume = {118},
	issn = {0006-3495},
	url = {https://www.sciencedirect.com/science/article/pii/S0006349520300382},
	doi = {10.1016/j.bpj.2020.01.012},
	number = {5},
	urldate = {2025-06-30},
	journal = {Biophysical Journal},
	author = {Sahli-Costabal, Francisco and Seo, Kinya and Ashley, Euan and Kuhl, Ellen},
	month = mar,
	year = {2020},
	pages = {1165--1176},
}

@article{lopez_impaired_2020,
	title = {Impaired {Myocardial} {Energetics} {Causes} {Mechanical} {Dysfunction} in {Decompensated} {Failing} {Hearts}},
	volume = {1},
	issn = {2633-8823},
	url = {https://doi.org/10.1093/function/zqaa018},
	doi = {10.1093/function/zqaa018},
	number = {2},
	urldate = {2025-06-30},
	journal = {Function},
	author = {Lopez, Rachel and Marzban, Bahador and Gao, Xin and Lauinger, Ellen and Van den Bergh, Françoise and Whitesall, Steven E and Converso-Baran, Kimber and Burant, Charles F and Michele, Daniel E and Beard, Daniel A},
	month = sep,
	year = {2020},
	pages = {zqaa018},
}

@article{axel_probing_2014,
	title = {Probing dynamic myocardial microstructure with cardiac magnetic resonance diffusion tensor imaging},
	volume = {16},
	issn = {1097-6647},
	url = {https://www.sciencedirect.com/science/article/pii/S1097664723001564},
	doi = {10.1186/s12968-014-0089-6},
	number = {1},
	urldate = {2025-06-30},
	journal = {Journal of Cardiovascular Magnetic Resonance},
	author = {Axel, Leon and Wedeen, Van J and Ennis, Daniel B},
	month = dec,
	year = {2014},
	pages = {89},
}

@misc{matsukubo_echocardiographic_1977,
	title = {Echocardiographic measurement of right ventricular wall thickness. {A} new application of subxiphoid echocardiography.},
	url = {https://www.ahajournals.org/doi/epdf/10.1161/01.CIR.56.2.278},
	doi = {10.1161/01.CIR.56.2.278},
	language = {en},
	urldate = {2025-06-28},
	author = {Matsukubo, Haruo and Matsuura, Tokru and Endo, Naoto and Asayama, Jun and Watanabe, Toshimitsu and Furukawa, Keizo and Kunishige, Hiroshi and Katsume, Hiroshi and Ijichi, Hamao},
	month = aug,
	year = {1977},
}

@article{luscher_what_2018,
	title = {What is a normal blood pressure?},
	volume = {39},
	issn = {0195-668X},
	url = {https://doi.org/10.1093/eurheartj/ehy330},
	doi = {10.1093/eurheartj/ehy330},
	number = {24},
	urldate = {2025-06-28},
	journal = {European Heart Journal},
	author = {Lüscher, Thomas F},
	month = jun,
	year = {2018},
	pages = {2233--2240},
}

@article{krishnamoorthi_simulation_2014,
	title = {Simulation {Methods} and {Validation} {Criteria} for {Modeling} {Cardiac} {Ventricular} {Electrophysiology}},
	volume = {9},
	issn = {1932-6203},
	url = {https://journals.plos.org/plosone/article?id=10.1371/journal.pone.0114494},
	doi = {10.1371/journal.pone.0114494},
	language = {en},
	number = {12},
	urldate = {2025-06-28},
	journal = {PLOS ONE},
	publisher = {Public Library of Science},
	author = {Krishnamoorthi, Shankarjee and Perotti, Luigi E. and Borgstrom, Nils P. and Ajijola, Olujimi A. and Frid, Anna and Ponnaluri, Aditya V. and Weiss, James N. and Qu, Zhilin and Klug, William S. and Ennis, Daniel B. and Garfinkel, Alan},
	month = dec,
	year = {2014},
	pages = {e114494},
}

@article{trayanova_whole-heart_2011,
	title = {Whole-{Heart} {Modeling}},
	volume = {108},
	url = {https://www.ahajournals.org/doi/full/10.1161/CIRCRESAHA.110.223610},
	doi = {10.1161/CIRCRESAHA.110.223610},
	number = {1},
	urldate = {2025-06-28},
	journal = {Circulation Research},
	publisher = {American Heart Association},
	author = {Trayanova, Natalia A.},
	month = jan,
	year = {2011},
	pages = {113--128},
}

@article{alnasser_advancements_2024,
	title = {Advancements in cardiac structures segmentation: a comprehensive systematic review of deep learning in {CT} imaging},
	volume = {11},
	issn = {2297-055X},
	shorttitle = {Advancements in cardiac structures segmentation},
	url = {https://www.frontiersin.org/journals/cardiovascular-medicine/articles/10.3389/fcvm.2024.1323461/full},
	doi = {10.3389/fcvm.2024.1323461},
	language = {English},
	urldate = {2025-06-28},
	journal = {Frontiers in Cardiovascular Medicine},
	publisher = {Frontiers},
	author = {Alnasser, Turki Nasser and Abdulaal, Lojain and Maiter, Ahmed and Sharkey, Michael and Dwivedi, Krit and Salehi, Mahan and Garg, Pankaj and Swift, Andrew James and Alabed, Samer},
	month = jan,
	year = {2024},
}

@article{dallarmellina_cardiac_2025,
	title = {Cardiac diffusion-weighted and tensor imaging: {A} consensus statement from the special interest group of the {Society} for {Cardiovascular} {Magnetic} {Resonance}},
	volume = {27},
	issn = {10976647},
	shorttitle = {Cardiac diffusion-weighted and tensor imaging},
	url = {https://linkinghub.elsevier.com/retrieve/pii/S1097664724011360},
	doi = {10.1016/j.jocmr.2024.101109},
	language = {en},
	number = {1},
	urldate = {2025-06-28},
	journal = {Journal of Cardiovascular Magnetic Resonance},
	author = {Dall’Armellina, Erica and Ennis, Daniel B. and Axel, Leon and Croisille, Pierre and Ferreira, Pedro F. and Gotschy, Alexander and Lohr, David and Moulin, Kevin and Nguyen, Christopher T. and Nielles-Vallespin, Sonja and Romero, William and Scott, Andrew D. and Stoeck, Christian and Teh, Irvin and Tunnicliffe, Elizabeth M. and Viallon, Magalie and Wang, Victoria and Young, Alistair A. and Schneider, Jürgen E. and Sosnovik, David E.},
	year = {2025},
	pages = {101109},
}

@article{si_tetgen_2015,
	title = {{TetGen}, a {Delaunay}-{Based} {Quality} {Tetrahedral} {Mesh} {Generator}},
	volume = {41},
	issn = {0098-3500},
	url = {https://dl.acm.org/doi/10.1145/2629697},
	doi = {10.1145/2629697},
	number = {2},
	urldate = {2025-05-15},
	journal = {ACM Trans. Math. Softw.},
	author = {Si, Hang},
	month = feb,
	year = {2015},
	pages = {11:1--11:36},
}

@article{karabelas_global_2022,
	title = {Global {Sensitivity} {Analysis} of {Four} {Chamber} {Heart} {Hemodynamics} {Using} {Surrogate} {Models}},
	volume = {69},
	issn = {1558-2531},
	url = {https://ieeexplore-ieee-org.stanford.idm.oclc.org/document/9745350},
	doi = {10.1109/TBME.2022.3163428},
	number = {10},
	urldate = {2025-05-11},
	journal = {IEEE Transactions on Biomedical Engineering},
	author = {Karabelas, Elias and Longobardi, Stefano and Fuchsberger, Jana and Razeghi, Orod and Rodero, Cristobal and Strocchi, Marina and Rajani, Ronak and Haase, Gundolf and Plank, Gernot and Niederer, Steven},
	month = oct,
	year = {2022},
	pages = {3216--3223},
}

@article{krishnamoorthi_numerical_2013,
	title = {Numerical {Quadrature} and {Operator} {Splitting} in {Finite} {Element} {Methods} for {Cardiac} {Electrophysiology}},
	volume = {29},
	issn = {2040-7939},
	url = {https://www.ncbi.nlm.nih.gov/pmc/articles/PMC4519349/},
	doi = {10.1002/cnm.2573},
	number = {11},
	urldate = {2025-05-11},
	journal = {International journal for numerical methods in biomedical engineering},
	author = {Krishnamoorthi, Shankarjee and Sarkar, Mainak and Klug, William S.},
	month = nov,
	year = {2013},
	pages = {1243--1266},
}

@article{sundnes_improved_2014,
	title = {Improved discretisation and linearisation of active tension in strongly coupled cardiac electro-mechanics simulations},
	volume = {17},
	issn = {1025-5842},
	url = {https://www.ncbi.nlm.nih.gov/pmc/articles/PMC4230300/},
	doi = {10.1080/10255842.2012.704368},
	number = {6},
	urldate = {2025-05-11},
	journal = {Computer methods in biomechanics and biomedical engineering},
	author = {Sundnes, J. and Wall, S. and Osnes, H. and Thorvaldsen, T. and McCulloch, A.D.},
	year = {2014},
	pages = {604--615},
}

@article{lee_multiphysics_2016,
	title = {Multiphysics {Computational} {Modeling} in \${\textbackslash}boldsymbol\{{\textbackslash}mathcal\{{C}\}\}{\textbackslash}mathbf\{{Heart}\}\$},
	volume = {38},
	issn = {1064-8275},
	url = {https://epubs.siam.org/doi/10.1137/15M1014097},
	doi = {10.1137/15M1014097},
	number = {3},
	urldate = {2025-05-11},
	journal = {SIAM Journal on Scientific Computing},
	publisher = {Society for Industrial and Applied Mathematics},
	author = {Lee, J. and Cookson, A. and Roy, I. and Kerfoot, E. and Asner, L. and Vigueras, G. and Sochi, T. and Deparis, S. and Michler, C. and Smith, N. P. and Nordsletten, D. A.},
	month = jan,
	year = {2016},
	pages = {C150--C178},
}

@article{balay_petsc_2019,
	title = {{PETSc} {Users} {Manual}},
	url = {https://ora.ox.ac.uk/objects/uuid:fa2b9e7c-1c58-429c-90fd-f780a3c3dc7d},
	language = {en},
	urldate = {2025-05-11},
	publisher = {Argonne National Laboratory},
	author = {Balay, S. and Abhyankar, S. and Adams, M. and Brown, J. and Brune, P. and Buschelman, K. and Dalcin, L. and Dener, A. and Eijkhout, V. and Gropp, W. and Karpeyev, D. and Kaushik, D. and Knepley, M. and May, D. and McInnes, L. and Mills, R. and Munson, T. and Rupp, K. and Sanan, P. and Smith, B. and Zampini, S. and Zhang, H.},
	year = {2019},
}

@article{pitt-francis_chaste_2009,
	series = {40 {YEARS} {OF} {CPC}: {A} celebratory issue focused on quality software for high performance, grid and novel computing architectures},
	title = {Chaste: {A} test-driven approach to software development for biological modelling},
	volume = {180},
	issn = {0010-4655},
	shorttitle = {Chaste},
	url = {https://www.sciencedirect.com/science/article/pii/S0010465509002604},
	doi = {10.1016/j.cpc.2009.07.019},
	number = {12},
	urldate = {2025-05-11},
	journal = {Computer Physics Communications},
	author = {Pitt-Francis, Joe and Pathmanathan, Pras and Bernabeu, Miguel O. and Bordas, Rafel and Cooper, Jonathan and Fletcher, Alexander G. and Mirams, Gary R. and Murray, Philip and Osborne, James M. and Walter, Alex and Chapman, S. Jon and Garny, Alan and van Leeuwen, Ingeborg M. M. and Maini, Philip K. and Rodríguez, Blanca and Waters, Sarah L. and Whiteley, Jonathan P. and Byrne, Helen M. and Gavaghan, David J.},
	month = dec,
	year = {2009},
	pages = {2452--2471},
}

@article{arndt_dealii_nodate,
	title = {The deal.{II} {Library}, {Version} 9.2, 2020},
	language = {en},
	author = {Arndt, Daniel and Bangerth, Wolfgang and Blais, Bruno and Clevenger, Thomas C and Fehling, Marc and Grayver, Alexander V and Heister, Timo and Heltai, Luca and Kronbichler, Martin and Munch, Peter and Maier, Matthias and Pelteret, Jean-Paul and Rastak, Reza and Turcksin, Bruno and Wang, Zhuoran and Wells, David},
}

@article{africa_lifex_2022,
	title = {lifex: {A} flexible, high performance library for the numerical solution of complex finite element problems},
	volume = {20},
	issn = {2352-7110},
	shorttitle = {lifex},
	url = {https://www.sciencedirect.com/science/article/pii/S2352711022001704},
	doi = {10.1016/j.softx.2022.101252},
	urldate = {2025-05-11},
	journal = {SoftwareX},
	author = {Africa, Pasquale Claudio},
	month = dec,
	year = {2022},
	pages = {101252},
}

@misc{noauthor_continuity_nodate,
	title = {Continuity – {A} {Problem} {Solving} {Environment} for {Multi}-{Scale} {Biology}},
	url = {https://continuity.ucsd.edu/},
	language = {en-US},
	urldate = {2025-05-11},
}

@article{wong_computational_2013,
	title = {Computational modeling of chemo-electro-mechanical coupling: {A} novel implicit monolithic finite element approach},
	volume = {29},
	issn = {2040-7939},
	shorttitle = {Computational modeling of chemo-electro-mechanical coupling},
	url = {https://www.ncbi.nlm.nih.gov/pmc/articles/PMC4567385/},
	doi = {10.1002/cnm.2565},
	number = {10},
	urldate = {2025-05-11},
	journal = {International journal for numerical methods in biomedical engineering},
	author = {Wong, J. and Göktepe, S. and Kuhl, E.},
	month = oct,
	year = {2013},
	pages = {1104--1133},
}

@article{dal_fully_2013,
	title = {A fully implicit finite element method for bidomain models of cardiac electromechanics},
	volume = {253},
	issn = {0045-7825},
	url = {https://www.sciencedirect.com/science/article/pii/S0045782512002241},
	doi = {10.1016/j.cma.2012.07.004},
	urldate = {2025-05-11},
	journal = {Computer Methods in Applied Mechanics and Engineering},
	author = {Dal, Hüsnü and Göktepe, Serdar and Kaliske, Michael and Kuhl, Ellen},
	month = jan,
	year = {2013},
	pages = {323--336},
}

@article{liu_robust_2019,
	title = {A robust and efficient iterative method for hyper-elastodynamics with nested block preconditioning},
	volume = {383},
	issn = {0021-9991},
	url = {https://www.sciencedirect.com/science/article/pii/S0021999119300440},
	doi = {10.1016/j.jcp.2019.01.019},
	urldate = {2025-05-11},
	journal = {Journal of Computational Physics},
	author = {Liu, Ju and Marsden, Alison L.},
	month = apr,
	year = {2019},
	pages = {72--93},
}

@article{guan_structure-preserving_2023,
	title = {A structure-preserving integrator for incompressible finite elastodynamics based on a grad-div stabilized mixed formulation with particular emphasis on stretch-based material models},
	volume = {414},
	issn = {0045-7825},
	url = {https://www.sciencedirect.com/science/article/pii/S0045782523002694},
	doi = {10.1016/j.cma.2023.116145},
	urldate = {2025-05-11},
	journal = {Computer Methods in Applied Mechanics and Engineering},
	author = {Guan, Jiashen and Yuan, Hongyan and Liu, Ju},
	month = sep,
	year = {2023},
	pages = {116145},
}

@article{liu_fluid-structure_2020,
	title = {Fluid-structure interaction modeling of blood flow in the pulmonary arteries using the unified continuum and variational multiscale formulation},
	volume = {107},
	issn = {0093-6413},
	url = {https://www.sciencedirect.com/science/article/pii/S0093641320300847},
	doi = {10.1016/j.mechrescom.2020.103556},
	urldate = {2025-05-11},
	journal = {Mechanics Research Communications},
	author = {Liu, Ju and Yang, Weiguang and Lan, Ingrid S. and Marsden, Alison L.},
	month = jul,
	year = {2020},
	pages = {103556},
}

@article{liu_unified_2018,
	title = {A unified continuum and variational multiscale formulation for fluids, solids, and fluid–structure interaction},
	volume = {337},
	issn = {0045-7825},
	url = {https://www.sciencedirect.com/science/article/pii/S0045782518301701},
	doi = {10.1016/j.cma.2018.03.045},
	urldate = {2025-05-11},
	journal = {Computer Methods in Applied Mechanics and Engineering},
	author = {Liu, Ju and Marsden, Alison L.},
	month = aug,
	year = {2018},
	pages = {549--597},
}

@article{karabelas_towards_2018,
	title = {Towards a {Computational} {Framework} for {Modeling} the {Impact} of {Aortic} {Coarctations} {Upon} {Left} {Ventricular} {Load}},
	volume = {9},
	issn = {1664-042X},
	url = {https://www.ncbi.nlm.nih.gov/pmc/articles/PMC5985756/},
	doi = {10.3389/fphys.2018.00538},
	urldate = {2025-05-11},
	journal = {Frontiers in Physiology},
	author = {Karabelas, Elias and Gsell, Matthias A. F. and Augustin, Christoph M. and Marx, Laura and Neic, Aurel and Prassl, Anton J. and Goubergrits, Leonid and Kuehne, Titus and Plank, Gernot},
	month = may,
	year = {2018},
	pages = {538},
}

@article{john_numerical_2010,
	series = {Turbulence {Modeling} for {Large} {Eddy} {Simulations}},
	title = {Numerical studies of finite element variational multiscale methods for turbulent flow simulations},
	volume = {199},
	issn = {0045-7825},
	url = {https://www.sciencedirect.com/science/article/pii/S0045782509000395},
	doi = {10.1016/j.cma.2009.01.010},
	number = {13},
	urldate = {2025-05-11},
	journal = {Computer Methods in Applied Mechanics and Engineering},
	author = {John, Volker and Kindl, Adela},
	month = feb,
	year = {2010},
	pages = {841--852},
}

@article{hughes_multiscale_1995,
	title = {Multiscale phenomena: {Green}'s functions, the {Dirichlet}-to-{Neumann} formulation, subgrid scale models, bubbles and the origins of stabilized methods},
	volume = {127},
	issn = {00457825},
	shorttitle = {Multiscale phenomena},
	url = {https://linkinghub.elsevier.com/retrieve/pii/0045782595008449},
	doi = {10.1016/0045-7825(95)00844-9},
	language = {en},
	number = {1-4},
	urldate = {2025-05-11},
	journal = {Computer Methods in Applied Mechanics and Engineering},
	author = {Hughes, Thomas J.R.},
	month = nov,
	year = {1995},
	pages = {387--401},
}

@article{pezzuto_orthotropic_2014,
	series = {Frontiers in {Finite}-{Deformation} {Electromechanics}},
	title = {An orthotropic active–strain model for the myocardium mechanics and its numerical approximation},
	volume = {48},
	issn = {0997-7538},
	url = {https://www.sciencedirect.com/science/article/pii/S0997753814000527},
	doi = {10.1016/j.euromechsol.2014.03.006},
	urldate = {2025-05-11},
	journal = {European Journal of Mechanics - A/Solids},
	author = {Pezzuto, S. and Ambrosi, D. and Quarteroni, A.},
	month = nov,
	year = {2014},
	pages = {83--96},
}

@article{franca_stabilized_1992,
	title = {Stabilized finite element methods: {II}. {The} incompressible {Navier}-{Stokes} equations},
	volume = {99},
	issn = {0045-7825},
	shorttitle = {Stabilized finite element methods},
	url = {https://www.sciencedirect.com/science/article/pii/004578259290041H},
	doi = {10.1016/0045-7825(92)90041-H},
	number = {2},
	urldate = {2025-05-11},
	journal = {Computer Methods in Applied Mechanics and Engineering},
	author = {Franca, Leopoldo P. and Frey, Sérgio L.},
	month = sep,
	year = {1992},
	pages = {209--233},
}

@article{hughes_new_1986,
	title = {A new finite element formulation for computational fluid dynamics: {V}. {Circumventing} the babuška-brezzi condition: a stable {Petrov}-{Galerkin} formulation of the stokes problem accommodating equal-order interpolations},
	volume = {59},
	issn = {0045-7825},
	shorttitle = {A new finite element formulation for computational fluid dynamics},
	url = {https://www.sciencedirect.com/science/article/pii/0045782586900253},
	doi = {10.1016/0045-7825(86)90025-3},
	number = {1},
	urldate = {2025-05-11},
	journal = {Computer Methods in Applied Mechanics and Engineering},
	author = {Hughes, Thomas J. R. and Franca, Leopoldo P. and Balestra, Marc},
	month = nov,
	year = {1986},
	pages = {85--99},
}

@article{nobile_active_2012,
	title = {An active strain electromechanical model for cardiac tissue},
	volume = {28},
	issn = {2040-7947},
	doi = {10.1002/cnm.1468},
	language = {eng},
	number = {1},
	journal = {International Journal for Numerical Methods in Biomedical Engineering},
	author = {Nobile, F. and Quarteroni, A. and Ruiz-Baier, R.},
	month = jan,
	year = {2012},
	pages = {52--71},
}

@article{chamberland_comparison_2010,
	title = {Comparison of the performance of some finite element discretizations for large deformation elasticity problems},
	volume = {88},
	issn = {0045-7949},
	url = {https://www.sciencedirect.com/science/article/pii/S0045794910000404},
	doi = {10.1016/j.compstruc.2010.02.007},
	number = {11},
	urldate = {2025-05-11},
	journal = {Computers \& Structures},
	author = {Chamberland, É. and Fortin, A. and Fortin, M.},
	month = jun,
	year = {2010},
	pages = {664--673},
}

@article{lipton_robustness_2010,
	series = {Computational {Geometry} and {Analysis}},
	title = {Robustness of isogeometric structural discretizations under severe mesh distortion},
	volume = {199},
	issn = {0045-7825},
	url = {https://www.sciencedirect.com/science/article/pii/S0045782509000346},
	doi = {10.1016/j.cma.2009.01.022},
	number = {5},
	urldate = {2025-05-11},
	journal = {Computer Methods in Applied Mechanics and Engineering},
	author = {Lipton, S. and Evans, J. A. and Bazilevs, Y. and Elguedj, T. and Hughes, T. J. R.},
	month = jan,
	year = {2010},
	pages = {357--373},
}

@article{gultekin_quasi-incompressible_2019,
	title = {On the quasi-incompressible finite element analysis of anisotropic hyperelastic materials},
	volume = {63},
	copyright = {2018 The Author(s)},
	issn = {1432-0924},
	url = {https://link.springer.com/article/10.1007/s00466-018-1602-9},
	doi = {10.1007/s00466-018-1602-9},
	language = {en},
	number = {3},
	urldate = {2025-05-11},
	journal = {Computational Mechanics},
	publisher = {Springer Berlin Heidelberg},
	author = {Gültekin, Osman and Dal, Hüsnü and Holzapfel, Gerhard A.},
	month = mar,
	year = {2019},
	note = {Company: Springer
Distributor: Springer
Institution: Springer
Label: Springer
Number: 3},
	pages = {443--453},
}

@article{gurev_high-resolution_2015,
	title = {A high-resolution computational model of the deforming human heart},
	volume = {14},
	copyright = {2015 Springer-Verlag Berlin Heidelberg (outside the USA)},
	issn = {1617-7940},
	url = {https://link.springer.com/article/10.1007/s10237-014-0639-8},
	doi = {10.1007/s10237-014-0639-8},
	language = {en},
	number = {4},
	urldate = {2025-05-11},
	journal = {Biomechanics and Modeling in Mechanobiology},
	publisher = {Springer Berlin Heidelberg},
	author = {Gurev, Viatcheslav and Pathmanathan, Pras and Fattebert, Jean-Luc and Wen, Hui-Fang and Magerlein, John and Gray, Richard A. and Richards, David F. and Rice, J. Jeremy},
	month = aug,
	year = {2015},
	note = {Company: Springer
Distributor: Springer
Institution: Springer
Label: Springer
Number: 4},
	pages = {829--849},
}

@book{brezzi_mixed_1991,
	address = {New York, NY},
	series = {Springer {Series} in {Computational} {Mathematics}},
	title = {Mixed and {Hybrid} {Finite} {Element} {Methods}},
	volume = {15},
	copyright = {http://www.springer.com/tdm},
	isbn = {978-1-4612-7824-5 978-1-4612-3172-1},
	url = {http://link.springer.com/10.1007/978-1-4612-3172-1},
	doi = {10.1007/978-1-4612-3172-1},
	urldate = {2025-05-11},
	publisher = {Springer},
	editor = {Brezzi, Franco and Fortin, Michel},
	year = {1991},
}

@article{yosibash_p-fems_2012,
	series = {Higher {Order} {Finite} {Element} and {Isogeometric} {Methods}},
	title = {\textit{p}-{FEMs} in biomechanics: {Bones} and arteries},
	volume = {249-252},
	issn = {0045-7825},
	shorttitle = {\textit{p}-{FEMs} in biomechanics},
	url = {https://www.sciencedirect.com/science/article/pii/S0045782512002812},
	doi = {10.1016/j.cma.2012.09.006},
	urldate = {2025-05-11},
	journal = {Computer Methods in Applied Mechanics and Engineering},
	author = {Yosibash, Zohar},
	month = dec,
	year = {2012},
	pages = {169--184},
}

@article{colli_franzone_numerical_2018,
	title = {A {Numerical} {Study} of {Scalable} {Cardiac} {Electro}-{Mechanical} {Solvers} on {HPC} {Architectures}},
	volume = {9},
	issn = {1664-042X},
	url = {https://www.ncbi.nlm.nih.gov/pmc/articles/PMC5895745/},
	doi = {10.3389/fphys.2018.00268},
	urldate = {2025-05-11},
	journal = {Frontiers in Physiology},
	author = {Colli Franzone, Piero and Pavarino, Luca F. and Scacchi, Simone},
	month = apr,
	year = {2018},
	pages = {268},
}

@inproceedings{chapelle_numerical_2009,
	title = {Numerical {Simulation} of the {Electromechanical} {Activity} of the {Heart}},
	isbn = {978-3-642-01932-6},
	issn = {1611-3349},
	url = {https://link.springer.com/chapter/10.1007/978-3-642-01932-6_39},
	doi = {10.1007/978-3-642-01932-6_39},
	language = {en},
	urldate = {2025-05-11},
	booktitle = {Functional {Imaging} and {Modeling} of the {Heart}},
	publisher = {Springer, Berlin, Heidelberg},
	author = {Chapelle, Dominique and Fernández, Miguel A. and Gerbeau, Jean-Frédéric and Moireau, Philippe and Sainte-Marie, Jacques and Zemzemi, Nejib},
	year = {2009},
	pages = {357--365},
}

@article{costa_three-dimensional_1996,
	title = {A {Three}-{Dimensional} {Finite} {Element} {Method} for {Large} {Elastic} {Deformations} of {Ventricular} {Myocardium}: {II}—{Prolate} {Spheroidal} {Coordinates}},
	volume = {118},
	issn = {0148-0731},
	shorttitle = {A {Three}-{Dimensional} {Finite} {Element} {Method} for {Large} {Elastic} {Deformations} of {Ventricular} {Myocardium}},
	url = {https://doi.org/10.1115/1.2796032},
	doi = {10.1115/1.2796032},
	number = {4},
	urldate = {2025-05-11},
	journal = {Journal of Biomechanical Engineering},
	author = {Costa, K. D. and Hunter, P. J. and Wayne, J. S. and Waldman, L. K. and Guccione, J. M. and McCulloch, A. D.},
	month = nov,
	year = {1996},
	pages = {464--472},
}

@article{deuflhard_adaptive_2009,
	title = {Adaptive finite element simulation of ventricular fibrillation dynamics},
	volume = {12},
	copyright = {2008 Springer-Verlag},
	issn = {1433-0369},
	url = {https://link.springer.com/article/10.1007/s00791-008-0088-y},
	doi = {10.1007/s00791-008-0088-y},
	language = {en},
	number = {5},
	urldate = {2025-05-11},
	journal = {Computing and Visualization in Science},
	publisher = {Springer-Verlag},
	author = {Deuflhard, Peter and Erdmann, Bodo and Roitzsch, Rainer and Lines, Glenn Terje},
	month = jun,
	year = {2009},
	note = {Company: Springer
Distributor: Springer
Institution: Springer
Label: Springer
Number: 5},
	pages = {201--205},
}

@article{augustin_anatomically_2016,
	title = {Anatomically accurate high resolution modeling of human whole heart electromechanics: {A} strongly scalable algebraic multigrid solver method for nonlinear deformation},
	volume = {305},
	issn = {0021-9991},
	shorttitle = {Anatomically accurate high resolution modeling of human whole heart electromechanics},
	url = {https://www.sciencedirect.com/science/article/pii/S0021999115007226},
	doi = {10.1016/j.jcp.2015.10.045},
	urldate = {2025-05-11},
	journal = {Journal of Computational Physics},
	author = {Augustin, Christoph M. and Neic, Aurel and Liebmann, Manfred and Prassl, Anton J. and Niederer, Steven A. and Haase, Gundolf and Plank, Gernot},
	month = jan,
	year = {2016},
	pages = {622--646},
}

@article{levrero-florencio_sensitivity_2020,
	title = {Sensitivity analysis of a strongly-coupled human-based electromechanical cardiac model: {Effect} of mechanical parameters on physiologically relevant biomarkers},
	volume = {361},
	issn = {0045-7825},
	shorttitle = {Sensitivity analysis of a strongly-coupled human-based electromechanical cardiac model},
	url = {https://www.sciencedirect.com/science/article/pii/S0045782519306541},
	doi = {10.1016/j.cma.2019.112762},
	urldate = {2025-05-08},
	journal = {Computer Methods in Applied Mechanics and Engineering},
	author = {Levrero-Florencio, F. and Margara, F. and Zacur, E. and Bueno-Orovio, A. and Wang, Z. J. and Santiago, A. and Aguado-Sierra, J. and Houzeaux, G. and Grau, V. and Kay, D. and Vázquez, M. and Ruiz-Baier, R. and Rodriguez, B.},
	month = apr,
	year = {2020},
	pages = {112762},
}

@article{guerrero-hurtado_efficient_2023,
	title = {Efficient multi-fidelity computation of blood coagulation under flow},
	volume = {19},
	issn = {1553-7358},
	url = {https://journals.plos.org/ploscompbiol/article?id=10.1371/journal.pcbi.1011583},
	doi = {10.1371/journal.pcbi.1011583},
	language = {en},
	number = {10},
	urldate = {2025-04-18},
	journal = {PLOS Computational Biology},
	publisher = {Public Library of Science},
	author = {Guerrero-Hurtado, Manuel and Garcia-Villalba, Manuel and Gonzalo, Alejandro and Martinez-Legazpi, Pablo and Kahn, Andrew M. and McVeigh, Elliot and Bermejo, Javier and Alamo, Juan C. del and Flores, Oscar},
	month = oct,
	year = {2023},
	pages = {e1011583},
}

@article{pashakhanloo_myofiber_2016,
	title = {Myofiber {Architecture} of the {Human} {Atria} as {Revealed} by {Submillimeter} {Diffusion} {Tensor} {Imaging}},
	volume = {9},
	url = {https://www.ahajournals.org/doi/10.1161/CIRCEP.116.004133},
	doi = {10.1161/CIRCEP.116.004133},
	number = {4},
	urldate = {2025-04-13},
	journal = {Circulation: Arrhythmia and Electrophysiology},
	publisher = {American Heart Association},
	author = {Pashakhanloo, Farhad and Herzka, Daniel A. and Ashikaga, Hiroshi and Mori, Susumu and Gai, Neville and Bluemke, David A. and Trayanova, Natalia A. and McVeigh, Elliot R.},
	month = apr,
	year = {2016},
	pages = {e004133},
}

@article{lombaert_human_2012,
	title = {Human {Atlas} of the {Cardiac} {Fiber} {Architecture}: {Study} on a {Healthy} {Population}},
	volume = {31},
	issn = {1558-254X},
	shorttitle = {Human {Atlas} of the {Cardiac} {Fiber} {Architecture}},
	url = {https://ieeexplore.ieee.org/document/6177265},
	doi = {10.1109/TMI.2012.2192743},
	number = {7},
	urldate = {2025-04-13},
	journal = {IEEE Transactions on Medical Imaging},
	author = {Lombaert, Herve and Peyrat, Jean-Marc and Croisille, Pierre and Rapacchi, Stanislas and Fanton, Laurent and Cheriet, Farida and Clarysse, Patrick and Magnin, Isabelle and Delingette, Hervé and Ayache, Nicholas},
	month = jul,
	year = {2012},
	pages = {1436--1447},
}

@article{zheng_four-chamber_2008,
	title = {Four-{Chamber} {Heart} {Modeling} and {Automatic} {Segmentation} for 3-{D} {Cardiac} {CT} {Volumes} {Using} {Marginal} {Space} {Learning} and {Steerable} {Features}},
	volume = {27},
	issn = {1558-254X},
	url = {https://ieeexplore.ieee.org/document/4601463},
	doi = {10.1109/TMI.2008.2004421},
	number = {11},
	urldate = {2025-04-11},
	journal = {IEEE Transactions on Medical Imaging},
	author = {Zheng, Yefeng and Barbu, Adrian and Georgescu, Bogdan and Scheuering, Michael and Comaniciu, Dorin},
	month = nov,
	year = {2008},
	pages = {1668--1681},
}

@article{baillargeon_living_2014,
	title = {The {Living} {Heart} {Project}: {A} robust and integrative simulator for human heart function},
	volume = {48},
	issn = {0997-7538},
	shorttitle = {The {Living} {Heart} {Project}},
	url = {https://www.ncbi.nlm.nih.gov/pmc/articles/PMC4175454/},
	doi = {10.1016/j.euromechsol.2014.04.001},
	urldate = {2025-04-11},
	journal = {European journal of mechanics. A, Solids},
	author = {Baillargeon, Brian and Rebelo, Nuno and Fox, David D. and Taylor, Robert L. and Kuhl, Ellen},
	year = {2014},
	pages = {38--47},
}

@techreport{zygote_media_group_inc_zygote_2014,
	title = {Zygote {Solid} {3D} {Heart} {Generations} {I} \& {II} {Development} {Report}},
	author = {Zygote Media Group Inc.},
	year = {2014},
}

@article{land_verification_2015,
	title = {Verification of cardiac mechanics software: benchmark problems and solutions for testing active and passive material behaviour},
	volume = {471},
	issn = {1364-5021},
	shorttitle = {Verification of cardiac mechanics software},
	doi = {10.1098/rspa.2015.0641},
	language = {eng},
	number = {2184},
	journal = {Proceedings. Mathematical, Physical, and Engineering Sciences},
	author = {Land, Sander and Gurev, Viatcheslav and Arens, Sander and Augustin, Christoph M. and Baron, Lukas and Blake, Robert and Bradley, Chris and Castro, Sebastian and Crozier, Andrew and Favino, Marco and Fastl, Thomas E. and Fritz, Thomas and Gao, Hao and Gizzi, Alessio and Griffith, Boyce E. and Hurtado, Daniel E. and Krause, Rolf and Luo, Xiaoyu and Nash, Martyn P. and Pezzuto, Simone and Plank, Gernot and Rossi, Simone and Ruprecht, Daniel and Seemann, Gunnar and Smith, Nicolas P. and Sundnes, Joakim and Rice, J. Jeremy and Trayanova, Natalia and Wang, Dafang and Jenny Wang, Zhinuo and Niederer, Steven A.},
	month = dec,
	year = {2015},
	pages = {20150641},
}

@article{zhu_svfsi_2022,
	title = {{svFSI}: {A} {Multiphysics} {Package} for {Integrated} {Cardiac} {Modeling}},
	volume = {7},
	issn = {2475-9066},
	shorttitle = {{svFSI}},
	url = {https://joss.theoj.org/papers/10.21105/joss.04118},
	doi = {10.21105/joss.04118},
	language = {en},
	number = {78},
	urldate = {2024-10-22},
	journal = {Journal of Open Source Software},
	author = {Zhu, Chi and Vedula, Vijay and Parker, Dave and Wilson, Nathan and Shadden, Shawn and Marsden, Alison},
	month = oct,
	year = {2022},
	pages = {4118},
}

@article{wang_modelling_2009,
	series = {Includes {Special} {Section} on the 12th {International} {Conference} on {Medical} {Imaging} and {Computer} {Assisted} {Intervention}},
	title = {Modelling passive diastolic mechanics with quantitative {MRI} of cardiac structure and function},
	volume = {13},
	issn = {1361-8415},
	url = {https://www.sciencedirect.com/science/article/pii/S1361841509000619},
	doi = {10.1016/j.media.2009.07.006},
	number = {5},
	urldate = {2025-03-21},
	journal = {Medical Image Analysis},
	author = {Wang, Vicky Y. and Lam, H. I. and Ennis, Daniel B. and Cowan, Brett R. and Young, Alistair A. and Nash, Martyn P.},
	month = oct,
	year = {2009},
	pages = {773--784},
}

@article{viola_high-fidelity_2023,
	title = {High-fidelity model of the human heart: {An} immersed boundary implementation},
	volume = {8},
	shorttitle = {High-fidelity model of the human heart},
	url = {https://link.aps.org/doi/10.1103/PhysRevFluids.8.100502},
	doi = {10.1103/PhysRevFluids.8.100502},
	number = {10},
	urldate = {2024-10-21},
	journal = {Physical Review Fluids},
	publisher = {American Physical Society},
	author = {Viola, Francesco and Del Corso, Giulio and Verzicco, Roberto},
	month = oct,
	year = {2023},
	pages = {100502},
}

@article{viola_gpu_2023,
	title = {{GPU} accelerated digital twins of the human heart open new routes for cardiovascular research},
	volume = {13},
	copyright = {2023 The Author(s)},
	issn = {2045-2322},
	url = {https://www.nature.com/articles/s41598-023-34098-8},
	doi = {10.1038/s41598-023-34098-8},
	language = {en},
	number = {1},
	urldate = {2024-10-23},
	journal = {Scientific Reports},
	publisher = {Nature Publishing Group},
	author = {Viola, Francesco and Del Corso, Giulio and De Paulis, Ruggero and Verzicco, Roberto},
	month = may,
	year = {2023},
	pages = {8230},
}

@article{verzicco_electro-fluid-mechanics_2022,
	title = {Electro-fluid-mechanics of the heart},
	volume = {941},
	issn = {0022-1120, 1469-7645},
	url = {https://www.cambridge.org/core/journals/journal-of-fluid-mechanics/article/electrofluidmechanics-of-the-heart/6BDA989AA6F798872F10624C6D786B09},
	doi = {10.1017/jfm.2022.272},
	language = {en},
	urldate = {2024-10-22},
	journal = {Journal of Fluid Mechanics},
	author = {Verzicco, R.},
	month = jun,
	year = {2022},
	pages = {P1},
}

@article{torre_isogeometric_2023,
	title = {Isogeometric mixed collocation of nearly-incompressible electromechanics in finite deformations for cardiac muscle simulations},
	volume = {411},
	issn = {0045-7825},
	url = {https://www.sciencedirect.com/science/article/pii/S0045782523001792},
	doi = {10.1016/j.cma.2023.116055},
	urldate = {2024-10-23},
	journal = {Computer Methods in Applied Mechanics and Engineering},
	author = {Torre, Michele and Morganti, Simone and Nitti, Alessandro and de Tullio, Marco D. and Pasqualini, Francesco S. and Reali, Alessandro},
	month = jun,
	year = {2023},
	pages = {116055},
}

@incollection{sugiura_ut-heart_2022,
	address = {New York, NY},
	title = {{UT}-{Heart}: {A} {Finite} {Element} {Model} {Designed} for the {Multiscale} and {Multiphysics} {Integration} of our {Knowledge} on the {Human} {Heart}},
	isbn = {978-1-0716-1831-8},
	shorttitle = {{UT}-{Heart}},
	url = {https://doi.org/10.1007/978-1-0716-1831-8_10},
	doi = {10.1007/978-1-0716-1831-8_10},
	language = {en},
	urldate = {2024-10-22},
	booktitle = {Computational {Systems} {Biology} in {Medicine} and {Biotechnology}: {Methods} and {Protocols}},
	publisher = {Springer US},
	author = {Sugiura, Seiryo and Okada, Jun-Ichi and Washio, Takumi and Hisada, Toshiaki},
	editor = {Cortassa, Sonia and Aon, Miguel A.},
	year = {2022},
	pages = {221--245},
}

@article{strocchi_simulating_2020,
	title = {Simulating ventricular systolic motion in a four-chamber heart model with spatially varying robin boundary conditions to model the effect of the pericardium},
	volume = {101},
	issn = {0021-9290},
	url = {https://www.sciencedirect.com/science/article/pii/S002192902030052X},
	doi = {10.1016/j.jbiomech.2020.109645},
	urldate = {2024-10-16},
	journal = {Journal of Biomechanics},
	author = {Strocchi, Marina and Gsell, Matthias A. F. and Augustin, Christoph M. and Razeghi, Orod and Roney, Caroline H. and Prassl, Anton J. and Vigmond, Edward J. and Behar, Jonathan M. and Gould, Justin S. and Rinaldi, Christopher A. and Bishop, Martin J. and Plank, Gernot and Niederer, Steven A.},
	month = mar,
	year = {2020},
	pages = {109645},
}

@article{shi_optimization_2024,
	title = {An optimization framework to personalize passive cardiac mechanics},
	volume = {432},
	issn = {0045-7825},
	url = {https://www.sciencedirect.com/science/article/pii/S004578252400656X},
	doi = {10.1016/j.cma.2024.117401},
	urldate = {2025-03-21},
	journal = {Computer Methods in Applied Mechanics and Engineering},
	author = {Shi, Lei and Chen, Ian Y. and Takayama, Hiroo and Vedula, Vijay},
	month = dec,
	year = {2024},
	pages = {117401},
}

@article{shavik_high_2018,
	title = {High {Spatial} {Resolution} {Multi}-{Organ} {Finite} {Element} {Modeling} of {Ventricular}-{Arterial} {Coupling}},
	volume = {9},
	issn = {1664-042X},
	doi = {10.3389/fphys.2018.00119},
	language = {eng},
	journal = {Frontiers in Physiology},
	author = {Shavik, Sheikh Mohammad and Jiang, Zhenxiang and Baek, Seungik and Lee, Lik Chuan},
	year = {2018},
	pages = {119},
}

@article{sharifi_multiscale_2024,
	title = {A multiscale finite element model of left ventricular mechanics incorporating baroreflex regulation},
	volume = {168},
	issn = {0010-4825},
	url = {https://www.sciencedirect.com/science/article/pii/S0010482523011551},
	doi = {10.1016/j.compbiomed.2023.107690},
	urldate = {2024-10-23},
	journal = {Computers in Biology and Medicine},
	author = {Sharifi, Hossein and Lee, Lik Chuan and Campbell, Kenneth S. and Wenk, Jonathan F.},
	month = jan,
	year = {2024},
	pages = {107690},
}

@article{sahli_costabal_importance_2017,
	title = {The importance of mechano-electrical feedback and inertia in cardiac electromechanics},
	volume = {320},
	issn = {0045-7825},
	url = {https://www.sciencedirect.com/science/article/pii/S0045782516317728},
	doi = {10.1016/j.cma.2017.03.015},
	urldate = {2025-04-03},
	journal = {Computer Methods in Applied Mechanics and Engineering},
	author = {Sahli Costabal, Francisco and Concha, Felipe A. and Hurtado, Daniel E. and Kuhl, Ellen},
	month = jun,
	year = {2017},
	pages = {352--368},
}

@article{rodriguez-padilla_impact_2022,
	title = {Impact of intraventricular septal fiber orientation on cardiac electromechanical function},
	volume = {322},
	issn = {0363-6135},
	url = {https://journals.physiology.org/doi/full/10.1152/ajpheart.00050.2022},
	doi = {10.1152/ajpheart.00050.2022},
	number = {6},
	urldate = {2024-12-19},
	journal = {American Journal of Physiology-Heart and Circulatory Physiology},
	publisher = {American Physiological Society},
	author = {Rodríguez-Padilla, Jairo and Petras, Argyrios and Magat, Julie and Bayer, Jason and Bihan-Poudec, Yann and El Hamrani, Dounia and Ramlugun, Girish and Neic, Aurel and Augustin, Christoph M. and Vaillant, Fanny and Constantin, Marion and Benoist, David and Pourtau, Line and Dubes, Virginie and Rogier, Julien and Labrousse, Louis and Bernus, Olivier and Quesson, Bruno and Haïssaguerre, Michel and Gsell, Matthias and Plank, Gernot and Ozenne, Valéry and Vigmond, Edward},
	month = jun,
	year = {2022},
	pages = {H936--H952},
}

@article{rodero_linking_2021,
	title = {Linking statistical shape models and simulated function in the healthy adult human heart},
	volume = {17},
	issn = {1553-7358},
	url = {https://journals.plos.org/ploscompbiol/article?id=10.1371/journal.pcbi.1008851},
	doi = {10.1371/journal.pcbi.1008851},
	language = {en},
	number = {4},
	urldate = {2024-10-22},
	journal = {PLOS Computational Biology},
	publisher = {Public Library of Science},
	author = {Rodero, Cristobal and Strocchi, Marina and Marciniak, Maciej and Longobardi, Stefano and Whitaker, John and O’Neill, Mark D. and Gillette, Karli and Augustin, Christoph and Plank, Gernot and Vigmond, Edward J. and Lamata, Pablo and Niederer, Steven A.},
	month = apr,
	year = {2021},
	pages = {e1008851},
}

@article{pfaller_importance_2019,
	title = {The importance of the pericardium for cardiac biomechanics: from physiology to computational modeling},
	volume = {18},
	issn = {1617-7940},
	shorttitle = {The importance of the pericardium for cardiac biomechanics},
	url = {https://doi.org/10.1007/s10237-018-1098-4},
	doi = {10.1007/s10237-018-1098-4},
	language = {en},
	number = {2},
	urldate = {2024-10-21},
	journal = {Biomechanics and Modeling in Mechanobiology},
	author = {Pfaller, Martin R. and Hörmann, Julia M. and Weigl, Martina and Nagler, Andreas and Chabiniok, Radomir and Bertoglio, Cristóbal and Wall, Wolfgang A.},
	month = apr,
	year = {2019},
	pages = {503--529},
}

@article{peirlinck_how_2022,
	title = {How drugs modulate the performance of the human heart},
	volume = {69},
	issn = {1432-0924},
	url = {https://doi.org/10.1007/s00466-022-02146-1},
	doi = {10.1007/s00466-022-02146-1},
	language = {en},
	number = {6},
	urldate = {2024-10-23},
	journal = {Computational Mechanics},
	author = {Peirlinck, M. and Yao, J. and Sahli Costabal, F. and Kuhl, E.},
	month = jun,
	year = {2022},
	pages = {1397--1411},
}

@article{motiwale_neural_2024,
	title = {A neural network finite element approach for high speed cardiac mechanics simulations},
	volume = {427},
	issn = {0045-7825},
	url = {https://www.sciencedirect.com/science/article/pii/S0045782524003165},
	doi = {10.1016/j.cma.2024.117060},
	urldate = {2024-10-22},
	journal = {Computer Methods in Applied Mechanics and Engineering},
	author = {Motiwale, Shruti and Zhang, Wenbo and Feldmeier, Reese and Sacks, Michael S.},
	month = jul,
	year = {2024},
	pages = {117060},
}

@article{mojumder_computational_2023,
	title = {Computational analysis of ventricular mechanics in hypertrophic cardiomyopathy patients},
	volume = {13},
	copyright = {2023 The Author(s)},
	issn = {2045-2322},
	url = {https://www.nature.com/articles/s41598-023-28037-w},
	doi = {10.1038/s41598-023-28037-w},
	language = {en},
	number = {1},
	urldate = {2024-10-23},
	journal = {Scientific Reports},
	publisher = {Nature Publishing Group},
	author = {Mojumder, Joy and Fan, Lei and Nguyen, Thuy and Campbell, Kenneth S. and Wenk, Jonathan F. and Guccione, Julius M. and Abraham, Theodore and Lee, Lik Chuan},
	month = jan,
	year = {2023},
	pages = {958},
}

@article{mendiola_mechanical_2022,
	title = {Mechanical {Interaction} of the {Pericardium} and {Cardiac} {Function} in the {Normal} and {Hypertensive} {Rat} {Heart}},
	volume = {13},
	issn = {1664-042X},
	url = {https://www.frontiersin.org/journals/physiology/articles/10.3389/fphys.2022.878861/full},
	doi = {10.3389/fphys.2022.878861},
	language = {English},
	urldate = {2024-12-11},
	journal = {Frontiers in Physiology},
	publisher = {Frontiers},
	author = {Mendiola, Emilio A. and Sacks, Michael S. and Avazmohammadi, Reza},
	month = may,
	year = {2022},
}

@article{lluch_calibration_2020,
	title = {Calibration of a fully coupled electromechanical meshless computational model of the heart with experimental data},
	volume = {364},
	issn = {0045-7825},
	url = {https://www.sciencedirect.com/science/article/pii/S0045782520300505},
	doi = {10.1016/j.cma.2020.112869},
	urldate = {2024-10-22},
	journal = {Computer Methods in Applied Mechanics and Engineering},
	author = {Lluch, Èric and Camara, Oscar and Doste, Rubén and Bijnens, Bart and De Craene, Mathieu and Sermesant, Maxime and Wang, Vicky Y. and Nash, Martyn P. and Morales, Hernán G.},
	month = jun,
	year = {2020},
	pages = {112869},
}

@article{kovacheva_causes_2021,
	title = {Causes of altered ventricular mechanics in hypertrophic cardiomyopathy: an in-silico study},
	volume = {20},
	issn = {1475-925X},
	shorttitle = {Causes of altered ventricular mechanics in hypertrophic cardiomyopathy},
	url = {https://doi.org/10.1186/s12938-021-00900-9},
	doi = {10.1186/s12938-021-00900-9},
	language = {en},
	number = {1},
	urldate = {2024-10-23},
	journal = {BioMedical Engineering OnLine},
	author = {Kovacheva, Ekaterina and Gerach, Tobias and Schuler, Steffen and Ochs, Marco and Dössel, Olaf and Loewe, Axel},
	month = jul,
	year = {2021},
	pages = {69},
}

@article{kariya_personalized_2020,
	title = {Personalized {Perioperative} {Multi}-scale, {Multi}-physics {Heart} {Simulation} of {Double} {Outlet} {Right} {Ventricle}},
	volume = {48},
	issn = {1573-9686},
	url = {https://doi.org/10.1007/s10439-020-02488-y},
	doi = {10.1007/s10439-020-02488-y},
	language = {en},
	number = {6},
	urldate = {2024-10-23},
	journal = {Annals of Biomedical Engineering},
	author = {Kariya, Taro and Washio, Takumi and Okada, Jun-ichi and Nakagawa, Machiko and Watanabe, Masahiro and Kadooka, Yoshimasa and Sano, Shunji and Nagai, Ryozo and Sugiura, Seiryo and Hisada, Toshiaki},
	month = jun,
	year = {2020},
	pages = {1740--1750},
}

@article{jung_integrated_2022,
	title = {An {Integrated} {Workflow} for {Building} {Digital} {Twins} of {Cardiac} {Electromechanics}—{A} {Multi}-{Fidelity} {Approach} for {Personalising} {Active} {Mechanics}},
	volume = {10},
	copyright = {http://creativecommons.org/licenses/by/3.0/},
	issn = {2227-7390},
	url = {https://www.mdpi.com/2227-7390/10/5/823},
	doi = {10.3390/math10050823},
	language = {en},
	number = {5},
	urldate = {2024-10-23},
	journal = {Mathematics},
	publisher = {Multidisciplinary Digital Publishing Institute},
	author = {Jung, Alexander and Gsell, Matthias A. F. and Augustin, Christoph M. and Plank, Gernot},
	month = jan,
	year = {2022},
	note = {Number: 5},
	pages = {823},
}

@article{jafari_framework_2019,
	title = {A {Framework} for {Biomechanics} {Simulations} {Using} {Four}-{Chamber} {Cardiac} {Models}},
	volume = {91},
	issn = {0021-9290},
	url = {https://www.ncbi.nlm.nih.gov/pmc/articles/PMC6579665/},
	doi = {10.1016/j.jbiomech.2019.05.019},
	urldate = {2025-04-05},
	journal = {Journal of biomechanics},
	author = {Jafari, Arian and Pszczolkowski, Edward and Krishnamurthy, Adarsh},
	month = jun,
	year = {2019},
	pages = {92--101},
}

@article{hadjicharalambous_investigating_2021,
	title = {Investigating the reference domain influence in personalised models of cardiac mechanics},
	volume = {20},
	issn = {1617-7940},
	url = {https://doi.org/10.1007/s10237-021-01464-2},
	doi = {10.1007/s10237-021-01464-2},
	language = {en},
	number = {4},
	urldate = {2024-10-22},
	journal = {Biomechanics and Modeling in Mechanobiology},
	author = {Hadjicharalambous, Myrianthi and Stoeck, Christian T. and Weisskopf, Miriam and Cesarovic, Nikola and Ioannou, Eleftherios and Vavourakis, Vasileios and Nordsletten, David A.},
	month = aug,
	year = {2021},
	pages = {1579--1597},
}

@article{guan_modelling_2021,
	title = {Modelling of fibre dispersion and its effects on cardiac mechanics from diastole to systole},
	volume = {128},
	issn = {1573-2703},
	url = {https://doi.org/10.1007/s10665-021-10102-w},
	doi = {10.1007/s10665-021-10102-w},
	language = {en},
	number = {1},
	urldate = {2024-10-23},
	journal = {Journal of Engineering Mathematics},
	author = {Guan, Debao and Zhuan, Xin and Holmes, William and Luo, Xiaoyu and Gao, Hao},
	month = apr,
	year = {2021},
	pages = {1},
}

@article{fritz_simulation_2014,
	title = {Simulation of the contraction of the ventricles in a human heart model including atria and pericardium},
	volume = {13},
	issn = {1617-7940},
	url = {https://doi.org/10.1007/s10237-013-0523-y},
	doi = {10.1007/s10237-013-0523-y},
	language = {en},
	number = {3},
	urldate = {2024-12-11},
	journal = {Biomechanics and Modeling in Mechanobiology},
	author = {Fritz, Thomas and Wieners, Christian and Seemann, Gunnar and Steen, Henning and Dössel, Olaf},
	month = jun,
	year = {2014},
	pages = {627--641},
}

@article{feng_whole-heart_2024,
	title = {Whole-heart modelling with valves in a fluid–structure interaction framework},
	volume = {420},
	issn = {0045-7825},
	url = {https://www.sciencedirect.com/science/article/pii/S0045782523008472},
	doi = {10.1016/j.cma.2023.116724},
	urldate = {2024-10-23},
	journal = {Computer Methods in Applied Mechanics and Engineering},
	author = {Feng, Liuyang and Gao, Hao and Luo, Xiaoyu},
	month = feb,
	year = {2024},
	pages = {116724},
}

@article{fan_transmural_2021,
	title = {Transmural {Distribution} of {Coronary} {Perfusion} and {Myocardial} {Work} {Density} {Due} to {Alterations} in {Ventricular} {Loading}, {Geometry} and {Contractility}},
	volume = {12},
	issn = {1664-042X},
	url = {https://www.frontiersin.org/journals/physiology/articles/10.3389/fphys.2021.744855/full},
	doi = {10.3389/fphys.2021.744855},
	language = {English},
	urldate = {2024-10-23},
	journal = {Frontiers in Physiology},
	publisher = {Frontiers},
	author = {Fan, Lei and Namani, Ravi and Choy, Jenny S. and Kassab, Ghassan S. and Lee, Lik Chuan},
	month = nov,
	year = {2021},
}

@article{fan_optimization_2022,
	title = {Optimization of cardiac resynchronization therapy based on a cardiac electromechanics-perfusion computational model},
	volume = {141},
	issn = {0010-4825},
	url = {https://www.sciencedirect.com/science/article/pii/S0010482521008441},
	doi = {10.1016/j.compbiomed.2021.105050},
	urldate = {2024-10-22},
	journal = {Computers in Biology and Medicine},
	author = {Fan, Lei and Choy, Jenny S. and Raissi, Farshad and Kassab, Ghassan S. and Lee, Lik Chuan},
	month = feb,
	year = {2022},
	pages = {105050},
}

@article{capuano_personalized_2024,
	title = {Personalized computational electro-mechanics simulations to optimize cardiac resynchronization therapy},
	issn = {1617-7940},
	url = {https://doi.org/10.1007/s10237-024-01878-8},
	doi = {10.1007/s10237-024-01878-8},
	language = {en},
	urldate = {2024-10-22},
	journal = {Biomechanics and Modeling in Mechanobiology},
	author = {Capuano, Emilia and Regazzoni, Francesco and Maines, Massimiliano and Fornara, Silvia and Locatelli, Vanessa and Catanzariti, Domenico and Stella, Simone and Nobile, Fabio and Greco, Maurizio Del and Vergara, Christian},
	month = aug,
	year = {2024},
}

@article{caforio_coupling_2022,
	title = {A coupling strategy for a first {3D}-{1D} model of the cardiovascular system to study the effects of pulse wave propagation on cardiac function},
	volume = {70},
	issn = {1432-0924},
	url = {https://doi.org/10.1007/s00466-022-02206-6},
	doi = {10.1007/s00466-022-02206-6},
	language = {en},
	number = {4},
	urldate = {2024-10-23},
	journal = {Computational Mechanics},
	author = {Caforio, Federica and Augustin, Christoph M. and Alastruey, Jordi and Gsell, Matthias A. F. and Plank, Gernot},
	month = oct,
	year = {2022},
	pages = {703--722},
}

@article{brown_modular_2024,
	title = {A modular framework for implicit {3D}–{0D} coupling in cardiac mechanics},
	volume = {421},
	issn = {0045-7825},
	url = {https://www.sciencedirect.com/science/article/pii/S0045782524000203},
	doi = {10.1016/j.cma.2024.116764},
	urldate = {2024-10-23},
	journal = {Computer Methods in Applied Mechanics and Engineering},
	author = {Brown, Aaron L. and Salvador, Matteo and Shi, Lei and Pfaller, Martin R. and Hu, Zinan and Harold, Kaitlin E. and Hsiai, Tzung and Vedula, Vijay and Marsden, Alison L.},
	month = mar,
	year = {2024},
	pages = {116764},
}

@article{barnafi_comparative_2023,
	title = {A comparative study of scalable multilevel preconditioners for cardiac mechanics},
	volume = {492},
	issn = {0021-9991},
	url = {https://www.sciencedirect.com/science/article/pii/S0021999123005168},
	doi = {10.1016/j.jcp.2023.112421},
	urldate = {2024-10-23},
	journal = {Journal of Computational Physics},
	author = {Barnafi, Nicolás A. and Pavarino, Luca F. and Scacchi, Simone},
	month = nov,
	year = {2023},
	pages = {112421},
}

@article{avazmohammadi_integrated_2017,
	title = {An {Integrated} {Inverse} {Model}-{Experimental} {Approach} to {Determine} {Soft} {Tissue} {Three}-{Dimensional} {Constitutive} {Properties}: {Application} to {Post}-{Infarcted} {Myocardium}},
	volume = {17},
	shorttitle = {An {Integrated} {Inverse} {Model}-{Experimental} {Approach} to {Determine} {Soft} {Tissue} {Three}-{Dimensional} {Constitutive} {Properties}},
	url = {https://pmc.ncbi.nlm.nih.gov/articles/PMC5809201/},
	doi = {10.1007/s10237-017-0943-1},
	language = {en},
	number = {1},
	urldate = {2024-11-19},
	journal = {Biomechanics and modeling in mechanobiology},
	author = {Avazmohammadi, Reza and Li, David S. and Leahy, Thomas and Shih, Elizabeth and Soares, João S. and Gorman, Joseph H. and Gorman, Robert C. and Sacks, Michael S.},
	month = aug,
	year = {2017},
	pages = {31},
}

@article{augustin_computationally_2021,
	title = {A computationally efficient physiologically comprehensive {3D}–{0D} closed-loop model of the heart and circulation},
	volume = {386},
	issn = {0045-7825},
	url = {https://www.sciencedirect.com/science/article/pii/S0045782521004230},
	doi = {10.1016/j.cma.2021.114092},
	urldate = {2024-10-22},
	journal = {Computer Methods in Applied Mechanics and Engineering},
	author = {Augustin, Christoph M. and Gsell, Matthias A. F. and Karabelas, Elias and Willemen, Erik and Prinzen, Frits W. and Lumens, Joost and Vigmond, Edward J. and Plank, Gernot},
	month = dec,
	year = {2021},
	pages = {114092},
}

@article{asner_patient-specific_2017,
	series = {Special {Issue} on {Biological} {Systems} {Dedicated} to {William} {S}. {Klug}},
	title = {Patient-specific modeling for left ventricular mechanics using data-driven boundary energies},
	volume = {314},
	issn = {0045-7825},
	url = {https://www.sciencedirect.com/science/article/pii/S0045782516308672},
	doi = {10.1016/j.cma.2016.08.002},
	urldate = {2024-12-11},
	journal = {Computer Methods in Applied Mechanics and Engineering},
	author = {Asner, L. and Hadjicharalambous, M. and Chabiniok, R. and Peressutti, D. and Sammut, E. and Wong, J. and Carr-White, G. and Razavi, R. and King, A. P. and Smith, N. and Lee, J. and Nordsletten, D.},
	month = feb,
	year = {2017},
	pages = {269--295},
}

@article{valen-sendstad_real-world_2018,
	title = {Real-{World} {Variability} in the {Prediction} of {Intracranial} {Aneurysm} {Wall} {Shear} {Stress}: {The} 2015 {International} {Aneurysm} {CFD} {Challenge}},
	volume = {9},
	issn = {1869-4098},
	shorttitle = {Real-{World} {Variability} in the {Prediction} of {Intracranial} {Aneurysm} {Wall} {Shear} {Stress}},
	doi = {10.1007/s13239-018-00374-2},
	language = {eng},
	number = {4},
	journal = {Cardiovascular Engineering and Technology},
	author = {Valen-Sendstad, Kristian and Bergersen, Aslak W. and Shimogonya, Yuji and Goubergrits, Leonid and Bruening, Jan and Pallares, Jordi and Cito, Salvatore and Piskin, Senol and Pekkan, Kerem and Geers, Arjan J. and Larrabide, Ignacio and Rapaka, Saikiran and Mihalef, Viorel and Fu, Wenyu and Qiao, Aike and Jain, Kartik and Roller, Sabine and Mardal, Kent-Andre and Kamakoti, Ramji and Spirka, Thomas and Ashton, Neil and Revell, Alistair and Aristokleous, Nicolas and Houston, J. Graeme and Tsuji, Masanori and Ishida, Fujimaro and Menon, Prahlad G. and Browne, Leonard D. and Broderick, Stephen and Shojima, Masaaki and Koizumi, Satoshi and Barbour, Michael and Aliseda, Alberto and Morales, Hernán G. and Lefèvre, Thierry and Hodis, Simona and Al-Smadi, Yahia M. and Tran, Justin S. and Marsden, Alison L. and Vaippummadhom, Sreeja and Einstein, G. Albert and Brown, Alistair G. and Debus, Kristian and Niizuma, Kuniyasu and Rashad, Sherif and Sugiyama, Shin-Ichiro and Owais Khan, M. and Updegrove, Adam R. and Shadden, Shawn C. and Cornelissen, Bart M. W. and Majoie, Charles B. L. M. and Berg, Philipp and Saalfield, Sylvia and Kono, Kenichi and Steinman, David A.},
	month = dec,
	year = {2018},
	pages = {544--564},
}

@article{salvador_digital_2024,
	title = {Digital twinning of cardiac electrophysiology for congenital heart disease},
	volume = {21},
	url = {https://royalsocietypublishing.org/doi/full/10.1098/rsif.2023.0729},
	doi = {10.1098/rsif.2023.0729},
	number = {215},
	urldate = {2025-04-08},
	journal = {Journal of The Royal Society Interface},
	publisher = {Royal Society},
	author = {Salvador, Matteo and Kong, Fanwei and Peirlinck, Mathias and Parker, David W. and Chubb, Henry and Dubin, Anne M. and Marsden, Alison L.},
	month = jun,
	year = {2024},
	pages = {20230729},
}

@article{buoso_personalising_2021,
	title = {Personalising left-ventricular biophysical models of the heart using parametric physics-informed neural networks},
	volume = {71},
	issn = {1361-8415},
	url = {https://www.sciencedirect.com/science/article/pii/S1361841521001122},
	doi = {10.1016/j.media.2021.102066},
	urldate = {2024-10-23},
	journal = {Medical Image Analysis},
	author = {Buoso, Stefano and Joyce, Thomas and Kozerke, Sebastian},
	month = jul,
	year = {2021},
	pages = {102066},
}

@article{peirlinck_precision_2021,
	title = {Precision medicine in human heart modeling},
	volume = {20},
	issn = {1617-7940},
	url = {https://doi.org/10.1007/s10237-021-01421-z},
	doi = {10.1007/s10237-021-01421-z},
	language = {en},
	number = {3},
	urldate = {2024-10-22},
	journal = {Biomechanics and Modeling in Mechanobiology},
	author = {Peirlinck, M. and Costabal, F. Sahli and Yao, J. and Guccione, J. M. and Tripathy, S. and Wang, Y. and Ozturk, D. and Segars, P. and Morrison, T. M. and Levine, S. and Kuhl, E.},
	month = jun,
	year = {2021},
	pages = {803--831},
}

@article{miller_implementation_2021,
	title = {An {Implementation} of {Patient}-{Specific} {Biventricular} {Mechanics} {Simulations} {With} a {Deep} {Learning} and {Computational} {Pipeline}},
	volume = {12},
	issn = {1664-042X},
	url = {https://www.frontiersin.org/journals/physiology/articles/10.3389/fphys.2021.716597/full},
	doi = {10.3389/fphys.2021.716597},
	language = {English},
	urldate = {2024-10-22},
	journal = {Frontiers in Physiology},
	publisher = {Frontiers},
	author = {Miller, Renee and Kerfoot, Eric and Mauger, Charlène and Ismail, Tevfik F. and Young, Alistair A. and Nordsletten, David A.},
	month = sep,
	year = {2021},
}

@article{barnafi_reconstructing_2024,
	title = {Reconstructing relaxed configurations in elastic bodies: {Mathematical} formulations and numerical methods for cardiac modeling},
	volume = {423},
	issn = {0045-7825},
	shorttitle = {Reconstructing relaxed configurations in elastic bodies},
	url = {https://www.sciencedirect.com/science/article/pii/S0045782524001014},
	doi = {10.1016/j.cma.2024.116845},
	urldate = {2024-10-23},
	journal = {Computer Methods in Applied Mechanics and Engineering},
	author = {Barnafi, N. A. and Regazzoni, F. and Riccobelli, D.},
	month = apr,
	year = {2024},
	pages = {116845},
}

@article{guan_updated_2023,
	title = {An updated {Lagrangian} constrained mixture model of pathological cardiac growth and remodelling},
	volume = {166},
	issn = {1742-7061},
	url = {https://www.sciencedirect.com/science/article/pii/S1742706123002787},
	doi = {10.1016/j.actbio.2023.05.022},
	urldate = {2025-04-08},
	journal = {Acta Biomaterialia},
	author = {Guan, Debao and Zhuan, Xin and Luo, Xiaoyu and Gao, Hao},
	month = aug,
	year = {2023},
	pages = {375--399},
}

@article{niestrawska_computational_2020,
	title = {Computational modeling of cardiac growth and remodeling in pressure overloaded hearts—{Linking} microstructure to organ phenotype},
	volume = {106},
	issn = {1742-7061},
	url = {https://www.sciencedirect.com/science/article/pii/S1742706120300891},
	doi = {10.1016/j.actbio.2020.02.010},
	urldate = {2024-10-22},
	journal = {Acta Biomaterialia},
	author = {Niestrawska, Justyna A. and Augustin, Christoph M. and Plank, Gernot},
	month = apr,
	year = {2020},
	pages = {34--53},
}

@article{schwarz_fluidsolid-growth_2023,
	series = {A {Special} {Issue} in {Honor} of the {Lifetime} {Achievements} of {T}. {J}. {R}. {Hughes}},
	title = {A fluid–solid-growth solver for cardiovascular modeling},
	volume = {417},
	issn = {0045-7825},
	url = {https://www.sciencedirect.com/science/article/pii/S004578252300436X},
	doi = {10.1016/j.cma.2023.116312},
	urldate = {2025-04-08},
	journal = {Computer Methods in Applied Mechanics and Engineering},
	author = {Schwarz, Erica L. and Pfaller, Martin R. and Szafron, Jason M. and Latorre, Marcos and Lindsey, Stephanie E. and Breuer, Christopher K. and Humphrey, Jay D. and Marsden, Alison L.},
	month = dec,
	year = {2023},
	pages = {116312},
}

@article{ambrosi_growth_2019,
	title = {Growth and remodelling of living tissues: perspectives, challenges and opportunities},
	volume = {16},
	shorttitle = {Growth and remodelling of living tissues},
	url = {https://royalsocietypublishing.org/doi/10.1098/rsif.2019.0233},
	doi = {10.1098/rsif.2019.0233},
	number = {157},
	urldate = {2025-04-08},
	journal = {Journal of The Royal Society Interface},
	publisher = {Royal Society},
	author = {Ambrosi, Davide and Ben Amar, Martine and Cyron, Christian J. and DeSimone, Antonio and Goriely, Alain and Humphrey, Jay D. and Kuhl, Ellen},
	month = aug,
	year = {2019},
	pages = {20190233},
}

@article{hirschvogel_effective_2025,
	title = {Effective block preconditioners for fluid dynamics coupled to reduced models of a non-local nature},
	volume = {435},
	issn = {0045-7825},
	url = {https://www.sciencedirect.com/science/article/pii/S0045782524007953},
	doi = {10.1016/j.cma.2024.117541},
	urldate = {2025-04-05},
	journal = {Computer Methods in Applied Mechanics and Engineering},
	author = {Hirschvogel, Marc and Bonini, Mia and Balmus, Maximilian and Nordsletten, David},
	month = feb,
	year = {2025},
	pages = {117541},
}

@article{regazzoni_stabilization_2025,
	title = {Stabilization of loosely coupled schemes for {0D}–{3D} fluid–structure interaction problems with application to cardiovascular modelling},
	volume = {157},
	issn = {0945-3245},
	url = {https://doi.org/10.1007/s00211-025-01452-z},
	doi = {10.1007/s00211-025-01452-z},
	language = {en},
	number = {1},
	urldate = {2025-04-05},
	journal = {Numerische Mathematik},
	author = {Regazzoni, Francesco},
	month = feb,
	year = {2025},
	pages = {249--306},
}

@article{kung_simulation_2014,
	title = {A {Simulation} {Protocol} for {Exercise} {Physiology} in {Fontan} {Patients} {Using} a {Closed} {Loop} {Lumped}-{Parameter} {Model}},
	volume = {136},
	issn = {0148-0731},
	url = {https://doi.org/10.1115/1.4027271},
	doi = {10.1115/1.4027271},
	number = {081007},
	urldate = {2025-04-05},
	journal = {Journal of Biomechanical Engineering},
	author = {Kung, Ethan and Pennati, Giancarlo and Migliavacca, Francesco and Hsia, Tain-Yen and Figliola, Richard and Marsden, Alison and Giardini, Alessandro and {MOCHA Investigators}},
	month = jun,
	year = {2014},
}

@article{vincent_understanding_2008,
	title = {Understanding cardiac output},
	volume = {12},
	issn = {1364-8535},
	url = {https://www.ncbi.nlm.nih.gov/pmc/articles/PMC2575587/},
	doi = {10.1186/cc6975},
	number = {4},
	urldate = {2025-04-05},
	journal = {Critical Care},
	author = {Vincent, Jean-Louis},
	year = {2008},
	pages = {174},
}

@article{van_de_bruaene_effect_2015,
	title = {Effect of respiration on cardiac filling at rest and during exercise in {Fontan} patients: {A} clinical and computational modeling study},
	volume = {9},
	issn = {2352-9067},
	shorttitle = {Effect of respiration on cardiac filling at rest and during exercise in {Fontan} patients},
	url = {https://www.sciencedirect.com/science/article/pii/S2352906715300257},
	doi = {10.1016/j.ijcha.2015.08.002},
	urldate = {2025-04-05},
	journal = {IJC Heart \& Vasculature},
	author = {Van De Bruaene, Alexander and Claessen, Guido and La Gerche, Andre and Kung, Ethan and Marsden, Alison and De Meester, Pieter and Devroe, Sarah and Bogaert, Jan and Claus, Piet and Heidbuchel, Hein and Budts, Werner and Gewillig, Marc},
	month = dec,
	year = {2015},
	pages = {100--108},
}

@incollection{peiro_reduced_2009,
	title = {Reduced models of the cardiovascular system},
	isbn = {978-88-470-1152-6},
	url = {https://link.springer.com/chapter/10.1007/978-88-470-1152-6_10},
	doi = {10.1007/978-88-470-1152-6_10},
	language = {en},
	urldate = {2025-04-05},
	booktitle = {Cardiovascular {Mathematics}},
	publisher = {Springer, Milano},
	author = {Peiró, Joaquim and Veneziani, Alessandro},
	year = {2009},
	pages = {347--394},
}

\end{document}